\newcommand{\arXiv}{}

\ifdefined\arXiv
   \documentclass[acmsmall,nonacm=true]{acmart}
\else
   \documentclass[acmsmall]{acmart}
\fi

\usepackage{draftwatermark}
\SetWatermarkText{Preprint version}
\SetWatermarkScale{0.6}
\SetWatermarkColor[gray]{0.8}
\usepackage{pgfplots}
\usepackage{graphicx}
\usepackage{booktabs}
\usepackage{listings}
\usepackage{caption}
\usepackage{subcaption}
\usepackage{makecell}
\usepackage{pifont}
\usepackage{forest}
\usepackage{multirow}
\forestset{qtree/.style={for tree={parent anchor=south, 
           child anchor=north,align=center,inner sep=0pt}}}
\usepackage{setspace}

\newcommand{\cmark}{\ding{51}}%
\newcommand{\xmark}{\ding{55}}%

\AtBeginDocument{%
  \providecommand\BibTeX{{%
    \normalfont B\kern-0.5em{\scshape i\kern-0.25em b}\kern-0.8em\TeX}}}

\setcopyright{acmcopyright}
\copyrightyear{2024}
\acmYear{2024}
\acmDOI{XXXXXXX.XXXXXXX}

\acmJournal{CSUR}

\pgfplotsset{compat=1.18}
\begin{document}

\title{A Survey on Deep Learning Hardware Accelerators for Heterogeneous HPC Platforms}

\author{Cristina Silvano} 
\email{cristina.silvano@polimi.it}
\orcid{0000-0003-1668-0883}
\author{Daniele Ielmini}
\email{daniele.ielmini@polimi.it}
\orcid{0000-0002-1853-1614}
\author{Fabrizio Ferrandi}
\email{fabrizio.ferrandi@polimi.it}
\orcid{0000-0003-0301-4419}
\author{Leandro Fiorin}
\email{leandro.fiorin@polimi.it}
\orcid{0000-0003-2630-1509}
\author{Serena Curzel}
\email{serena.curzel@polimi.it}
\orcid{0000-0002-8202-1627}
\affiliation{%
  \institution{Politecnico di Milano}
  \country{Italy}
}

\author{Luca Benini}
\email{luca.benini@unibo.it}
\orcid{0000-0001-8068-3806}
\author{Francesco Conti}
\email{f.conti@unibo.it}
\orcid{0000-0002-7924-933X}
\author{Angelo Garofalo}
\email{angelo.garofalo@unibo.it}
\orcid{0000-0002-7495-6895}
\affiliation{
  \institution{Università di Bologna}
  \country{Italy}
}

\author{Cristian Zambelli}
\email{cristian.zambelli@unife.it}
\orcid{0000-0001-8755-0504}
\author{Enrico Calore}
\email{enrico.calore@fe.infn.it}
\orcid{0000-0002-2301-3838}
\author{Sebastiano Fabio Schifano}
\email{sebastiano.fabio.schifano@unife.it}
\orcid{0000-0002-0132-9196}
\affiliation{
  \institution{Università degli Studi di Ferrara}
  \country{Italy}
}

\author{Maurizio Palesi}
\email{maurizio.palesi@unict.it}
\orcid{0000-0003-3129-0664}
\author{Giuseppe Ascia}
\email{giuseppe.ascia@unict.it}
\orcid{0000-0001-7452-5828}
\author{Davide Patti}
\email{davide.patti@unict.it}
\orcid{0000-0003-0874-7793}
\affiliation{
  \institution{Università degli Studi di Catania}
  \country{Italy}
}

\author{Nicola Petra}
\email{nicola.petra@unina.it}
\orcid{0000-0001-7167-9530}
\author{Davide De Caro}
\email{dadecaro@unina.it}
\orcid{0000-0003-0204-0949}
\affiliation{%
  \institution{Universit\`a degli Studi di Napoli Federico II}
  \country{Italy}
}

\author{Luciano Lavagno}
\email{luciano.lavagno@polito.it}
\orcid{0000-0002-9762-6522}
\author{Teodoro Urso}
\email{teodoro.urso@polito.it}
\orcid{0009-0005-4366-1102}
\affiliation{
    \institution{Politecnico di Torino}
    \country{Italy}
}

\author{Valeria Cardellini}
\email{cardellini@ing.uniroma2.it}
\orcid{0000-0002-6870-7083}
\author{Gian Carlo Cardarilli}
\email{g.cardarilli@uniroma2.it}
\orcid{0000-0002-7444-876X}
\affiliation{
    \institution{Università di Roma ``Tor Vergata''}
    \country{Italy}
}

\author{Robert Birke}
\email{robert.birke@unito.it}
\orcid{0000-0003-1144-3707}
\affiliation{
  \institution{Universit\`a degli Studi di Torino}
  \country{Italy}
}

\author{Stefania Perri}
\email{s.perri@unical.it}
\orcid{0000-0003-1363-9201}
\affiliation{%
  \institution{Universit\`a degli Studi della Calabria}
  \country{Italy}
}

\renewcommand{\shortauthors}{Cristina Silvano et al.}

\begin{abstract}

Recent trends in deep learning (DL) have made hardware accelerators essential for various high-performance computing (HPC) applications, including image classification, computer vision, and speech recognition.
This survey summarizes and classifies the most recent developments in DL accelerators, focusing on their role in meeting the performance demands of HPC applications. We explore cutting-edge approaches to DL acceleration, covering not only GPU- and TPU-based platforms but also specialized hardware such as FPGA- and ASIC-based accelerators, Neural Processing Units, open hardware RISC-V-based accelerators, and co-processors. This survey also describes accelerators leveraging emerging memory technologies and computing paradigms, including 3D-stacked Processor-In-Memory, non-volatile memories like Resistive RAM and Phase Change Memories used for in-memory computing, as well as Neuromorphic Processing Units, and Multi-Chip Module-based accelerators. Furthermore, we provide insights into emerging quantum-based accelerators and photonics. Finally, this survey categorizes the most influential architectures and technologies from recent years, offering readers a comprehensive perspective on the rapidly evolving field of deep learning acceleration. 
  

\end{abstract}

\ccsdesc[500]{Computer systems organization~Architectures}
\ccsdesc[500]{Hardware~Reconfigurable logic and FPGAs}
\ccsdesc[500]{Hardware~Emerging technologies}
\ccsdesc[500]{Hardware~Very large scale integration design}
\ccsdesc[300]{Hardware~Power and energy}
\ccsdesc[500]{Computing methodologies~Machine learning}

\keywords{Hardware Accelerators, High-Performance Computing, Deep Learning, Deep Neural Networks, Emerging Memory Technologies.}



\maketitle

\section{Introduction}
\label{sec:introduction}

With the advent of the Exascale era, we have witnessed a growing convergence between High-Performance Computing (HPC) and Artificial Intelligence (AI). 
The increasing computing power of HPC systems, combined with their ability to manage vast amounts of data, has driven the development of more and more sophisticated machine learning (ML) techniques. 
Deep Learning (DL), a subset of ML, utilizes Deep Neural Networks (DNNs) with multiple layers of artificial neurons to mimic the human brain behavior by learning from large datasets.
Thanks to advancements in technology and system architecture, HPC nodes now integrate not only an increasing number of high-end parallel processors but also specialized co-processors such as Graphics Processing Units (GPUs) and vector/tensor computing units.
This supercomputing power has significantly accelerated both the training and inference phases of DNN models used in several application scenarios.
The introduction of the pioneering AlexNet~\cite{Krizhevsky2012} model at the ImageNet challenge in 2012 marked a turning point, demonstrating the power of GPU acceleration in deep learning.
Since then, numerous DNN models have been developed for various tasks including image recognition and classification, Natural Language Processing (NLP), and Generative AI. These applications demand specialized \emph{hardware accelerators}, to efficiently handle the heavy computational workload of DNN algorithms.
Today, DL accelerators are deployed across a wide range of computing systems spanning from ultra-low-power resource-constrained devices to high-performance servers, HPC infrastructures, and large-scale data centers.

\textbf{Scope of the survey.} 
This survey is an attempt to provide an extensive overview of the most influential architectures to accelerate Deep Learning 
for high-performance applications. 
The survey highlights various approaches that support DL acceleration including GPU-based accelerators, Tensor Processor Units, FPGA-based accelerators, and ASIC-based accelerators, such as Neural Processing Units and specialized co-processors based on the open-hardware RISC-V architecture. 
The survey also includes accelerators based on emerging technologies and computing paradigms, such as 3D-stacked 
PIM, emerging non-volatile memories such as the Resistive Random Access Memory (RRAM) and the Phase Change Memory (PCM), Neuromorphic Processing Units, and Multi-Chip Modules.

Overall, we have reviewed the research on DL accelerators from the past two decades, covering a significant period of literature in this field. 
Being DL acceleration a prolific and rapidly evolving field, we do not claim to cover exhaustively all the research works appeared so far, but we focused on the most influential contributions.
Moreover, this survey can be leveraged as a connecting point for some previous surveys on AI and DL accelerators  ~\cite{chen2020engineering, Hassanpour2022, gao2023acm, rathi2023acm} and other surveys focused on some more specific aspects of DL, including the architecture-oriented optimization of sparse matrices ~\cite{reuther_hpec22} and the Neural Architecture Search ~\cite{Chitty2022ACMSUR}. Another research trend in state-of-the-art AI architecture design addresses transformer models. A recent survey on the full stack of optimizations on transformer inference has recently been published in \cite{kim2023stackoptimizationtransformerinference}. 

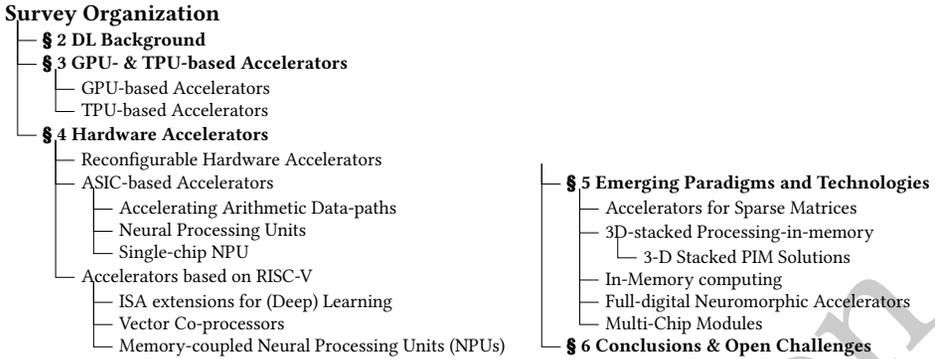
\begin{figure}[t!]
\centering
\small
\subfloat{
\label{fig:contribution_left}
\scalebox{0.75}{
\begin{forest}
  dir tree
  [\textbf{\large{Survey Organization}}
    [\textbf{\pmb{\S}~\ref{sec:introduction} Introduction}
    ]
    [\textbf{\pmb{\S}~\ref{sec:background} DL Background}
    ]    
    [\textbf{\pmb{\S}~\ref{sec:gpu_tpu} GPU- \& TPU-based Accelerators}
     [GPU-based accelerators
      ]
      [TPU-based accelerators
      ]
    ]
    [\textbf{\pmb{\S}~\ref{sec:hw_accelerators} Hardware Accelerators}
      [Reconfigurable Hardware Accelerators
      ]
      [ASIC-based Accelerators]
    ]
]
\end{forest}}}
\subfloat{
\label{fig:contribution_right}
\scalebox{0.75}{
\begin{forest}
    dir tree
    [
    [
      [Accelerators for Sparse Matrices
      ]
      [Accelerators based on open-hardware RISC-V 
       ]
    ]
    [\textbf{\pmb{\S}~\ref{sec:emerging} Accelerators based on Emerging Technologies}
      [Processing-in-Memory 
      ]
      [In-Memory Computing
      ]
      [Neuromorphic Accelerators
      ] 
      [Multi-Chip Modules
      ] 
      [Quantum and Photonic Computing
      ]
    ]
    [\textbf{\pmb{\S}~\ref{sec:conclusion} Conclusions}
    ]
    ]
\end{forest}}}
\caption{Organization of the survey}
\label{fig:contribution}
\end{figure}

\textbf{Organization of the survey.} The survey is structured in different sections and sub-sections belonging to the areas of computer architecture and hardware design, as shown in Figure~\ref{fig:contribution}, 
To this aim, we organized the material in a way that all research papers corresponding to multiple types of sections are cited under each section.
Moreover, for each section, we have selectively chosen the most notable and influential works and, for each work, we focused on its most innovative contributions.

To conclude, we hope this survey could be useful for a wide range of readers, including computer architects, hardware developers, HPC engineers, researchers, and technical professionals. A major effort was spent to use a clear and concise technical writing style: we hope this effort could be useful in particular to the young generations of master's and Ph.D. students. To facilitate the reading, a list of acronyms is reported in Table \ref{tab:acronyms}.

\forestset{
  dir tree/.style={
    for tree={
      parent anchor=south west,
      child anchor=west,
      anchor=mid west,
      inner ysep=-3.5pt,
      grow'=0,
      align=left,
      edge path={
        \noexpand\path [draw, \forestoption{edge}] (!u.parent anchor) ++(1em,0) |- (.child anchor)\forestoption{edge label};
      },
      if n children=0{}{
        delay={
          prepend={[,phantom, calign with current]}
        }
      },
      fit=rectangle,
      before computing xy={
        l=2em
      }
    },
  }
}

\begin{table}[b!]
\caption{List of acronyms}
\tiny
{
\resizebox{13.8cm}{!}{%
\begin{tabular}{@{}lll@{}}
\toprule
\textbf{Acronym} & \textbf{Acronym} & \textbf{Acronym}\\ \midrule
\textbf{AI:} Artificial Intelligence & 
\textbf{ASIC:} Application Specific Integrated Circuit &
\textbf{BRAM:} Block Random Access Memory \\
\textbf{CMOS:} Complementary Metal Oxide Semiconductor & 
\textbf{CNN:} Convolutional Neural Network & 
\textbf{CPU:} Central Processing Unit \\ 
\textbf{DL:} Deep Learning &
\textbf{DP:} Double Precision  &
\textbf{DNN:} Deep Neural Network \\ 
\textbf{DRAM:} Dynamic Random Access Memory &
\textbf{EDA:} Electronic Design Automation &
\textbf{FLOPS:} Floating Point Operations per Second \\ 
\textbf{FMA:} Fused Multiply-Add & 
\textbf{FPGA:} Field-Programmable Gate Array & 
\textbf{GEMM:} General Matrix Multiply \\ 
\textbf{GP-GPU:} General-Purpose Graphics Processing Unit &
\textbf{GPU:} Graphics Processing Unit &
\textbf{HBM:} High Bandwidth Memory \\
\textbf{HDL:} Hardware Description Language &
\textbf{HLS:} High Level Synthesis &
\textbf{HMC:} Hybrid Memory Cube \\
\textbf{HPC:} High-Performance Computing &
\textbf{MLP:} Multi-Layer Perceptron &
\textbf{NPU:} Neural Processing Unit \\
\textbf{IMC:} In-Memory Computing & 
\textbf{IoT:} Internet of Things &
\textbf{ISA:} Instruction Set Architecture \\
\textbf{MCM:} Multi-Chip Module &
\textbf{ML:} Machine Learning &
\textbf{PCM:} Phase Change Memory \\
\textbf{PIM:} Processing In-Memory &
\textbf{QC:} Quantum Computing &
\textbf{QNN:} Quantized Neural Network \\ 
\textbf{QPU:} Quantum Processing Unit &
\textbf{RAM:} Random Access Memory &
\textbf{RRAM:} Resistive RAM \\
\textbf{RISC:} Reduced Instruction Set Computer &
\textbf{RNN:} Recurrent Neural Network & 
\textbf{SNN:} Spiking Neural Network \\
\textbf{SoC:} System on Chip & 
\textbf{SRAM:} Static Random Access Memory &
\textbf{TPU:} Tensor Processing Unit \\
\bottomrule
\end{tabular}
}
}
\label{tab:acronyms}
\end{table}
\section{Deep Learning Background}
\label{sec:background}
Deep Learning \cite{lecun2015,Schmidhuber_2015} is a subset of ML methods that can automatically discover the representations needed for feature detection or classification from large data sets, by employing multiple layers of processing to extract progressively higher-level features. 
The most recent works in literature clearly show that two main DL topologies have emerged as dominant: DNNs and Transformers.

Concerning DNNs, there are three types of DNNs mostly used today: Multi-Layer Perceptrons (MLPs), Convolutional Neural Networks (CNNs), and  Recurrent Neural Networks (RNNs).
MLPs \cite{Rosenblatt1957} are feed-forward ANNs composed of a series of fully connected layers, where each layer is a set of nonlinear functions of a weighted sum of all outputs of the previous one. On the contrary, in a CNN \cite{Lecun1998}, a convolutional layer extracts the simple features from the inputs by executing convolution operations. Each layer is a set of nonlinear functions of weighted sums of different subsets of outputs from the previous layer, with each subset sharing the same weights. Each convolutional layer in the model can capture a different high-level representation of input data, allowing the system to automatically extract the features of the inputs to complete a specific task, e.g., image classification, face authentication, and image semantic segmentation. Finally, RNNs \cite{Schmidhuber_2015} address the time-series problem of sequential input data. Each RNN layer is a collection of nonlinear functions of weighted sums of the outputs of the previous layer and the previous state, calculated when processing the previous samples and stored in the RNN’s internal memory. RNN models are widely used in NLP for natural language modeling, word embedding, and machine translation. More details on concepts and terminology related to DNNs are provided in Section A.1 of Appendix A. 



Each type of DNN is especially effective for a specific subset of cognitive applications. Depending on the target application, and the resource constraints of the computing system, different DNN models have been deployed. 
Besides DNNs, Transformer-based models~\cite{vaswaniAttentionAllYou2017} recently captured a great deal of attention.
Transformers were originally proposed for NLP~\cite{vaswaniAttentionAllYou2017}, and are designed to recognize long-distance dependencies between data by  \emph{attention layers}, where the weights used to linearly transform input data are computed dynamically based on the input data itself.
%
While DNNs use convolutional layers to perform “local” operations on small portions of the input, Transformers use attention layers to perform  “global” operations on the whole input. Although quite different,
DNNs and Transformers share many underlying principles (such as gradient descent training, and reliance on linear algebra), and many of the DL-dedicated architectures described in this survey address both types of topologies.
\section{GPU- and TPU-based accelerators}
\label{sec:gpu_tpu}

\subsection{GPU-based accelerators}

Graphics Processing Units (GPUs) are specific-purpose processors introduced to compute efficiently graphics-related tasks, such as 3D rendering.
They became widely used since the nineties as co-processors, working alongside Central Processing Units (CPUs) to offload graphics-related computations.
The introduction of programmable shaders into GPU architectures increased their flexibility paving the way for their adoption to perform general-purpose computations. Despite being specifically designed for computer graphics, their highly parallel architecture is well suited to tackle a wide range of applications.
Consequently, in the early 2000s, GPUs started to be used to accelerate data-parallel computations not necessarily related to graphics. This practice is commonly referred to as General-Purpose computing on Graphics Processing Units GPUs (GP-GPU) and started to be increasingly popular in the early 2010s with the advent of the CUDA language.
The technological development of the last ten years significantly increased the computing power of GPUs, which, due to their highly parallel nature, are incidentally very well suited to accelerate neural network training algorithms. The availability of such computing power allowed more complex neural network models to become practically usable, fostering the development of DNNs. 

The impressive results obtainable with DNNs, followed by significant investments in this market sector, induced hardware manufacturers to modify GPU architectures in order to be even more optimized to compute such workloads, as an example implementing the support for lower-precision computations. This led to a de-facto co-design of GPU architectures and neural network algorithms implementations, which is nowadays significantly boosting the performance, accuracy, and energy efficiency of AI applications. The basic features of GPU architectures able to boost the performance of HPC and DL applications are briefly reviewed in Section A.2 of the Appendix A.

GPUs can execute multiple, simultaneous computations. This enables the distribution of training processes and can significantly speed up ML operations. With GPUs, it is possible to cumulate many cores that use fewer resources without sacrificing neither efficiency nor power.

The performance of GPU accelerators could be compared in different ways. 
As a first approximation, their theoretical peak performance and memory bandwidth could be used.
Anyhow several other architectural characteristics could affect the final performance of actual algorithm implementation. To get a better overview of their expected performance, running a specific workload, it could be preferable to use reference benchmarks, possibly made of representative sets of commonly used algorithm implementations.
For this reason, different benchmarks have been developed, each of them able to test the obtainable performance concerning a given workload characteristic, or a given set of application kernels. In the context of ML, one of the most used benchmarks is MLPerf~\cite{mlperf}, which has a specific set of training phase tasks~\cite{mlperf-training}.
Its results on two different systems, embedding the latest GPU architecture and its predecessor (i.e., NVIDIA Hopper and Ampere) are shown in Table~\ref{gpu-mlperf-train}, highlighting on average an approximate $2\times$ factor of performance improvement.
Different vendors, like AMD and Intel, have also developed GP-GPU architectures mostly oriented to HPC and more recently to AI computing. Yet the terminology used by different vendors is not the same, they share most of the hardware details. For example, AMD names Compute Unit which NVIDIA calls Streaming Multiprocessor, and Intel calls Compute Slice or Execution-Unite (EU). Further, NVIDIA names Warp the set of instructions scheduled and executed at each cycle, while AMD uses the term Wavefront, and Intel uses the term EU-Thread. Concerning the execution model, NVIDIA uses the Single Instruction Multiple Thread (SIMT), while AMD and Intel use the Single Instruction Multiple Data (SIMD)~\cite{khairy2019}. In Table~\ref{tab:differentvendors}, we report the main hardware features of the three most recent GP-GPU architectures developed by NVIDIA~\cite{h100}, AMD~\cite{mi250x} and Intel~\cite{arc770}. We compare the peak performance related to the 32-bit single- and 64-bit double-precision, and the peak performance achieved using half-precision.
\begin{table}
\caption{MLPerf Training v2.1 Benchmark Results (minutes)}\label{gpu-mlperf-train}
\centering
\resizebox{0.8\textwidth}{!}{
\begin{tabular}{lllllllll}
\toprule
 & ImageNet & KiTS19   & OpenImages & COCO       & LibriSpeech	& Wikipedia & Go \\
 & ResNet   & 3D U-Net & RetinaNet  & Mask R-CNN & RNN-T       & BERT      & Minigo \\
\midrule
8 $\times$ A100 &  30.8 & 25.6 & 89.1 & 43.1 & 32.5 & 24.2 & 161.6 \\
8 $\times$ H100 &  14.7 & 13.1 & 38.0 & 20.3 & 18.2 & 6.4 & 174.6 \\
\bottomrule
\end{tabular}
}
\vskip -0.5cm
\end{table}
\begin{table}[b]
\caption{Selected features of the most recent GP-GPU systems.}
\centering 
\resizebox{0.8\textwidth}{!}{
\begin{tabular}{l r r r} 
\toprule 
Model          & NVIDIA H100  & AMD Instinct MI250X & Intel Arc 770 \\  
\midrule 
Clock [GHz]           & 1.6            & 1.7              & 2.4        \\ 
Peak Performance in Double Precision [TFLOPS]& 30             & 47.9             & 4.9        \\  
Peak Performance in Single Precision [TFLOPS]& 60             & 95.8             & 19.7       \\  
Peak Performance in FP16 [TFLOPS]  & 120            & 383              & 39.3       \\  
Max Memory [GB]       &  80 HBM2e      & 128GB HBM2e      & 16GB GDDR6 \\ 
Mem BW [TB/s]         & 2.0            & 3.2              & 0.56       \\ 
Thermal Design Power (TDP) [Watt]      & 350            & 560              & 225        \\   
\midrule
\end{tabular} 
}
\label{tab:differentvendors}
\end{table}
The comparison evidences that the higher the memory bandwidth provided to sustain DL workloads like model training, the higher the power consumption. Further, a huge number of parallel resources intended as physical cores are mandatory to achieve high computing performance to reduce model training time, however at the expense of reduced energy efficiency.

\subsection{TPU-based accelerators}
Tensor Processing Units (TPUs) dedicated to training and inference have been proposed very early after the emergence of the first large CNN-based applications. This is due to the observation that these workloads are dominated by linear algebra kernels that can be refactored as matrix multiplications (particularly if performed in batches) and that their acceleration is particularly desirable for high-margin applications in data centers. More recently, the emergence of exponentially larger models with each passing year (e.g., the GPT-2, GPT-3, GPT-4 Transformer-based large language models) required a continuous investment in higher-performance training architectures.

Google showcased the first TPU~\cite{Jouppi_2017,jouppiMotivationEvaluationFirst2018} at ISCA in 2017, but according to the original paper the first deployment occurred in 2015 -- just three years after the ``AlexNet revolution''. Their last TPU v4 implementation outperforms the previous TPU v3 by 2.1x and improves performance/Watt by 2.7x \cite{jouppi2023isca}.
The architecture of the TPU presented in ~\cite{Jouppi_2017,jouppiMotivationEvaluationFirst2018} is centered on a large (256$\times$256) systolic array operating on 8-bit integers and targeting exclusively data center inference applications; this is coupled with a large amount of on-chip SRAM for activations (24 MiB) and a high-bandwidth (30 GiB/s) dedicated path to off-chip L3 DRAM for weights.
The next design iterations (TPUv2, TPUv3)~\cite{jouppiDomainspecificSupercomputerTraining2020} forced to move from an inference-oriented design to a more general engine tuned for both inference and training, employing the 16-bit BF16 floating-point format, more cores (2 per chip) using each one or two 4$\times$ smaller arrays than TPUv1 (128$\times$128, to reduce under-usage inefficiencies). TPUv2/v3 also introduced high-bandwidth memory support, which results in more than 20$\times$ increase in the available off-chip memory bandwidth.
In 2019, Habana Labs and Intel proposed Goya and Gaudi as microarchitectures for the acceleration of inference \cite{medina2019hotchips}. 
Goya \cite{medina2019hotchips} relies on PCIe 4.0 to interface to a host processor and exploits a heterogenous design approach comprising of a large General Matrix Multiply (GEMM) engine, TPUs, and a large shared DDR4 memory pool. Each TPU also incorporates its local memory that can be either hardware-managed or fully software-managed, allowing the compiler to optimize the residency of data and reduce movement. Each of the individual TPUs is a VLIW design optimized for AI applications and especially for training. The TPU supports mixed-precision operations including 8-bit, 16-bit, and 32-bit SIMD vector operations for both integer and floating-point. Gaudi has an enhanced version of the TPUs and uses HBM global memories rather than the DDR used in Goya, increasing the support towards bfloat16 data types and including more operations and functionalities dedicated to training operations. 
While Google and Intel rely on a mixture of in-house designs and GPUs, the other main data center providers typically rely on NVIDIA GPUs
to serve 
DL workloads.
Starting from the Volta architecture~\cite{choquetteVoltaPerformanceProgrammability2018} and continuing with Ampere~\cite{choquetteA100DatacenterGPU2021} and Hopper~\cite{choquetteNVIDIAHopperH1002023,elsterNvidiaHopperGPU2022}, NVIDIA has embedded inside the GPU Streaming Multiprocessors the counterpart of smaller TPUs, i.e., \textit{TensorCores}.

GraphCore Colossus Mk1 and Mk2 IPUs~\cite{knowlesGraphcore2021,jiaDissectingGraphcoreIPU2019} target 
the GNNs, DNNs, and Transformers training employing a tiled many-core architecture.
GraphCore focuses on a high power- and cost-efficient memory hierarchy that does not rely on high-bandwidth off-chip HBM, but on cheaper DRAM chips combined with a large amount of on-chip SRAM (in the order of 1 GiB per chip). 

IBM Research focused on reducing the data precision used for training~\cite{agrawal20219,venkataramaniRaPiDAIAccelerator2021}, by introducing Hybrid-FP8 formats in training ASICs and tensor processors.
Further improvements were achieved with Cambricon-Q~\cite{zhaoCambriconQHybridArchitecture2021}, which exploits the statistical properties of tensors to minimize bandwidth and maximize efficiency.
Finally, Gemmini~\cite{gonzalez16mm106GOPS2021,gencGemminiEnablingSystematic2021} and RedMulE~\cite{tortorellaRedMulECompactFP162022a,tortorellaRedMuleMixedPrecisionMatrixMatrix2023} introduce tensor processor hardware IPs (respectively, generated from a template and hand-tuned) that can be integrated inside System-on-Chips, similarly to what NVIDIA does with TensorCores.
Further details on TPU architectures are provided in Section A.2 of the Appendix A.

\section{Hardware Accelerators}
\label{sec:hw_accelerators}

Typical HPC workloads, like genomics, astrophysics, finance, and cyber security, require the elaboration of massive amounts of data and they can take advantage of DL methods with results that can surpass human ability \cite{Bengio_2009, Schmidhuber_2015, Goodfellow_2016,SPAGNOLO20201}.
However, an ever-increasing computing power, a rapid change of the data analysis approaches, and the introduction of novel computational paradigms are needed.
DL models can be efficiently supported by optimized hardware platforms providing high levels of parallelism and a considerable amount of memory resources.
These platforms can be developed using CPUs, GPUs, FPGAs, and ASICs \cite{Goodfellow_2016, DHILLESWARARAO_2022, Sze_2017, Machupalli_2022, Du_2018, Wang_2019, Leibo2019ACMSUR}.

\begin{figure}[t]
 \centering
 \includegraphics[width=0.75\columnwidth]{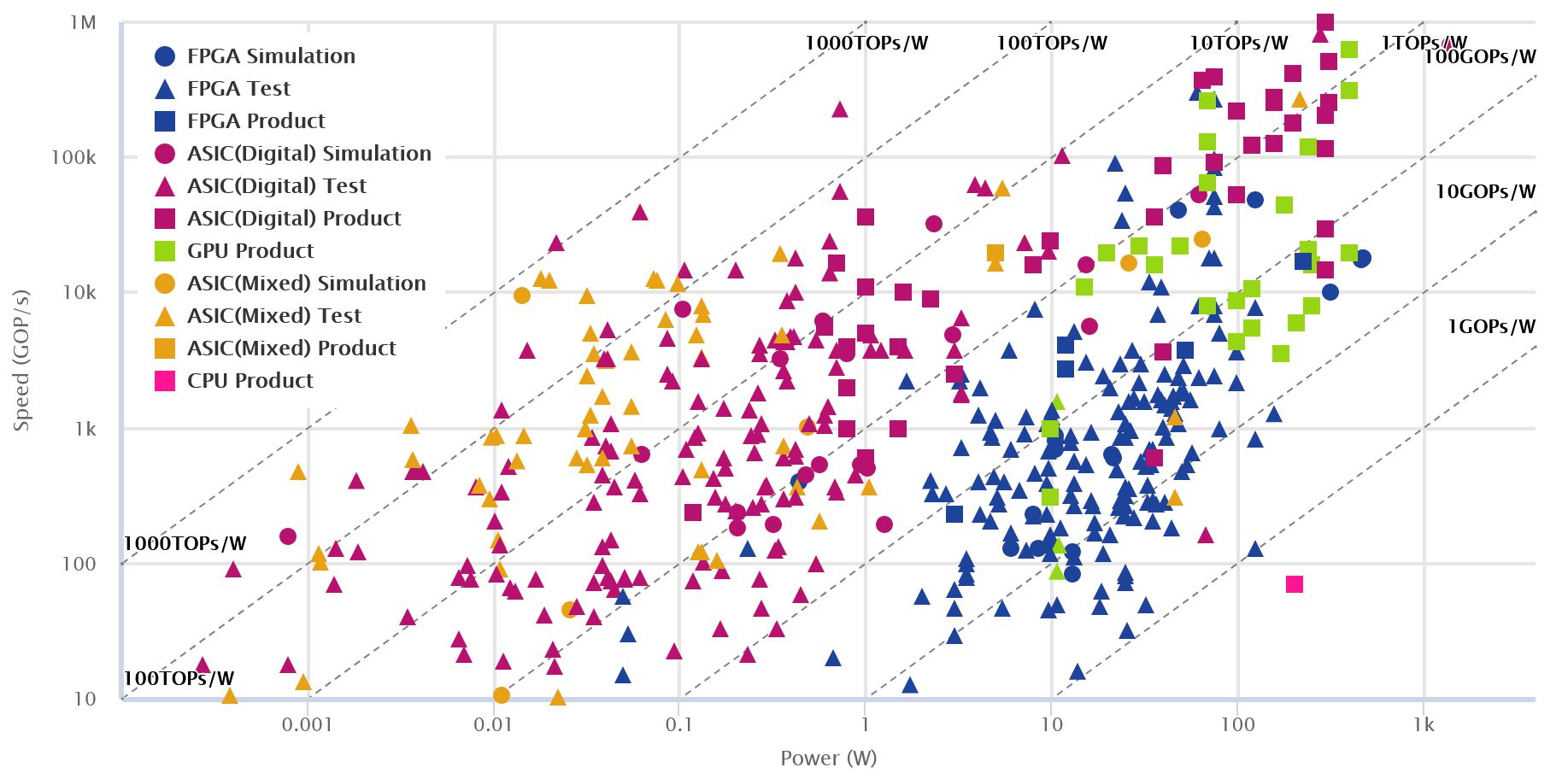}
 \caption{Overview on state-of-the-art Neural Network accelerators based on available data collected in \cite{Guo_Online}. Legenda: \textit{Simulation} means GOPS/W values collected from post-layout simulation; \textit{Test} means from prototype devices; \textit{Product} means from off-the-shelf devices.}
\label{fig:graph_accel}
\end{figure}

Figure~\ref{fig:graph_accel} presents a comparison of state-of-the-art architectures in terms of speed \textit{(Giga Operations per Second)} versus power consumption \textit{(Watt).} The diagonal dashed lines represent energy efficiency levels in \textit{(GOPS/W)} and \textit{TOPS/W} (Tera Operations per Second per Watt), with higher slopes indicating better energy efficiency.
The most energy-efficient devices are ASICs and GPUs clustering in the high range of energy efficiency (1–100 TOPS/W) and mainly located in the top right region characterized by the highest computational throughput. Powerful GPUs are generally preferred for heavier tasks like training and running large, complex models built on large datasets.
Conversely, FPGAs are well suited to accelerate specific inference tasks that privilege lower power consumption over processing speeds. From the figure, FPGAs (represented in blue) are well distributed across different power and performance levels, mostly clustering in the lower-to-mid range of energy efficiency (from 10 GOPS/W to 1 TOPS/W).



\begin{figure}[t]
  \centering
    \begin{subfigure}[c]{0.325\textwidth}
        \centering
        \includegraphics[width=\textwidth]{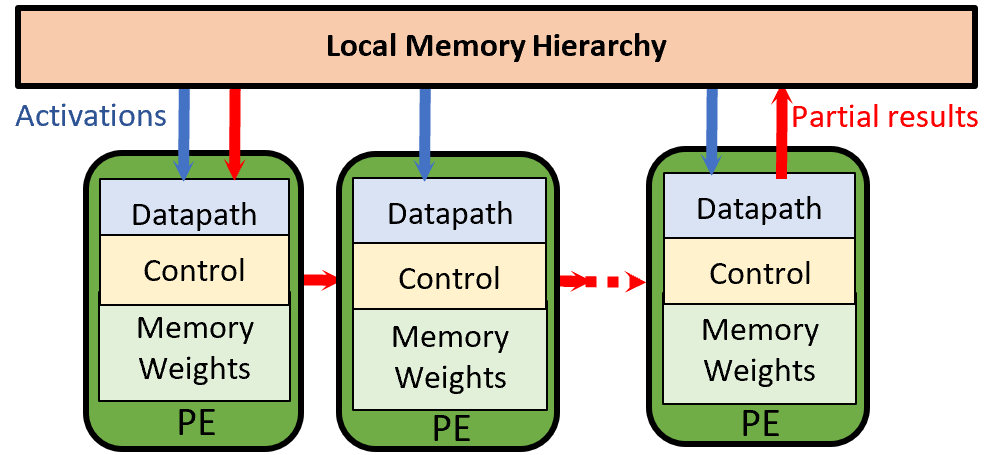}
         \caption{}
    \end{subfigure}
    \begin{subfigure}[c]{0.325\textwidth}
        \centering
        \includegraphics[width=\textwidth]{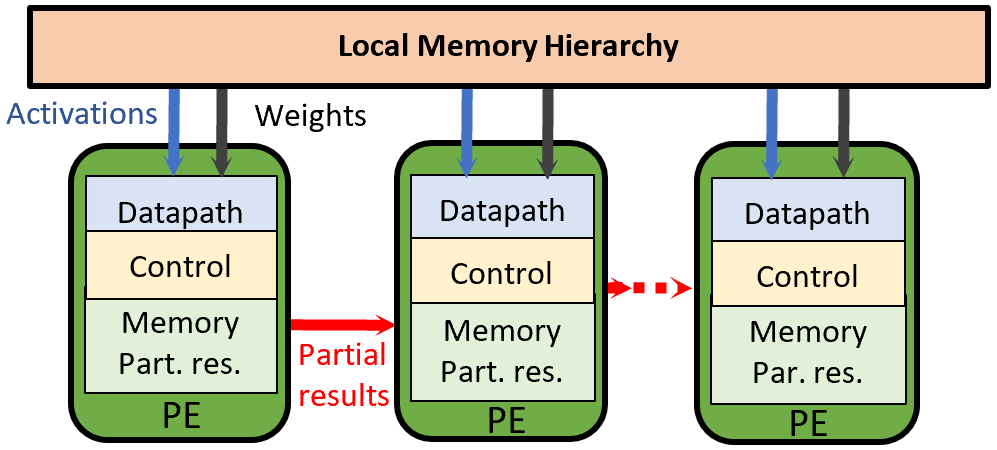}
         \caption{}
    \end{subfigure}
    \begin{subfigure}[c]{0.325\textwidth}
         \centering
         \includegraphics[width=\textwidth]{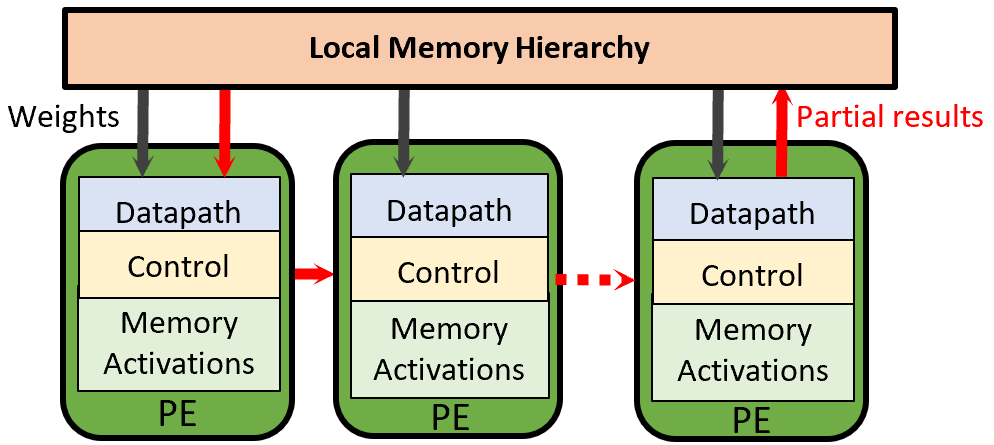}
         \caption{}
    \end{subfigure}
    \caption{Dataflows in DL accelerators: (a) Weights stationary; (b) Output stationary; (c) Input stationary.}
    \label{fig:FPGA_Dataflow}
\end{figure}

Independently of the technology, a common problem in the design of accelerators is the energy and latency cost of accessing the off-chip DRAM memory, in particular considering the significant amount of data that HPC applications need to process. As sketched in Figure~\ref{fig:FPGA_Dataflow}, different data reuse and stationary strategies can be exploited to reduce the number of accesses \cite{Park2015, Peemen2013, Sankaradas2009, Sriam2011, Cavigelli2017, Gupta2015}.
In weight stationary dataflow, convolutional weights 
are fixed and stored in the local memory of the Processing Elements (PEs) and reused on input activations uploaded step-by-step from the external DRAM.
Conversely, in output stationary dataflow, partial outputs from PEs are stored locally and reused step-by-step until the computation is completed. Then, just the final results are moved to the external DRAM.
An efficient alternative is input stationary dataflow: input activations are stored in the PEs' local memory, while weights are uploaded from the external DRAM and sent to the PEs.
Another approach common to many accelerators is the introduction of \emph{quantization} to reduce the data type width \cite{quant2021, Liu2022}.
Integer or fixed-point data formats are generally preferred to the more computationally intensive floating-point ones. This guarantees better memory occupation and lower computational cost \cite{jin2022fnet}.
Extreme quantization techniques that use only one bit for the data stored (Binary Neural Networks \cite{BNN2020}) are widely used for very large networks but to get a comparable accuracy they require 2-11$\times$ the number of parameters and operations \cite{Umuroglu2017}, making them not suitable for complex problems.

\subsection{Reconfigurable Hardware Accelerators}




FPGAs and Coarse-Grained Reconfigurable Arrays (CGRAs) are highly sought-after solutions to hardware accelerate a wide range of applications. The main feature of such platforms is the ability to support different computational requirements by repurposing the underlying hardware accelerators at deploy-time and also at runtime.
More details on FPGA technologies and related EDA frameworks are respectively provided in Sections A.3 and A.4 of Appendix A.
Several FPGA-based hardware accelerators for DL are structured as heterogeneous embedded systems \cite{Ma2018,Yazdanbakhsh2018,Aimar2019,Perri2020,Li2021} that mainly consist of a general-purpose processor, for running the software workload; a computational module, designed to speed up common DL operators, like convolutions \cite{Qiu2016,Venieris2019}, de-convolutions \cite{Chang2020, Sestito2021}, pooling, fully connected operations, activation, and softmax functions \cite{Spagnolo2022_1, Spagnolo2022_2}; and a memory hierarchy to optimize data movement to/from the external DRAM to store data to be processed and computational results. A typical approach to accelerate convolutions consists of a systolic array architecture (SA), a regular pattern that can be easily replicated \cite{Xuechao2017}. 
Each PE in the array is a SIMD vector accumulation module where inputs and weights are supplied at each cycle by shifting them from the horizontally and vertically adjacent PEs. 
The use of pipelined groups of PEs with short local communication and regular architecture enables a high clock frequency and limited global data transfer.
Although FPGAs have traditionally been proposed as accelerators for edge applications, they are starting to be adopted also in data centers.
Microsoft's Project Brainwave \cite{Brainwave} uses several FPGA boards to accelerate the execution of RNNs in the cloud, exploiting the reconfigurability to adapt the platform to different DL models.
One way to face the limitations imposed by the capability of FPGAs to effectively map very large DL models is to use a deeply pipelined multi-FPGA design. Recent studies focus on optimizing this type of architecture and maximizing the overall throughput \cite{Zhang2016,Rahman2017,Junnan2019}.
In these application contexts, CGRAs represent an alternative to FPGAs, providing reconfigurability with coarser-grained functional units.
They are based on an array of PEs, performing the basic arithmetic, logic, and memory operations at the word level and using a small register file as temporary data storage. 
Neighboring PEs are connected through reconfigurable routing that allows transferring intermediate results of the computations towards the proper neighbors for the next computational step. 
CGRAs can represent a vaible solution to accelerate dense linear algebra applications, such as ML, image processing, and computer vision \cite{amber2022vlsi, tangram}.
\subsection{ASIC-based Accelerators}


To comply with the computational capabilities required by AI workloads, new powerful processing architectures are upcoming. Among them, there are two different types of Neural Processing Units: single-chip NPUs and NPUs integrated in the general purpose CPU. One of the main trends toward the next generation of laptops follows the second option by pushing the performance of AI workloads by integrating into the general-purpose CPU not only a GPU to accelerate graphics but also an NPU. This is the case of the recent Lunar Lake Intel processor architecture \cite{lunarlake}. 
Table \ref{tab:npu_survey} is an attempt to offer a common ground of different types of AI-accelerators in terms of process technology node, area, peak performance, energy efficiency and maturity level.

The purpose of an integrated NPU is to accelerate the performance and improve the energy efficiency of specific AI-tasks offloaded from the CPU \cite{song2019isscc}. In particular, NPUs are designed to accommodate a reasonable amount of multiply/accumulate (MAC) units, which are the PEs devised in the convolutional and fully-connected layers of DNNs \cite{Chen_2017,desoli17isscc}. 

\begin{table}[b]
\caption{Summary of ASIC-based AI-accelerators.}
\resizebox{8cm}{!}{
\begin{tabular}{@{}lccccr@{}}
\toprule
\textbf{Design} & \textbf{Process} & \textbf{Area} & \textbf{Peak Perf.} & \textbf{Energy Eff.} & \textbf{Maturity-level}\\ 
& \textbf{[nm]}& \textbf{[mm$^2$]} & \textbf{[TOPS]} & \textbf{[TOPS/W]} & \\
\midrule
Samsung \cite{song2019isscc} & 8 & 5.5 & 933 & 6.9 & Silicon\\
UM+NVIDIA \cite{zhang2019vlsi} & 16 & 2.4 & 480 & - & Silicon\\
MediaTek \cite{lin2020isscc} & 7 & 3.04 & 880 & 3.6 & Silicon\\
Alibaba \cite{jiao2020isscc} & 12 & 709 & 700 & 825 & Silicon\\
Samsung \cite{park2021isscc} & 5 & 5.46 & 1196 & 29.4 & Silicon\\
Samsung \cite{park2022isscc} & 4 & 4.74 & 1197 & 39.3 & Silicon\\
DaDianNao \cite{chen2016commacm} & 28 & 67.7 & 5.59 & - & Layout\\
ShiDianNao \cite{chen2016commacm} & 65 & 4.86 & 0.19 & - & Layout\\
Cambricon \cite{zhang2016micro} & 65 & 6.38 & 1.1 & - & Layout\\
EIE \cite{han2016isca} &  28 & 63.8 & 0.002 & 0.18 & Simulation\\
Eyeriss \cite{Chen_2017} & 65 & 16 & 0.03 & 0.07 & Layout\\
STM \cite{desoli17isscc} & 28 & 34.8 & 0.75 & 2.9 & Silicon\\
IBM \cite{oh2020vlsic} & 14 & 9.84 & 3 & 1.1 & Silicon\\
IBM \cite{lee2022jsscc} & 7 & 19.6 & 16.3 & 3.58 & Silicon\\
\bottomrule
\end{tabular}
}
\label{tab:npu_survey}
\end{table}

Each PE contains a synaptic weight buffer and MAC units to perform the computation of a neuron, namely, multiplication, accumulation, and an activation function (e.g., sigmoid). A PE can be realized with full-CMOS optimized circuits to trade off speed and power consumption. One of the most popular approaches adopted to this aim is referred to as \emph{approximate computing paradigm} to approximate the design at the cost of an acceptable accuracy loss. Representative approximate computing techniques suitable to design efficient arithmetic data paths are overviewed in Section A.5 of the Appendix A. 

An alternative method to design PEs consists in using emerging non-volatile memories such as RRAM and PCM to perform \textit{in situ} matrix-vector multiplication as in the RENO chip \cite{liu2015dac} or as in the MAC units proposed in \cite{xue2019isscc,narayanan2021ted}. The advantage of these architectures is that only the input and final output are digital; the intermediate results are all analog and are coordinated by analog routers. Data converters (DACs and ADCs) are required only when transferring data between the NPU and the CPU with an advantage in terms of energy efficiency (the work in \cite{xue2019isscc} reports an energy efficiency of 53.17 TOPS/W), although there are insufficient experimental data to support this evidence in comparison with full-digital NPUs.

In the DNN landscape, single-chip domain-specific accelerators achieved great success in both cloud and edge scenarios. These custom architectures offer better performance and energy efficiency concerning CPUs/GPUs thanks to an optimized data flow (or data reuse pattern) that reduces off-chip memory accesses, while improving the system efficiency \cite{chen2020engineering}.
The DianNao series represents a full digital stand-alone DNN accelerator that introduces a customized design to minimize the memory transfer latency and enhance the system efficiency. DaDianNao \cite{chen2016commacm} targets the datacenter scenario and integrates a large on-chip embedded dynamic random access memory (eDRAM) to avoid the long main memory access time. The same principle applies to the embedded scenario. ShiDianNao \cite{chen2016commacm} is a DNN accelerator dedicated to CNN applications. Using a weight-sharing strategy, its footprint is much smaller than the previous design. It is possible to map all of the CNN parameters onto a small on-chip static random access memory (SRAM) when the CNN model is small. In this way, ShiDianNao avoids expensive off-chip DRAM access time and achieves 60 times more energy efficiency compared to DianNao. 
Furthermore, domain-specific instruction set architectures (ISAs) have been proposed to support a wide range of NN applications. Cambricon \cite{zhang2016micro} and EIE \cite{han2016isca} are examples of architectures that integrate scalar, vector, matrix, logical, data transfer, and control instructions. Their ISA considers data parallelism and the use of customized vector/matrix instructions.

Eyeriss is another notable accelerator \cite{Chen_2017} that can support high throughput inference and optimize system-level energy efficiency, also including off-chip DRAMs. The main features of Eyeriss are a spatial architecture based on an array of 168 PEs that creates a four-level memory hierarchy, a dataflow that reconfigures the spatial architecture to map the computation of a given CNN and optimize towards the best energy efficiency, a network-on-chip (NoC) architecture that uses both multi-cast and point-to-point single-cycle data delivery, and run-length compression (RLC) and PE data gating that exploit the statistics of zero data in CNNs to further improve EE.

In \cite{desoli17isscc}, STMicroelectronics introduced the Orlando system-on-chip, a 28nm FDSOI-based CNN accelerator integrating an SRAM-based architecture with low-power features and adaptive circuitry to support a wide voltage range. Such a DNN processor provides an energy-efficient set of convolutional accelerators supporting kernel compression, an on-chip reconfigurable data-transfer fabric, a power-efficient array of DSPs to support complete real-world computer vision applications, an ARM-based host subsystem with peripherals, a range of high-speed I/O interfaces, and a chip-to-chip multilink to pair multiple accelerators together.

IBM presented in \cite{oh2020vlsic} a processor core for AI training and inference tasks applicable to a broad range of neural networks (such as CNN, LSTM, and RNN). High compute efficiency is achieved for robust FP16 training via efficient heterogeneous 2-D systolic array-SIMD compute engines that leverage DLFloat16 FPUs. A modular dual-corelet architecture with a shared scratchpad memory and a software-controlled network/memory interface enables scalability to many-core SoCs for scale-out paradigms. In 2022, IBM also presented a 7-nm four-core mixed-precision AI chip \cite{lee2022jsscc} that demonstrates leading-edge power efficiency for low-precision training and inference without model accuracy degradation. The chip is based on a high-bandwidth ring interconnect to enable efficient data transfers, while workload-aware power management with clock frequency throttling maximizes the application performance within a given power envelope.

Qualcomm presented an AI core that is a scalar 4-way VLIW architecture that includes vector/tensor units and lower precision to enable high-performance inference \cite{karam2021hotchip}. The design uses a 7 nm technology and is sought to be integrated into the AI 100 SoC to reach up to 149 TOPS with a power efficiency of 12.37 TOPS/W.
\subsection{Accelerators for Sparse Matrices}

Network pruning and zero-valued activations introduce sparsity that can be exploited by hardware accelerators to achieve compute and data reduction. 
This section overviews accelerator architectures designed to manage sparse matrices. Definitions, storage formats appropriate for sparse matrices, and their impacts on the computational complexity of DL models are discussed in Appendix A.6.

Eyeriss~\cite{Chen_2017} targets CNN acceleration 
by storing in DRAM only nonzero-valued activations in Compressed Sparse Columns (CSC) format 
and by skipping 
zero-valued activations to save energy. 
Eyeriss v2~\cite{eyerissv2:2019}, which targets DNNs on mobile devices, also supports sparse network models. 
It utilizes the CSC format to store weights and activations, which are kept compressed not only in memory but also during processing. 
To improve flexibility, it uses a hierarchical mesh for the PEs interconnections. 
By means of these optimizations, Eyeriss v2 is  significantly 
faster and 
more energy-efficient than the original Eyeriss. 

Cnvlutin~\cite{cnvlutin:isca2016} uses hierarchical data-parallel units, 
skips computation cycles for zero-valued activations 
and employs a co-designed data storage format based on Compressed Sparse
Rows (CSR) to compress the activations in DRAM. 
However, it does not consider the sparsity of the weights. 
On the contrary, Cambricon-X architecture~\cite{zhang2016micro} enables the PEs to store the compressed weights in CSR format 
for asynchronous computation. However, it does not exploit activation sparsity.  
EIE~\cite{han2016isca}, besides compressing the weights through a variant of CSC sparse matrix representation and skipping zero-valued activations,
employs a scalable array of PEs, each storing a partition of the DNN in SRAM
that allows obtaining significant energy savings with respect to DRAM.  
NullHop~\cite{Aimar2019} applies the Compressed Image Size (CIS) format to the weights and skips the null activations, 
similarly to EIE.
Sparse CNN (SCNN)~\cite{SCNN:isca2017} is an accelerator architecture for inference in CNNs. 
It employs a cluster of asynchronous PEs 
comprising several multipliers and accumulators.  
SCNN exploits sparsity in both weights and activations, which are stored in the classic CSR representation. 
It employs a Cartesian product-based computation architecture that maximizes the reuse of weights and activations 
within the cluster of PEs; 
the values are delivered to an array of multipliers, 
and the resulting scattered products are summed using a dedicated interconnection mesh. 
By exploiting two-sided sparsity, SCNN improves performance and energy over dense architectures. 
SparTen~\cite{sparten:micro2019} is based on SCNN~\cite{SCNN:isca2017}. 
It addresses some considerable overheads of SCNN in performing the sparse vector-vector dot product 
by improving the distribution of the operations to the multipliers 
and allows using any convolutional stride. 
It also addresses unbalanced sparsity distribution across the PEs employing an offline software scheme. 
The PermDNN architecture~\cite{perm:micro2018} uses permuted diagonal matrices to not generate load imbalance which is caused by the irregularity of unstructured sparse DNN models. 

SqueezeFlow~\cite{squeezeflow:2019} exploits concise convolution rules to benefit from the reduction of computation and memory accesses 
as well as the acceleration of existing dense CNN architectures without intrusive PE modifications. 
The Run Length Compression (RLC) format is used to compress activations and weights.
A different strategy is pursued by the Unique Weight CNN (UCNN) accelerator~\cite{ucnn:isca2018}, which proposes a generalization of the sparsity problem. 
Rather than considering only the repetition of zero-valued weights, UCNN exploits repeated weights with any value by reusing CNN sub-computations 
and reducing the model size in memory.
SIGMA~\cite{sigma:2020} is characterized by a flexible and scalable architecture 
that offers high utilization of its PEs regardless of kernel shape (i.e., matrices of arbitrary dimensions) and sparsity pattern. 
It targets the acceleration of GEMMs with unstructured sparsity. 
Bit-Tactical~\cite{bittactical:2019} uses a 
static scheduling middleware and a co-designed hardware front-end, with a lightweight sparse shuffling network that comprises 
two multiplexers per activation input. 
Unlike SIGMA and other accelerators, Bit-tactical leverages scheduling in software to align inputs and weights. 
Flexagon~\cite{flexagon:aplos2023} is a reconfigurable accelerator 
capable of performing sparse-sparse matrix multiplication computation 
by using the particular data flow that best matches each case. 

Besides the design of specialized hardware accelerators to exploit model sparsity, 
a parallel trend is to use GPU architectures. 
Pruned sparse models with unstructured sparse patterns introduce  
irregular memory accesses that are unfriendly on commodity GPU architectures. 
The first direction to tackle this issue is at the software layer, 
using pruning algorithms that enforce a particular sparsity pattern, such as tile sparsity~\cite{guo:sc2020}, 
on the model that allows leveraging existing GEMM accelerators. 
A second direction is to introduce new architectural support, such as Sparse Tensor Cores~\cite{sparsetensorcore:2021}. 
The NVIDIA Ampere architecture introduces this design with a fixed
50\% weight pruning target and achieves a better accuracy and performance trade-off. 
However, sparsity from activations, which are dynamic and unpredictable, 
is challenging to leverage on GPUs.  
Indeed, the current Sparse Tensor Core can take advantage of weight sparsity.
Reconfigurability appears to be a keyword for the design of new sparse accelerators 
because some network models exhibit \textit{dynamic sparsity}~\cite{fedus:jmlr2022}, where the position of non-zero elements changes over time.  
\subsection{Accelerators based on open-hardware RISC-V}
\label{sec:riscv}
RISC-V is an open-source, modular instruction set architecture (ISA) that is gaining popularity in computer architecture research due to its flexibility and suitability for integration with acceleration capabilities for DL.
The RISC-V ISA is designed with a small, simple core that can be extended with optional instruction set extensions (ISEs) to support various application domains.
RISC-V offers several advantages for DL acceleration research.
First, the modular nature of the ISA allows researchers to easily integrate acceleration capabilities as ISEs, which can be customized to suit the specific needs of different DL models.
Second, RISC-V supports a set of standard interfaces, such as AXI4, that can be used to interface with external acceleration units integrated on the same System-on-Chip at various levels of coupling.
This makes it easy to integrate specialized DL hardware accelerators into RISC-V-based systems.
Moreover, the defining feature of the RISC-V ISA is its openness, meaning that anybody can design a RISC-V implementation without paying royalties or needing a particular license.
Thanks to this non-technical advantage against other ISAs (such. as ARM, x86), RISC-V has gained significant attention from academia and emerging startups.
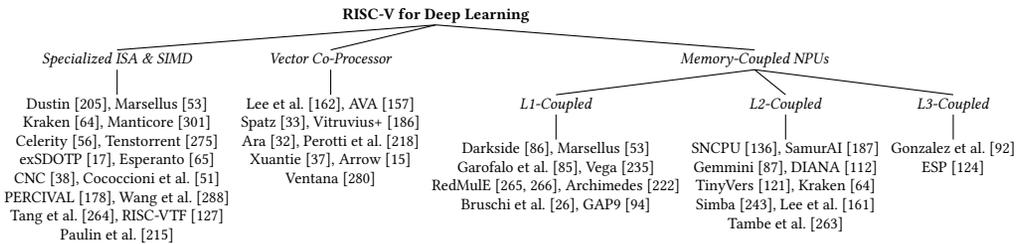
\begin{figure}[t]
    \centering
    {
        \tiny
        \scalebox{0.9}{
        \begin{forest}
            rounded/.style={ellipse,draw},
            squared/.style={rectangle,draw},
            qtree,
            [{\textbf{RISC-V for Deep Learning}}
              [{\textit{Specialized ISA \& SIMD}}
                [{
                  Dustin~\cite{ottaviDustin16CoresParallel2023},                  Marsellus~\cite{conti22124TOPS2023},
                  Kraken~\cite{dimauroKrakenDirectEvent2022}\\
                  Manticore~\cite{zarubaManticore4096CoreRISCV2021},
                  Celerity~\cite{davidsonCelerityOpenSource511Core2018}, 
                  Tenstorrent~\cite{vasiljevicComputeSubstrateSoftware2021}\\
                  exSDOTP~\cite{bertacciniMiniFloatNNExSdotpISA2022},
                  Esperanto~\cite{ditzelAcceleratingMLRecommendation2022},
                  CNC~\cite{chenEightCoreRISCVProcessor2022}\\
                  Cococcioni~\cite{cococcioniLightweightPositPorcessing2021},
                  PERCIVAL~\cite{mallasenPERCIVAL2022},
                  Wang~\cite{wangWinogradBasedConvolution2021}\\                Tang~\cite{tangGeneralPurposeGraphConvolutionNeuralAccelerator2022},
                  RISC-VTF~\cite{JiaoRISCVExtensionTransformer2021},
                  Paulin~\cite{paulinRNNBasedRadioResource2021}\\
                  Occamy~\cite{paulinOccamy432Core282024},
                  Zhou~\cite{zhouRISCVBasedFullyParallel2023},
                  speedAI240~\cite{snelgroveSpeedAI2402Petaflop30Teraflops2023}\\
                  DECADES~\cite{gaoDECADES67mm246TOPS2023a},
                  CIFER~\cite{liCIFERCacheCoherent12nm2023},
                  RDCIM~\cite{yiRDCIMRISCVSupported2024d}\\
                  Shaheen~\cite{valenteHeterogeneousRISCVBased2024},
                  FlexNN~\cite{nadaliniTOPSRISCVParallel2023a}\\
                }]
              ],
              [{\textit{Vector Co-Processor}}
                [{
                    Lee~\cite{lee201445nm}, 
                    AVA~\cite{lazoAdaptableRegisterFile2022}\\
                    Spatz~\cite{cavalcanteSpatzCompactVector2022}, 
                    Vitruvius+~\cite{minerviniVitruviusAreaEfficientRISCV2023}\\
                    Ara~\cite{cavalcanteAra1GHzScalable2020},
                    Perotti~\cite{perottiNewAraVector2022}\\
                    Xuantie~\cite{chenXuantie910CommercialMultiCore2020}, 
                    Arrow~\cite{assirArrowRISCVVector2021}\\
                    Ventana~\cite{ventanaProduct},
                    Vecim~\cite{wang30Vecim2892024}\\
                    MANIC~\cite{gobieskiMANIC19Mu2023},
                    YUN~\cite{perottiYunOpenSource64Bit2023}\\
                    Xvpfloat~\cite{guthmullerXvpfloatRISCVISA2024}\\
                }]
              ],
              [{\textit{Memory-Coupled NPUs}}
                [{\textit{L1-Coupled}}
                  [{
                    Darkside~\cite{garofaloDARKSIDEHeterogeneousRISCV2022}, 
                    Marsellus~\cite{conti22124TOPS2023}\\
                    Garofalo~\cite{garofaloHeterogeneousInMemoryComputing2022}, 
                    Vega~\cite{rossiVegaTenCoreSoC2022}\\                 RedMulE~\cite{tortorellaRedMuleMixedPrecisionMatrixMatrix2023,tortorellaRedMulECompactFP162022a}, 
                    Archimedes~\cite{prasadSpecializationMeetsFlexibility}\\
                    Bruschi~\cite{bruschiEndtoEndDNNInference2022}, 
                    GAP9~\cite{gap9Product}\\
                    Siracusa~\cite{prasadSiracusa16Nm2024}\\
                  }]
                ],
                [{\textit{L2-Coupled}}
                  [{
                    SNCPU~\cite{juSystolicNeuralCPU2023}, 
                    SamurAI~\cite{miro-panadesSamurAIVersatileIoT2022}\\
                    Gemmini~\cite{gencGemminiEnablingSystematic2021}, 
                    DIANA~\cite{houshmandDIANAEndtoEndHybrid2023}\\
                    TinyVers~\cite{jainTinyVersTinyVersatile2023}, 
                    Kraken~\cite{dimauroKrakenDirectEvent2022}\\
                    Simba~\cite{shao_micro19}, 
                    Lee~\cite{lee64TOPSEnergyEfficientTensor2022}\\
                    Tambe~\cite{tambe2212nm182023}\\
                    Metis~AIPU~\cite{hager11MetisAIPU2024}\\
                  }]
                ],
                [{\textit{L3-Coupled}}
                  [{
                    Gonzalez~\cite{gonzalez16mm106GOPS2021}\\
                    ESP~\cite{jia12nmAgileDesignedSoC2022}\\
                    Dos~Santos~\cite{dossantos1412nmLinuxSMPCapable2024}\\
                  }]
                ],
              ]
            ]
        \end{forest}
        } 
        }
    \caption{Taxonomy of RISC-V based acceleration units discussed in Section~\ref{sec:riscv}}
    \label{fig:riscv_taxonomy}
\end{figure}
Figure~\ref{fig:riscv_taxonomy} reports a synthetic taxonomy of representative RISC-V-based accelerators for DL. 

\subsubsection{RISC-V ISA extensions for (Deep) Learning}
Works in \cite{cococcioniLightweightPositPorcessing2021,mallasenPERCIVAL2022} propose ISA extensions for \emph{posit} numbers which can be used to do weight compression. 
Posit numbers need fewer bits to obtain the same precision or dynamic range of IEEE floats allowing them to store more weights in a same-sized memory. For example, the work in \cite{cococcioniLightweightPositPorcessing2021} provides an efficient conversion between 8- or 16-bit posits and 32-bit IEEE floats or fixed point formats with little loss in precision leading to a 10x speedup in inference time. 
Other works directly address the compute-intensive parts of different neural networks, in particular CNNs, GCNs, and transformers.
The new Winograd-based convolution instruction proposed in \cite{wangWinogradBasedConvolution2021} 
enables to compute a 
convolution producing a $2\times2$ output using a $3\times3$ kernel on a $4\times 4$ input in a single instruction using 19 clock cycles instead of multiple instructions totaling 140 cycles using the standard RISC-V ISA.
The set of general-purpose instructions for GCNs designed in \cite{tangGeneralPurposeGraphConvolutionNeuralAccelerator2022} mitigate the compute inefficiencies in aggregating and combining feature vectors.  
As such the authors combine the software programmability given by the RISC-V ISA with the compute efficiency of GCN accelerators.
Similarly, \cite{JiaoRISCVExtensionTransformer2021} focuses on transformer models. Notably, the extension comprises instructions to accelerate the well-known ReLU activation and softmax functions.
Paulin~et~al.~\cite{paulinRNNBasedRadioResource2021} performs a similar task but focuses on RNNs.

Many ISA extensions focus on low-bit-width arithmetics to accelerate inference of  Quantized Neural Networks (QNNs), often combined with multi-core parallel execution to further boost performance and efficiency.
Several developments augment the PULP RI5CY core used in Vega~\cite{rossiVegaTenCoreSoC2022} to improve its energy efficiency on QNNs.
Marsellus~\cite{conti22124TOPS2023} (16 cores) and Kraken~\cite{dimauroKrakenDirectEvent2022} (8 cores) use \textit{Xpulpnn}, an ISA extension for low-bitwidth (2/4-bit) integer dot-products used to accelerate symmetric precision QNNs, which is further extended in \textit{FlexNN}~\cite{nadaliniTOPSRISCVParallel2023a}.
Dustin~\cite{ottaviDustin16CoresParallel2023} (16 cores) also exploits a similar concept, but it also introduces a lockstep mechanism to operate all the cores in a SIMD fashion, further increasing their efficiency.
Shaheen~\cite{valenteHeterogeneousRISCVBased2024} exploits the same techniques in a more powerful SoC dedicated to applications for unmanned aerial vehicles.
Many architectures, like Manticore~\cite{zarubaManticore4096CoreRISCV2021}, Occamy~\cite{paulinOccamy432Core282024}, CIFER~\cite{liCIFERCacheCoherent12nm2023}, and DECADES~\cite{gaoDECADES67mm246TOPS2023a}, exploit ISA extensions for faster RISC-V based DL workload execution in the context of many-core architectures where a large number of cores cooperate
On the other hand, this approach is also popular with emerging commercial platforms, such as Celerity~\cite{davidsonCelerityOpenSource511Core2018}, Esperanto~\cite{ditzelAcceleratingMLRecommendation2022},  Tenstorrent~\cite{vasiljevicComputeSubstrateSoftware2021}, and speedAI240~\cite{snelgroveSpeedAI2402Petaflop30Teraflops2023}.
All these architectures are targeted at datacenter-based training and batch inference of large DNNs, Transformers, and Large Language Models (LLMs): hence, they typically focus on floating point multiply-accumulate and dot-product operations, such as exSDOTP~\cite{bertacciniMiniFloatNNExSdotpISA2022}.
Finally, a growing trend is to integrate digital in-memory computing (IMC) devices inside the pipeline of RISC-V processors as instruction set extensions. Two notable early examples at the silicon prototype maturity level are given by RDCIM~\cite{yiRDCIMRISCVSupported2024d} and Zhou~et~al.~\cite{zhouRISCVBasedFullyParallel2023}.

\subsubsection{RISC-V Vector Co-processors}
Vector co-processors represent a sort of natural architectural target for DL-oriented RISC-V acceleration.
AVA~\cite{lazoAdaptableRegisterFile2022}, Vitruvius+~\cite{minerviniVitruviusAreaEfficientRISCV2023}, Ara~\cite{cavalcanteAra1GHzScalable2020,perottiNewAraVector2022,perottiYunOpenSource64Bit2023}, Xvpfloat~\cite{guthmullerXvpfloatRISCVISA2024}, MANIC~\cite{gobieskiMANIC19Mu2023} are academic vector co-processors meant to accelerate the full RISC-V \textit{V} extension for vectorizable applications, including DL.
Commercial RISC-V vector processors mainly targeted at HPC markets, such as 
Xuantie~\cite{chenXuantie910CommercialMultiCore2020}, and Ventana~\cite{ventanaProduct}, have recently started appearing. 
In addition, vector co-processors explicitly tailored for DL, like Spatz~\cite{cavalcanteSpatzCompactVector2022} and Arrow~\cite{assirArrowRISCVVector2021}, have been developed.
The former, in particular, focuses only on a subset of the \textit{V} extension and on 32-bit data, capturing better opportunities for energy efficiency.
Further pushing this trend, Vecim~\cite{wang30Vecim2892024} uses an IMC array to implement part of a reduced-precision (FP16) DL-dedicated vector extension for RISC-V.

\subsubsection{RISC-V Memory-coupled Neural Processing Units (NPUs)}
Concerning the tightest kind of memory coupling, at L1, most proposals in the state-of-the-art are based on the Parallel Ultra-Low Power (PULP) template, and devote significant effort to enabling fast communication between RISC-V cores and accelerators.
Representative system architectures designed in this way are available at several levels of maturity, like the prototypes Vega~\cite{rossiVegaTenCoreSoC2022} and 
Darkside~\cite{garofaloDARKSIDEHeterogeneousRISCV2022}, the commercial products
GreenWaves Technologies GAP9~\cite{gap9Product}, Archimedes~\cite{prasadSpecializationMeetsFlexibility} and
Siracusa~\cite{prasadSiracusa16Nm2024}, and the simulation templates
~\cite{garofaloHeterogeneousInMemoryComputing2022} and 
~\cite{bruschiEndtoEndDNNInference2022}.


\begin{figure}[t]
    \resizebox{\columnwidth}{!}{%
    \centering
    \includegraphics{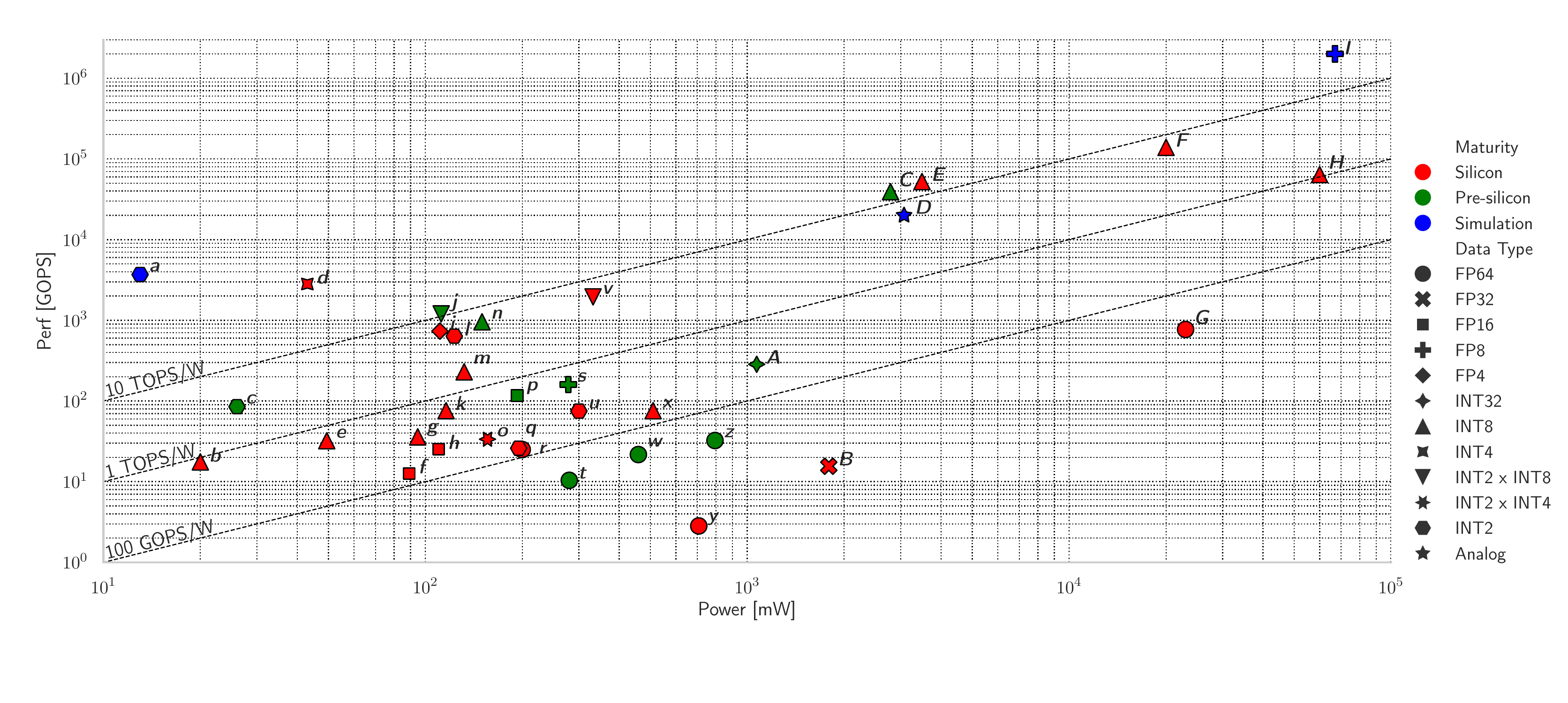}
    }
    \begin{minipage}{\textwidth}\setstretch{0.6}
    \vspace{-2em}
    { \tiny \textit{\bfseries{a}}: Zhou~et~al.~\cite{zhouRISCVBasedFullyParallel2023},
\textit{\bfseries{b}}:~TinyVers~\cite{jainTinyVersTinyVersatile2023},
\textit{\bfseries{c}}:~FlexNN~\cite{nadaliniTOPSRISCVParallel2023a},
\textit{\bfseries{d}}:~RDCIM~\cite{yiRDCIMRISCVSupported2024d},
\textit{\bfseries{e}}:~Vega~\cite{rossiVegaTenCoreSoC2022},
\textit{\bfseries{f}}:~Darkside~\cite{garofaloDARKSIDEHeterogeneousRISCV2022},
\textit{\bfseries{g}}:~SamurAI~\cite{miro-panadesSamurAIVersatileIoT2022},
\textit{\bfseries{h}}:~Vecim~\cite{wang30Vecim2892024},
\textit{\bfseries{i}}:~Tambe~et~al.~\cite{tambe2212nm182023},
\textit{\bfseries{j}}:~Archimedes~\cite{prasadSpecializationMeetsFlexibility},
\textit{\bfseries{k}}:~SNCPU~\cite{juSystolicNeuralCPU2023},
\textit{\bfseries{l}}:~Marsellus~\cite{conti22124TOPS2023},
\textit{\bfseries{m}}:~DIANA~\cite{houshmandDIANAEndtoEndHybrid2023},
\textit{\bfseries{n}}:~Garofalo~et~al.~\cite{garofaloHeterogeneousInMemoryComputing2022},
\textit{\bfseries{o}}:~Dustin~\cite{ottaviDustin16CoresParallel2023},
\textit{\bfseries{p}}:~RedMulE~\cite{tortorellaRedMuleMixedPrecisionMatrixMatrix2023,tortorellaRedMulECompactFP162022a},
\textit{\bfseries{q}}:~Shaheen~\cite{valenteHeterogeneousRISCVBased2024},
\textit{\bfseries{r}}:~Manticore~\cite{zarubaManticore4096CoreRISCV2021},
\textit{\bfseries{s}}:~exSDOTP~\cite{bertacciniMiniFloatNNExSdotpISA2022},
\textit{\bfseries{t}}:~NewAra~\cite{perottiNewAraVector2022},
\textit{\bfseries{u}}:~Kraken~\cite{dimauroKrakenDirectEvent2022},
\textit{\bfseries{v}}:~Siracusa~\cite{prasadSiracusa16Nm2024},
\textit{\bfseries{w}}:~Vitruvius+~\cite{minerviniVitruviusAreaEfficientRISCV2023},
\textit{\bfseries{x}}:~CNC~\cite{chenEightCoreRISCVProcessor2022},
\textit{\bfseries{y}}:~YUN~\cite{perottiYunOpenSource64Bit2023},
\textit{\bfseries{z}}:~Ara~\cite{cavalcanteAra1GHzScalable2020},
\textit{\bfseries{A}}:~Spatz~\cite{cavalcanteSpatzCompactVector2022},
\textit{\bfseries{B}}:~CIFER~\cite{liCIFERCacheCoherent12nm2023},
\textit{\bfseries{C}}:~Axelera~AI~\cite{ward-foxtonAxeleraDemosAI},
\textit{\bfseries{D}}:~Bruschi~et~al.~\cite{bruschiEndtoEndDNNInference2022},
\textit{\bfseries{E}}:~Metis~AIPU~\cite{hager11MetisAIPU2024},
\textit{\bfseries{F}}:~Esperanto~\cite{ditzelAcceleratingMLRecommendation2022},
\textit{\bfseries{G}}:~Occamy~\cite{paulinOccamy432Core282024},
\textit{\bfseries{H}}:~Lee~et~al.~\cite{lee64TOPSEnergyEfficientTensor2022},
\textit{\bfseries{I}}:~speedAI240~\cite{snelgroveSpeedAI2402Petaflop30Teraflops2023}}
    \end{minipage}
    \par
    \caption{Performance and power consumption of SotA DL accelerators based on open-HW RISC-V.} 
    \label{fig:riscv_perf_power}
\end{figure}
\begin{table}[bt]
\caption{Summary of RISC-V Deep Learning acceleration architectures}
\label{tab:riscv}
\resizebox{\columnwidth}{!}{%
\begin{tabular}{@{}lllrrrrrrrll@{}}
\toprule
\textbf{Category} & \textbf{Accelerator}                                                                             & \textbf{Tech [nm]} & \multicolumn{1}{l}{\textbf{Area [mm2]}} & \multicolumn{1}{l}{\textbf{Freq [MHz]}} & \multicolumn{1}{l}{\textbf{Voltage [V]}} & \multicolumn{1}{l}{\textbf{Power [mW]}} & \multicolumn{1}{l}{\textbf{Perf [GOPS]}} & \multicolumn{1}{l}{\textbf{Eff [GOPS/W]}} & \multicolumn{1}{l}{\textbf{\# MAC units}} & \textbf{Data Type} & \textbf{Maturity} \\ \midrule
\multirow{16}{*}{\textbf{ISA}} & Dustin~\cite{ottaviDustin16CoresParallel2023}                                                    & 65               & 10                                      & 205                                     & 1.2                                      & 156                                     & 33.6                                     & 215                                                     & 128                                      & INT2 x INT4        & Silicon           \\
& Kraken (RISC-V cores)~\cite{dimauroKrakenDirectEvent2022}                                                      & 22               & 9                                       & 330                                     & 0.8                                      & 300                                     & 75                                       & 750                                                     & 128                                      & INT2               & Silicon           \\
& Manticore~\cite{zarubaManticore4096CoreRISCV2021}                                                & 22               & 888                                     & 500                                     & 0.6                                      & 200                                     & 25                                       & 188                                                     & 24                                       & FP64               & Pre-silicon       \\
& Celerity~\cite{davidsonCelerityOpenSource511Core2018}                                            & 16               & 25                                      & 1050                                    & -                                        & 1900                                    & -                                        & -                                                       & 496                                      & INT32              & Silicon           \\
& Tenstorrent~\cite{vasiljevicComputeSubstrateSoftware2021}                                        & 12               & 477                                     & -                                       & -                                        & -                                       & 92000                                    & -                                                       & -                                        & FP16               & Silicon           \\
& exSDOTP~\cite{bertacciniMiniFloatNNExSdotpISA2022}                                               & 12               & 0.52                                    & 1260                                    & 0.8                                      & \textit{278}                            & 160                                      & 575                                                     & 16                                       & FP8                & Pre-silicon       \\
& Esperanto~\cite{ditzelAcceleratingMLRecommendation2022}                                          & 7                & 570                                     & 1000                                    & -                                        & 20000                                   & 139000                                   & 6.95                                                    & 69632                                    & INT8               & Silicon           \\
& CNC~\cite{chenEightCoreRISCVProcessor2022}                                                       & 4                & 1.92                                    & 1150                                    & 0.85                                     & 510                                     & 75.8                                     & 149                                                     & 512                                      & INT8               & Silicon           \\
& Occamy~\cite{paulinOccamy432Core282024}                                                          & 12               & 146                                     & 1000                                    & 0.8                                      & 23000                                   & 770                                      & 28.1                                                    & 432                                      & FP64               & Silicon           \\
& Zhou~et~al.~\cite{zhouRISCVBasedFullyParallel2023}                                               & 28               & -                                       & 50                                      & -                                        & 13                                      & 3690                                     & 285                                                     & 144000 1--bit                   & INT2 (IMC)              & Simulation        \\
& speedAI240~\cite{snelgroveSpeedAI2402Petaflop30Teraflops2023}                                    & 7                & -                                       & 1300                                    & -                                        & 67000                                   & 2000000                                  & 30000                                                   & 372000                                   & FP8                & Simulation        \\
& DECADES~\cite{gaoDECADES67mm246TOPS2023a}                                                        & 12               & 62                                      & 911                                     & 1.2                                      & -                                       & 1460                                     & -                                                       & 60                                       & INT64              & Silicon           \\
& CIFER~\cite{liCIFERCacheCoherent12nm2023}                                                        & 12               & 16                                      & 1195                                    & 0.8                                      & 1792                                    & 15.54                                    & 6.63                                                    & 14                                       & FP32               & Silicon           \\
& RDCIM~\cite{yiRDCIMRISCVSupported2024d}                                                          & 55               & 9.8                                     & 200                                     & 1.2                                      & 43                                      & 2820                                     & 66300                                                   & 524288 1--bit                   & INT4 (IMC)              & Silicon           \\
& Shaheen~\cite{valenteHeterogeneousRISCVBased2024}                                                & 22               & 9                                       & 500                                     & 0.8                                      & 195                                     & 26                                       & 133                                                     & 8                                        & INT2               & Silicon           \\
& FlexNN~\cite{nadaliniTOPSRISCVParallel2023a}                                                     & 22               & 0.55                                    & 463                                     & 0.65                                     & 26                                      & 85                                       & 3260                                                    & 128                                      & INT2               & Pre-silicon       \\ \midrule
\multirow{10}{*}{\textbf{Vector}} & Lee~et~al.~\cite{lee64TOPSEnergyEfficientTensor2022}                                             & 14               & 181                                     & 2000                                    & 0.8                                      & 60000                                   & 64000                                    & 1450                                                    & 16384                                    & INT8               & Silicon           \\
& AVA~\cite{lazoAdaptableRegisterFile2022}                                                         & 22               & 3.9                                     & -                                       & -                                        & -                                       & -                                        & -                                                       & -                                        & FP64               & Pre-silicon       \\
& Spatz~\cite{cavalcanteSpatzCompactVector2022}                                                    & 22               & 20                                      & 594                                     & 0.8                                      & 1070                                    & 285                                      & 266                                                     & 256                                      & INT32              & Pre-silicon       \\
& Vitruvius+~\cite{minerviniVitruviusAreaEfficientRISCV2023}                                       & 22               & 1.3                                     & 1400                                    & 0.8                                      & 459                                     & 21.7                            & 47.3                                                    & 8                                        & FP64               & Pre-silicon       \\
& Ara~\cite{cavalcanteAra1GHzScalable2020}                                                         & 22               & 10735 kGE                               & 1040                                    & 0.8                                      & 794                                     & 32.4                                     & 40.8                                                    & 16                                       & FP64               & Pre-silicon       \\
& Perotti~et~al.~\cite{perottiNewAraVector2022}                                                    & 22               & 0.81                                    & 1340                                    & 0.8                                      & 280                                     & 10.4                                     & 37.1                                                    & 4                                        & FP64               & Pre-silicon       \\
& Vecim~\cite{wang30Vecim2892024}                                                                  & 65               & 4                                       & 250                                     & 1                                        & 110                                     & 25.3                                     & 230                                                     & 4                                        & FP16 (IMC)         & Silicon           \\
& MANIC~\cite{gobieskiMANIC19Mu2023}                                                               & 22               & 0.57                                    & 48.9                                    & 1.05                                     & 2                                       & 0.512                                    & 256                                                     & 1                                        & INT32              & Silicon           \\
& YUN~\cite{perottiYunOpenSource64Bit2023}                                                         & 65               & 6                                       & 280                                     & 1.2                                      & 707                                     & 2.83                                     & 4                                                       & 4                                        & FP64               & Silicon           \\
& Xvpfloat~\cite{guthmullerXvpfloatRISCVISA2024}                                                   & 7                & 0.14                                    & 1250                                    & 0.675                                    & -                                       & -                                        & -                                                       & 1                                        & FP64               & Pre-silicon       \\ \midrule
\multirow{8}{*}{\textbf{L1 NPU}} & Darkside~\cite{garofaloDARKSIDEHeterogeneousRISCV2022}                                           & 65               & 3.85                                    & 200                                     & 1.2                                      & 89.1                                    & 12.6                                     & 152                                                     & 32                                       & FP16               & Silicon           \\
& Marsellus (NPU)~\cite{conti22124TOPS2023}                                                        & 22               & 18.7                        & 420                                     & 0.8                                      & 123                                     & 637                                      & 7600                                                    & 10368 1--bit                             & INT2               & Silicon           \\
& Garofalo~et~al.~\cite{garofaloHeterogeneousInMemoryComputing2022}                                & 22               & 30                                      & 500                                     & 0.8                                      & 150                                     & 958                                      & 6390                                                    & 36 (DW)                                  & INT8               & Pre-silicon       \\
& Vega~\cite{rossiVegaTenCoreSoC2022}                                                              & 22               & 12                                      & 450                                     & 0.8                                      & 49.4                                    & 32.2                                     & 651                                                     & 27                                       & INT8               & Silicon           \\
& RedMulE~\cite{tortorellaRedMuleMixedPrecisionMatrixMatrix2023,tortorellaRedMulECompactFP162022a} & 22               & 0.73                                    & 613                                     & 0.8                                      & 193                                     & 117                                      & 608                                                     & 96                                       & FP16               & Pre-silicon       \\
& Archimedes~\cite{prasadSpecializationMeetsFlexibility}                                           & 22               & 3.38                                    & 270                                     & 0.65                                     & 112                                     & 1198                                     & 10.6                                                    & 5184                                     & INT2 x INT8        & Pre-silicon       \\
& Bruschi~et~al.~\cite{bruschiEndtoEndDNNInference2022}                                            & 5                & 480                                     & -                                       & -                                        & 3070                                    & 20000                                    & 6500                                                    & 3.35$\times$10${}^{7}$                   & Analog (IMC)       & Simulation        \\
& Siracusa~\cite{prasadSiracusa16Nm2024}                                                           & 16               & 16                                      & 360                                     & 0.8                                      & 332                                     & 1950                                     & 7000                                                    & 10368 1$\times$8--bit                    & INT2 x INT8        & Silicon           \\ \midrule
\multirow{9}{*}{\textbf{L2 NPU}} & SNCPU~\cite{juSystolicNeuralCPU2023}                                                             & 65               & 4.47                                    & 400                                     & 1                                        & 116                                     & 75.9                                     & 655                                                     & 100                                      & INT8               & Silicon           \\
& SamurAI~\cite{miro-panadesSamurAIVersatileIoT2022}                                               & 28               & 4.52                                    & 350                                     & 0.9                                      & 94.7                                    & 36                                       & 380                                                     & 64                                       & INT8               & Silicon           \\
& Gemmini~\cite{gencGemminiEnablingSystematic2021}                                                 & 22               & 1.03                                    & 1000                                    & -                                        & -                                       & -                                        & -                                                       & 256                                      & INT8               & Pre-silicon       \\
& DIANA~(digital)~\cite{houshmandDIANAEndtoEndHybrid2023}                                          & 22               & 10.24                                   & 280                                     & 0.8                                      & 132                                     & 230                                      & 1740                                                    & 256                                      & INT8               & Silicon           \\
& DIANA~(analog)~\cite{houshmandDIANAEndtoEndHybrid2023}                                           & 22               & 10.24                                   & 350                                     & 0.8                                      & 132                                     & 18100                                    & 176000                                                  & 256                                      & Analog (IMC)       & Silicon           \\
& TinyVers~\cite{jainTinyVersTinyVersatile2023}                                                    & 22               & 6.25                                    & 150                                     & 0.8                                      & 20                                      & 17.6                                     & 863                                                     & 64                                       & INT8               & Silicon           \\
& Simba~\cite{shao_micro19}                                                    & 16               & 6                                       & 161                                     & 0.42                                     & -                                       & -                                        & 9100                                                    & 1024                                     & INT8               & Silicon           \\
& Metis~AIPU~\cite{hager11MetisAIPU2024}                                                           & 12               & 144                                     & 800                                     & 0.68                                     & 3490                                    & 52400                                    & 15000                                                   & -                                        & INT8 (IMC)         & Silicon           \\
& Tambe~et~al.~\cite{tambe2212nm182023}                                                            & 12               & 4.59                                    & 717                                     & 1                                        & 111                                     & 734                                      & 6612                                                    & -                                        & FP4                & Silicon           \\ \midrule
\multirow{3}{*}{\textbf{L3 NPU}} & Gonzalez~et~al.~\cite{gonzalez16mm106GOPS2021}                                                   & 22               & 16                                      & 961                                     & -                                        & -                                       & -                                        & 106.1                                                   & 256                                      & INT8               & Silicon           \\
& ESP~\cite{jia12nmAgileDesignedSoC2022}                                                           & 12               & 21.6                                    & 1520                                    & 1                                        & 1830                                    & -                                        & -                                                       & 3x NVDLA                                 & INT8               & Silicon           \\
& Dos~Santos~et~al.~\cite{dossantos1412nmLinuxSMPCapable2024}                                      & 12               & 64                                      & 1600                                    & 1                                        & 4330                                    & -                                        & -                                                       & 4x NVDLA                                 & INT8               & Silicon           \\
\bottomrule
\end{tabular}
}
\end{table}
Moving the shared memory from L1 to L2/L3, there are other NPU solutions in the state-of-the-art.
For example, SNCPU~\cite{juSystolicNeuralCPU2023}, 
can act as either a set of 10 RISC-V cores or be reconfigured in a systolic NPU.
In 
~\cite{gonzalez16mm106GOPS2021} and
~\cite{gencGemminiEnablingSystematic2021}, 
systolic arrays generated by Gemmini 
are coupled with a RISC-V core by exploiting a shared L3 or L2 memory, respectively.
Simba~\cite{shao_micro19} 
is also meant to be scaled towards server-grade performance using the integration of chiplets on multi-chip modules.
ESP~\cite{giriESP4MLPlatformBasedDesign2020,giriAcceleratorIntegrationOpenSource2021} and 
~\cite{tambe2212nm182023} also focus on integrating hardware accelerators and NPUs in large-scale Network-on-Chips using RISC-V cores as computing engines.
Axelera~AI propose a so-far unique architecture 
that uses a L2 shared-memory accelerator exploiting digital SRAM-based IMC, called Metis~AIPU~\cite{hager11MetisAIPU2024}.
SamurAI~\cite{miro-panadesSamurAIVersatileIoT2022}, TinyVers~\cite{jainTinyVersTinyVersatile2023}, and DIANA~\cite{houshmandDIANAEndtoEndHybrid2023} build up AI-IoT systems composed of a microcontroller and L2-coupled NPUs.
Kraken~\cite{dimauroKrakenDirectEvent2022} couples the 
RISC-V ISA-extended cluster with specialized L2-coupled Spiking Neural Network (SNN) and Ternary Neural Network (TNN) accelerators.

\subsubsection{Summary}

Figure~\ref{fig:riscv_perf_power} 
clearly shows that RISC-V-based solutions occupy essentially the full spectrum of DL architectures ranging from 10~mW microcontrollers up to 100~W SoCs meant to be integrated as part of HPC systems.
So far, most of the research has focused on the lower end of this spectrum, striving for the best energy efficiency.
We can observe how efficiency is strongly correlated with architectural techniques yielding accuracy (e.g., data bit-width reduction \& quantization) and with the usage of emerging computational paradigms such as in-memory computing. Table~\ref{tab:riscv} compares the above discussed architectures quantitatively and
reports 
their highest performance 
and energy efficiency values.

\section{Accelerators based on Emerging Technologies}
\label{sec:emerging}
To design efficient DNN hardware accelerators, combining optimized memory architectures and processing modules is crucial to achieve high speed at reasonable costs and power dissipation.
Such architectures must be designed taking into account the large amount of memory necessary to store the input feature maps, weights, and intermediate results generated by the convolutional layers in a DNN. Moreover, managing DNN computational models causes a large number of data movements between the memory and the processing elements, often posing a challenge in terms of achievable speed performance, energy consumption, and memory bandwidth. For these reasons, several innovative memory architectures and technologies have recently emerged to increase memory capacity and data bandwidth, to reduce memory access latency, and potentially to improve the power efficiency. In this Section, we discuss several technologies: Processing-in-Memory and In-Memory Computing (see Fig. \ref{fig:neuro_taxonomy}), Neuromorphic accelerators, approaches based on Multi-Chip Modules, and Quantum and Photonic computing. 


\begin{figure}[!t]
    \centering
    {
        \tiny
        \begin{forest}
            rounded/.style={ellipse,draw},
            squared/.style={rectangle,draw},
            qtree,
            [{\textbf{Processing-in-Memory and In-memory Computing}}
              [{\textit{PIM}}
                [{\textit{In-subarray (DRAM)}}
                    [{
                    DRISA/DrAcc\cite{Li_2017,Deng_2018},
                    IMI\cite{Finkbeiner_2017} \\
                    SIMDRAM\cite{Hajinazar_2021},
                    PIM-DRAM\cite{Roy_2021} \\
                    }]
                ]
                [{\textit{Logic die (3D-DRAM)}}
                    [{
                    Neurocube\cite{Kim2016},
                    Tetris\cite{Gao2017} \\
                    NeuralHMC\cite{Min2019}, 
                    VIMA\cite{Cordeiro2021} \\
                    }]
              ]
               [{\textit{Bank (3D-DRAM)}}
                    [{
                    Newton\cite{He2020} \\
                    HBM-PIM\cite{Kwon2021} \\ 
                    }]
              ]
              ],
              [{\textit{IMC}}
                [{\textit{SRAM}}
                    [{Orlando \cite{desoli17isscc},\\
                      STM-IMC \cite{desoli2023isscc}
                    }]
                ]
                [{\textit{PCM}}
                    [{
                    Hermes \cite{khaddam2022hermes},\\
                    IBM \cite{gallo202264}, PCM-AIMC \cite{garofaloHeterogeneousInMemoryComputing2022}
                    }]
                ]
                [{\textit{RRAM}}
                    [{ISAAC \cite{shafiee2016isca},              PipeLayer \cite{song2017hpca} \\
                    NeuralPIM \cite{cao2022tcomp},
                    NeuRRAM \cite{wan2022nature} \\
                    }]
                ]
              ],
              ],
            ]
        \end{forest}
        }
    \caption{Taxonomy of accelerators based on the emerging memory technologies discussed in Sections~\ref{ssec:PIM}, \ref{ssec:NPU3}}
    \label{fig:neuro_taxonomy}
\end{figure}

\subsection{Processing-in-Memory}
\label{ssec:PIM}

Processing-in-Memory (PIM) solutions are mostly implemented on DRAM modules.
PIMs' computing elements can compute in parallel in all subarrays/banks, accessing data through the internal DRAM buses, and reducing the amount of data transferred between host and memory.
Depending on where the computation is performed, we can identify three main categories of PIMs \cite{Roy_2021}: (1) In-subarray PIMs (the compute occurs at the local sense amplifiers), (2) bank-level PIMs (processing logic is integrated into each DRAM die at the level of the memory banks, after the column decoder and selector blocks), and (3) logic-die level PIMs (compute cores are embedded into the logic die of a 3D-stacked memory block).
Table~\ref{tab:pim_survey} presents a summary of the three types of PIM accelerators.

3D-stacked memory blocks rely on the possibility of stacking layers of conventional 2D DRAM or other memory types together with one or more optional layers of logic circuits. 
These logic layers are often implemented with different process technologies and can include buffer circuitry, test logic, and PEs. 
Two main 3D stacked memory standards have been recently proposed: the Hybrid Memory Cube (HMC) and the High Bandwidth Memory (HBM). 
They both provide highly parallel access to the memory, a sought-after characteristic in the highly parallel architecture of the DNN accelerators. 
The PEs of 3D stacked DNN accelerators can be embedded in the logic die or in the memory dies, significantly reducing the latency of accessing data in main memory, and improving the system energy efficiency.
However, as detailed in Section A.7 of Appendix A, there are some challenges and limitations to consider when using this technology \cite{Kim2022}.

\begin{table}[t]
\caption{Summary of Processing-in-memory DNN accelerators.}
\resizebox{\columnwidth}{!}{%
\begin{tabular}{@{}lccccccccc@{}}
\toprule
\textbf{PIM} & \textbf{Year} & \textbf{Integration Level} & \textbf{Mem. Tech.} & \textbf{Functions} & \textbf{Data Type} & \textbf{Tech. Node} & \textbf{Performance [GOPs/s]} & \textbf{Power [W]} & \textbf{Maturity}\\ \midrule
DRISA/DrAcc\cite{Li_2017,Deng_2018} & 2017 & In-subarray & 2D DRAM & XOR & variable & - & - & - & Simulation \\  
IMI\cite{Finkbeiner_2017} & 2017 & In-subarray & 2D DRAM & boolean & variable & - & - & - & Simulation \\ 
SIMDRAM\cite{Hajinazar_2021} & 2021 & In-subarray & 2D DRAM & MAJ/NOT & variable & - & - & - & Simulation \\ 
PIM-DRAM\cite{Roy_2021} & 2021 & In-subarray & 2D DRAM & ADD/AND & variable & - & - & - & Simulation \\ 
Neurocube\cite{Kim2016} & 2016 & Logic die & HMC & MAC & 16-bit fixed point & 15nm & 132 & 3.4 + HMC & Layout \\
Tetris\cite{Gao2017} & 2017 & Logic die & HMC & ALU/MAC & 16-bit fixed point & 45nm & - & 8.42 & Simulation \\
NeuralHMC\cite{Min2019} & 2019 & Logic die & HMC & MAC & 32-bit floating point & - & - & - & Simulation \\
VIMA\cite{Cordeiro2021} & 2021 & Logic die & HMC & ALU/MULT/DIV & 32-bit integer/floating point & - & - & 3.2 + HMC & Simulation \\
Newton\cite{He2020} & 2020 & Bank & HBM & MAC & bfloat16 & - & - & - & Simulation \\
HBM-PIM\cite{Kwon2021} & 2020 & Bank & HBM & ALU/MAC & 16-bit floating point & 20nm & 1200 & - & Silicon \\
\bottomrule
\end{tabular}
}
\label{tab:pim_survey}
\end{table}

Most in-subarray PIMs for DNNs rely on solutions similar to Ambit \cite{Seshadri_2017} and RowClone \cite{Seshadri_2013} for implementing the computing elements.
Ambit 
exploits the analog operation of DRAM technology to perform bit-wise AND, OR and NOT operations completely inside the DRAM. 
RowClone is a mechanism that efficiently copies rows inside the same DRAM subarray by exploiting the vast internal DRAM bandwidth without CPU intervention.

DRISA\cite{Li_2017} leverages these technologies by implementing bit-wise XORs, 
and by expressing more complex functions as sequences of such a basic operation.
Additional logic 
(e.g., shifters) 
and modifications in the memory controllers are needed for driving the execution of operation opcodes.
Higher bit-widths are supported,
but with  the execution time increasing exponentially.
However, 
multiple subarrays and banks provide large parallelism and large computational throughput.
While DRISA evaluates the implementation of CNNs with binary weights, DrAcc \cite{Deng_2018} focuses on CNNs with ternary weights.

The Micron In-Memory Intelligence (IMI) architecture \cite{Finkbeiner_2017} is built on simple bit-serial computing elements placed below standard DRAM array’s sense-amplifiers and provides the memory block with the ability for massive SIMD parallelism by supporting vector instructions over an entire bank.
Complex operations are implemented as serial sequences of basic logic functions, such as XOR and AND.
A control unit is attached to each DRAM bank and translates the IMI instruction to be executed into row-cycles that control the SIMD computing elements.

SIMDRAM \cite{Hajinazar_2021} is a flexible general-purpose processing-using-DRAM framework that 
provides a flexible mechanism to support the implementation of arbitrary user-defined operations 
as sequences of basic functions including MAJ and NOT.
The sequence of DRAM commands generated by the framework are executed by a control unit located inside the memory controller, which manages the computation 
from start to end.

PIM-DRAM\cite{Roy_2021} is 
an in-subarray PIM, which
performs multiplications 
as sequences of bit-wise in-subarray addition and AND operations
while the accumulation and activation functions are performed in the bank architecture.
The processing happens by enabling multiple wordlines at the same time to leverage the large available internal DRAM bandwidth.


\begin{figure}[b]
  \centering
    \begin{subfigure}[c]{0.35\textwidth}
        \centering
        \includegraphics[width=\textwidth]{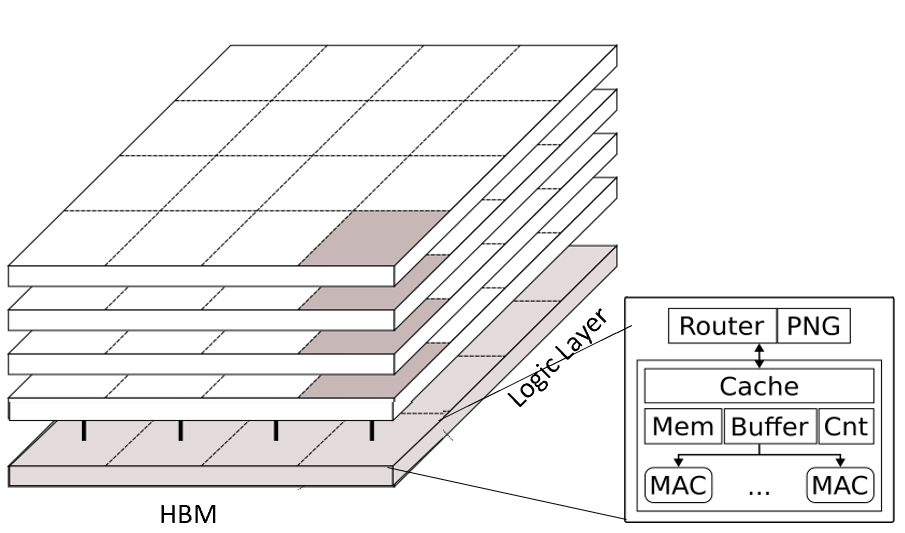}
         \caption{}
         \label{fig:Neurocube}
    \end{subfigure}
    \qquad
    \qquad
    \qquad
    \begin{subfigure}[c]{0.4\textwidth}
         \centering
         \includegraphics[width=\textwidth]{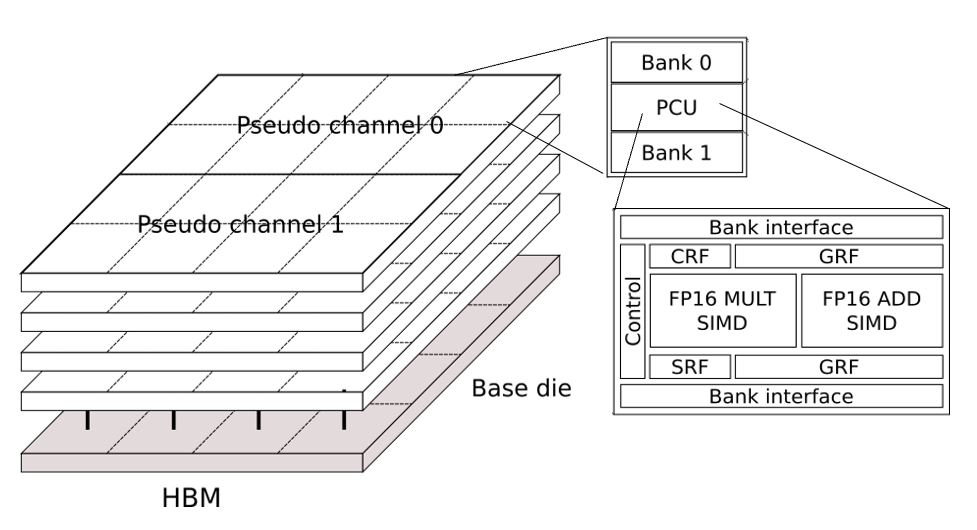}
         \caption{}
         \label{fig:HBM-PIM}
    \end{subfigure}
    \caption{(a) Neurocube architecture. (b) HBM-PIM architecture.}
    \vskip -0.5cm
    \label{fig:nonew}
\end{figure}



A first example of a logic die-level PIM implementation is Neurocube \cite{Kim2016} that, as shown in Figure~\ref{fig:Neurocube}, is embedded into the logic die of an HMC, and consists of a cluster of  PEs connected by a 2D mesh Network-on-Chip (NoC).
The PE is composed of a row of multiply accumulator (MAC) units, a cache memory, a temporal
buffer, and a memory module for storing shared synaptic weights.
Each PE is associated with a single memory vault and can operate independently and communicate through the TSVs and the vault controller.
A host communicates with the Neurocube through the external links of the HMC to configure the Neurocube for different neural network architectures.
Each vault controller in the HMC has an associated programmable neuro sequence generator (PNG), i.e., a programmable state machine that controls the data movements required for neural computation.
Neurocube implements an output stationary dataflow.

Tetris \cite{Gao2017} uses an HMC memory stack organized into 16 vaults. Each vault is associated with a PE, connected to the vault controller, and composed of a systolic array of 14 $\times$14 PEs and a small SRAM buffer, shared among the PEs.
A 2D mesh NoC connects all the PEs.
The dimension of the buffers in the logic layer is reduced and optimized to take into account the lower cost of accessing the DRAM layers, as well as the area constraints of the 3D package.
Each PE has a register file and a MAC locally storing the inputs/weights and performing computations.
Tetris implements a row stationary dataflow that maps 1D convolutions onto a single PE.
A 2D convolution is orchestrated on the 2D array interconnect so that the data propagation among PEs remains local.
In \cite{Gao2017}, an optimal scheduling is discussed to maximize on-chip reuse of weights and/or activations, and resource utilization.
However, a programming model is not presented.

NeuralHMC \cite{Min2019} adopts 
a weight-sharing pipelined MAC 
to lower the cost of accessing weight data, by reducing the original 32-bit floating-point weights to a 5 or 8-bit cluster index, saving memory consumption.
Moreover, it allows
reducing and optimizing packet scheduling and on-chip communication in multi-HMC architectures.

The HIVE architecture \cite{Alves2016} extends the HMC ISA for performing common vector operations directly inside the HMC. By migrating ML kernels on near-data processing (NDP) architectures capable of large-vector operations, the Vector-In-Memory Architecture (VIMA) proposed in \cite{Cordeiro2021} supports all ARM NEON Integer and floating-point instructions, operating over vectors of 8 KB of data by fetching data over the 32 channels (vaults) of the HMC in parallel. In this way, it leads to a significant speed-up and energy reduction with respect to an x86 baseline.


Several accelerators adopting the bank-level PIM approach can be found in the literature. 
The Newton fixed data flow accelerator proposed in \cite{He2020} employs only MAC units, buffers, and a DRAM-like command interface with the host CPU, avoiding the overhead and granularity issues of launching the kernel and switching between the PIM/non-PIM operational modes. The output vector write traffic is reduced by means of an unusually wide interleaved layout (DRAM row-wide). Moreover, input/output vectors have high reuse while the matrix has no reuse.



HBM-PIM \cite{Kwon2021} implements a function-in-memory DRAM (FIMDRAM) that integrates a 16-wide 
SIMD engine within the memory banks exploits bank-level parallelism to provide 4 $\times$ higher processing bandwidth than an off-chip memory solution (Figure~\ref{fig:HBM-PIM}).
Each computing unit (PCU) is shared among two banks, and there are 8 PCUs per pseudo-channel.
The PCU is divided into a register group, an execution unit, a decoding unit for parsing instructions needed to perform operations, and interface units to control data flow.
The register group consists of a command-register file for instruction memory (CRF), a general-purpose register file for weight and accumulation (GRF), and a scalar register file to store constants for MAC operations (SRF).
The PIM controller is integrated to support the programmability of the PCU and
the seamless integration with the host by determining the switching between the PIM/non-PIM operational modes.
If the PIM mode is asserted, the PCUs execute the instructions pre-stored in the CRF, incrementing the program counter every time a DRAM’s read command is issued.
3D-stacked PIM has also been proposed for accelerating applications loosely related to DNNs. We present a brief overview of these accelerators in Section A.7 of the Appendix A.
\begin{table}[t]
\caption{Summary of IMC accelerators based on RRAM and PCM memories}
\resizebox{\columnwidth}{!}{%
\begin{tabular}{@{}lllllrrrrr@{}}
\toprule
\textbf{Accelerator} & \textbf{Technology} & \textbf{Process} & \textbf{Application} &\textbf{Area [mm$^2$]} & \textbf{Power [mW]} & \textbf{Performance [GOPS]} & \textbf{EE [GOPS/W]} & \textbf{AE [GOPS/mm$^2$]} & Maturity-level\\ \midrule
ISAAC \cite{shafiee2016isca} & RRAM+CMOS & 32 nm & CNN & 85.4 & 65800 & - & 380.7 & 466.8 & Simulation\\
PipeLayer \cite{song2017hpca} & RRAM+CMOS & - & CNN & 82.63 & - & - & 140 & 1485 & Simulation\\
NeuralPIM \cite{cao2022tcomp} & RRAM+CMOS & 32 nm & CNN+RNN & 86.4 & 67700 & - & 2040.6 & 1904 & Simulation\\
PRIME \cite{chi2016isca} & RRAM+CMOS & 65 nm & MLP+CNN & - & - & - & 2100 & 1230 & Simulation\\
NeuRRAM \cite{wan2022nature} & RRAM+CMOS & 130 nm & CNN+RNN+RBN & 159 & 49.7 & 2135 & 43000 & - &Layout\\
Hermes \cite{khaddam2022hermes} & PCM+CMOS & 14 nm & MLP+CNN+LSTM & - & - & - & 10500 & 1590 &Silicon\\
\bottomrule
\end{tabular}
}
\label{tab:dnn_accelerators_rram_pcm}
\end{table}

\subsection{In-Memory Computing}
\label{ssec:NPU3}
In-memory computing (IMC) has been proposed to break both the memory and the compute wall in data-driven AI workloads, using either SRAM or emerging memory technologies (such as PCM and RRAM described in Section A.8 of Appendix A) integrated in a dedicated accelerator (Table \ref{tab:dnn_accelerators_rram_pcm}).

Full-digital IMC designs offer a fast path for the integration of the next generation of neural processing systems like NPUs. 
Recently, STMicroelectronics proposed a scalable and design time parametric IMC-NPU relying on digital SRAM IMC for edge AI \cite{desoli2023isscc}. This architecture is the evolution of the Orlando SoC \cite{desoli17isscc} and is specialized in accelerating the inference workloads. When manufactured in 18 nm FDSOI technology, this IMC-NPU achieves an energy efficiency of 77 TOPS/W and an area efficiency of 13.6 TOPS/mm$^2$. With its four key features and dedicated hardware able to lower the activity of the memory early terminating the operations when needed, NeuroCIM \cite{kim2022vlsi} achieves 310.4 TOPS/W.
The ISAAC non-volatile inference-based machine on RRAM technology   \cite{shafiee2016isca} is a tile-based architecture 
for CNN processing which combines the data encoding and the processing steps within \textit{in situ} MAC units (IMA). The design is pipelined fetching data from an external eDRAM chip to the computing tile. The data format in ISAAC is fixed at 16-bit. During computation, at each clock cycle, 1-bit is given as input to the IMA, whose result is converted to the digital format, thus requiring 16 clock cycles to process the input. Such a design allows implementing the computation on different tiles in a fully pipelined approach to increase computing performance and throughput. The PipeLayer \cite{song2017hpca} architecture introduces intra-layer parallelism and an inter-layer pipeline for tiled architecture, using duplicates of processing units featuring the same weights to process multiple data in parallel.
Designs like PRIME \cite{chi2016isca} take part of the RRAM memory arrays to serve as acceleration instead of adding an extra processing unit for computation. 
As outlined in \cite{cao2022tcomp}, existing PIM RRAM accelerators suffer from frequent and energy-intensive analog-to-digital (A/D) conversions, severely limiting their performance. To efficiently accelerate DL tasks by minimizing the required A/D conversions, a new architecture was presented with analog accumulation and neural-approximated peripheral circuits. 
The new dataflow introduced in \cite{cao2022tcomp} remarkably reduces the required A/D conversions for matrix-vector multiplications by extending shift-and-add operations to the analog domain before the final quantization.
A summary of the technological features in major RRAM accelerators can be found in \cite{smagulova2023pieee}.

The first PCM-based silicon demonstrator for DNN inference is Hermes \cite{khaddam2022hermes} which consists of a 256x256 PCM cross-bar and optimized ADC circuitry to reduce the read-out latency and energy penalty. The SoC is implemented in 14nm technology, showing 10.5 TOPS/W energy efficiency and performance density of 1.59 TOPS/mm$^2$. The same 256x256 PCM cross-bar has been integrated into a scaled-up mixed-signal architecture that targets the inference of long short-term memory (LSTM) and ResNet-based neural networks~\cite{gallo202264}. The chip, implemented in the same 14nm technology, consists of 64 analog cores interconnected via an on-chip communication network and complemented with digital logic to execute activation functions, normalization, and other kernels than Matrix-Vector Multiplications (MVMs). The accelerator achieves a peak throughput of 63.1 TOPS with an energy efficiency of 9.76 TOPS/W for 8-bit input/8-bit output MVM operations.

Besides silicon stand-alone demonstrators, the PCM technology is evaluated from a broader perspective in heterogeneous architectures that target different classes of devices, from IoT end nodes to many-core HPC systems. Such studies aim to highlight and overcome the system-level challenges that arise when PCM technology is integrated into more complex mixed-signal systems. For example, Garofalo et al.~\cite{garofaloHeterogeneousInMemoryComputing2022} analyze the limited flexibility of AIMC cores that can only sustain MVM-oriented workloads, but they are inefficient in executing low-reuse kernels and other ancillary functions such as batch-normalization and activation functions. To better balance Amdahl’s effects that show up on the execution of end-to-end DNN inference workloads, they propose as a solution an analog-digital edge system that complements the computing capabilities of PCM-based accelerators with the flexibility of general-purpose cores. The architecture, benchmarked on a real-world MobileNetV2 model, demonstrates significant advantages over purely digital solutions. 
Bruschi et al.~\cite{bruschiEndtoEndDNNInference2022} leave the edge domain to study the potentiality of PCM-based AIMC in much more powerful HPC many-core systems. The work presents a general-purpose chipset-oriented architecture of 512 processing clusters, each composed of RISC-V cores for digital computations and nvAIMC cores for analog-amenable operations, such as 2D convolutions. This system is benchmarked on a ResNet18 DNN model, achieving 20.2 TOPS and 6.5 TOPS/W. 

\subsection{Neuromorphic Accelerators}
\label{ssec:NPU4}
Neuromorphic computing represents a paradigm shift from Von Neumann-based architectures to distributed and co-integrated memory and PEs \cite{frenkel2021arxiv}. 
Neuromorphic chip architectures enable the hardware implementation of spiking neural networks (SNNs) \cite{rathi2023acm} and advanced bio-inspired computing systems that have the potential to achieve even higher energy efficiency with respect to DNN stand-alone accelerators described so far \cite{akopyan2015tcad}. 

SpiNNaker chip \cite{painkras2013jsscc} is a digital architecture designed on a 130 nm technology for SNN and neuroscience simulation acceleration. It is based on a distributed von Neumann approach using a globally asynchronous locally synchronous (GALS) design for efficient handling of asynchronous spike data. 
The SpiNNaker 2 system \cite{liu2018frontiers} uses a 22 nm technology and embeds 4 ARM Cortex M4F cores out of the planned 152 per chip. The objective is to simulate two orders of magnitude more neurons per chip compared to \cite{painkras2013jsscc}. However, it has been demonstrated that GPU-based accelerators compare favorably to a SpiNNaker system when it comes to large SNN and cortical-scale simulations \cite{knight2018frontiers}.

In comparison with the above-described accelerators, full-custom digital hardware leads to
higher-density and more energy-efficient neuron and synapse integration for SNN \cite{frenkel2021arxiv}. The 45 nm design in \cite{seo2011cicc} is a small-scale architecture embedding 256 Leaky-Integration-Fire (LIF) neurons and up to 64k synapses based on the Stochastic Synaptic Time Dependant Plasticity (S-STDP) concept. Due to its reasonably high neuron and synapse densities and energy-efficiency, this design is an ideal choice for edge computing scenarios. At the same integration scale, the ODIN chip embeds 256 neurons and 64k Spike Driven Synaptic Plasticity (SDSP)-based 4-bit synapses in a 28 nm CMOS process \cite{frenkel2018tbiocas}. A first attempt to scale up the NPU for SNN is represented by the 65 nm MorphIC chip, which is based on the ODIN core integrated into a quadcore design \cite{frenkel2019tbiocas}. 

Two large-scale neuromorphic platforms required for cognitive computing applications, are currently offered: the 28 nm IBM TrueNorth \cite{akopyan2015tcad} and the 14 nm Intel Loihi \cite{davies2018micro}. TrueNorth is a GALS design embedding as high as 1M neurons and 256M binary non-plastic synapses per chip, where neurons rely on a custom model that allows modifying their behaviors by combining up to three neurons \cite{cassidy2013ijcnn}. Loihi is a fully asynchronous design embedding up to 180k neurons and 114k (9-bit) to 1M (binary) synapses per chip. Neurons rely on a LIF model with a configurable number of compartments to which several functionalities such as axonal and refractory delays, spike latency, and threshold adaptation have been added. The spike-based plasticity rule used for synapses is programmable. Loihi will evolve then in the new Loihi 2 neuromorphic chip and TrueNorth into the NorthPole platform \cite{northpole}.

Digital designs for neuromorphic chips can obtain versatility with a joint optimization of power and area efficiencies. This flexibility is highlighted with platforms going from versatility-driven (e.g., SpiNNaker) to efficiency-driven (e.g., ODIN and MorphIC), through platforms aiming at a well-balanced trade-off on both sides (e.g., Loihi). Table \ref{tab:neuromorphic_chip} summarizes the main characteristics of the neuromorphic chips described so far with particular insight on the Energy per spike operation (SOP).

\begin{table}[t]
\caption{Summary of neuromorphic accelerators.}
\resizebox{\columnwidth}{!}{%
\begin{tabular}{@{}llrrrrlrr@{}}
\toprule
\textbf{Chip name} & \textbf{Technology} & \textbf{Cores} & \textbf{Core Area [mm$^2$]} & \textbf{Neurons per core} & \textbf{Synapses per core} & \textbf{Weights storage} & \textbf{Supply Voltage [V]} & \textbf{Energy per SOP [J]}\\ \midrule
SpiNNaker \cite{painkras2013jsscc} & 0.13 $\mu$m & 18 & 3.75 & 1000 & - & Off-chip & 1.2 & $>$11.3n/26.6n\\
\cite{seo2011cicc} & 45 nm SOI & 1 & 0.8 & 256 & 64k & 1-bit SRAM & 0.53 - 1.0 & -\\
ODIN \cite{frenkel2018tbiocas} & 28 nm FDSOI & 1 & 0.086 & 256 & 64k & (3+1)-bits (SRAM) & 0.55 - 1.0 & 8.4p/12.7p\\
MorphIC \cite{frenkel2019tbiocas} & 65 nm LP & 4 & 0.715 & 512 & 528k & 1-bit (SRAM) & 0.8 - 1.2 & 30p/51p\\ 
TrueNorth \cite{akopyan2015tcad} & 28 nm & 4096 & 0.095 & 256 & 64k & 1-bit (SRAM) & 0.7 - 1.05 & 26p\\
Loihi \cite{davies2018micro} & 14 nm FinFET & 128 & 0.4 & 1024 & 1M & 1- to 9 bits (SRAM) & 0.5 - 1.25 & $>$23.6p\\
\bottomrule
\end{tabular}
}
\label{tab:neuromorphic_chip}
\end{table}

\subsection{Accelerators based on Multi-Chip Modules}
\label{sec:mcm}

The alternate multichip-module (MCM) silicon interposer-based integration technology, described in Section A.9 of Appendix A, offers several advantages over single-chip designs, including increased functionality, reduced power consumption, higher performance, improved reliability, and cost savings.
By utilizing MCM, designers can combine multiple chips and functionalities into a single package, resulting in a reduced overall footprint and cost. 
Furthermore, MCM-based designs can utilize off-the-shelf components and existing manufacturing processes and technology, contributing to cost savings in overall manufacturing.

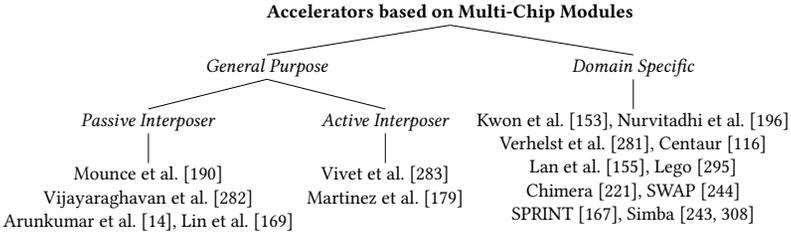
\begin{figure}[b]
    \centering
    {
        \small
        \scalebox{0.7}{
        \begin{forest}
            rounded/.style={ellipse,draw},
            squared/.style={rectangle,draw},
            qtree,
            [{\textbf{Accelerators based on Multi-Chip Modules}}
              [{\textit{General Purpose}}
                    [{
                      Mounce~\cite{mounce_ac16},
                      Vijayaraghavan~\cite{vijayaraghavan_hpca17}\\
                      Arunkumar~\cite{arunkumar_isca17},
                      Lin~\cite{lin_vlsi19}\\
                      Vivet~\cite{vivet_jssc21},
                      Martinez~\cite{martinez_vlsi20}
                     }]
              ],
              [{\textit{Domain-Specific}}
                [{
                  Kwon~\cite{kwon_iceic23}, Nurvitadhi~\cite{nurvitadhi_fccm19},       
                  Verhelst~\cite{verhelst2022ml}\\
                  Arvon~\cite{tang_jssc24}, 
                  Lan~\cite{lan_eptc21}, Lego~\cite{xuan_isocc22}
                  Chimera~\cite{prabhu2022chimera}\\
                  SWAP~\cite{sharma2022swap},
                  SPRINT~\cite{li2021sprint}, 
                  Simba~\cite{zimmer_jssc20,shao_micro19},
                  Instinct~\cite{smith_isscc24}, DOJO~\cite{talpes_micro23}
                }]
              ]
            ]
        \end{forest}}
        }
    \caption{Taxonomy of MCM based accelerators discussed in Section~\ref{sec:mcm}}
    \label{fig:mcm_taxonomy}
\end{figure}

Figure~\ref{fig:mcm_taxonomy} illustrates a comprehensive taxonomy of the MCM-based designs explored in this survey. Specifically, Section A.9.2 of Appendix A discusses a collection of representative general-purpose MCM-based designs, while this section concentrates on MCM-based DNN accelerators.

In the realm of DL, chiplet-based design is utilized to create hardware accelerators that are both efficient and scalable.
The chiplet-based design proposed in~\cite{kwon_iceic23} is a viable solution to provide higher performance at a lower cost compared to IP-based design. In~\cite{kwon_iceic23}, various aspects of designing a chiplet AI processor are considered, including incorporating NPU chiplets, HBM chiplets, and 2.5D interposers, ensuring signal integrity for high-speed interconnections, power delivery network for chiplets, bonding reliability, thermal stability, and interchiplet data transfer on heterogeneous integration architecture.

At the aim of balancing both data movement and compute capabilities of data-intensive DL algorithms, keeping the entire DL model on-chip is becoming the new norm for real-time services to avoid expensive off-chip memory accesses. In~\cite{nurvitadhi_fccm19} it is shown how the integration of FPGA with ASIC chiplets enhances on-chip memory capacity and bandwidth and provides compute throughput that outperforms GPU-based platforms (NVIDIA Volta). Specifically, the GPU and chiplet-based FPGA computing capabilities are 6\% and 57\% of their peak, respectively. Moreover, the FPGA achieves a delay that is 1/16 and energy efficiency that is 34x better than the GPU. 

In accordance with the recent trend in DL accelerators, chiplet integration is considered a promising implementation strategy for both homogeneous and heterogeneous multi-core accelerators to further increase throughput and match the ever-growing computational demands ~\cite{verhelst2022ml}. . 
In~\cite{lan_eptc21}, a chiplet-based architecture is proposed for a multi-core neuromorphic processor with a chip-package co-design flow. The proposed design is reusable for different neuromorphic computing applications by scaling the number of chips in a package and by reusing existing IPs from different technology nodes with 2.5D integration technology.
The MCM architecture presented in~\cite{xuan_isocc22} is a promising approach to address the issue of using modern DNN accelerators in multi-tenant DNN data centers, but it leaves the challenge of distributing DNN model layers with different parameters across chiplets still open. When  a dynamic scheduler is used to comply with the size of DNN model layers and increase chiplet utilization, the Lego MCM architecture achieves a 1.51x speedup over a monolithic DNN accelerator.
Chimera~\cite{prabhu2022chimera} is a non-volatile chip for DNN training and inference that does not require off-chip memory. Multiple Chimera accelerator chiplets can be combined in a multi-chip system to enable inference on models larger than the single-chip memory with only $5\%$ energy overhead.
The Arvon accelerator~\cite{tang_jssc24} is a heterogeneous System-in-Package that integrates a 14-nm FPGA chiplet with two 22-nm DSP chiplets using embedded multidie interconnect bridge technology. Arvon is designed to support various workloads, including neural network processing through a specific compilation procedure that optimizes workload distribution across the FPGA and DSP chiplets.

Notable examples of commercial chiplet-based hardware accelerators include AMD's Instinct~\cite{smith_isscc24} and Tesla's DOJO~\cite{talpes_micro23}. The AMD Instinct MI300 series are designed for HPC and AI at exascale levels. They utilize a modular chiplet architecture that integrates data center-class CPUs, GPU accelerated compute, AMD Infinity Cache, and 8-stack HBM3 memory in a single package. Specifically, the MI300X model targets traditional dual-processor CPU servers with eight GPU accelerators, host DDR, and device HBM for large AI model training and inference. The MI300A model combines three Zen 4 CPU chiplets with six CDNA 3 GPU chiplets for high-density HPC systems, facilitating seamless CPU-GPU interaction without explicit data transfers. The Tesla's DOJO system is an exascale computer designed for ML training applications. It features a homogeneous modular architecture with a hierarchical chiplet-based organization. Each chiplet, named D1 die, contains 354 DOJO nodes, each functioning as a full-fledged computer with dedicated processing pipelines, local memory, and network interfaces. A training tile is formed by integrating 25 D1 dies within a single package.

SWAP~\cite{sharma2022swap} is a DNN inference accelerator based on the 2.5D integration of multiple resistive RAM chiplets. In ~\cite{sharma2022swap},  a design space exploration flow is proposed to optimize the interconnection Network-on-Package, minimizing inter-chiplet communications and enabling link pruning. Further improvements are achieved in SPRINT~\cite{li2021sprint}, where a photonic-based interconnects is used as an alternative to metallic-based inter-chiplet networks.

\begin{figure}[t]
    \centering
    \includegraphics[width=0.8\textwidth]{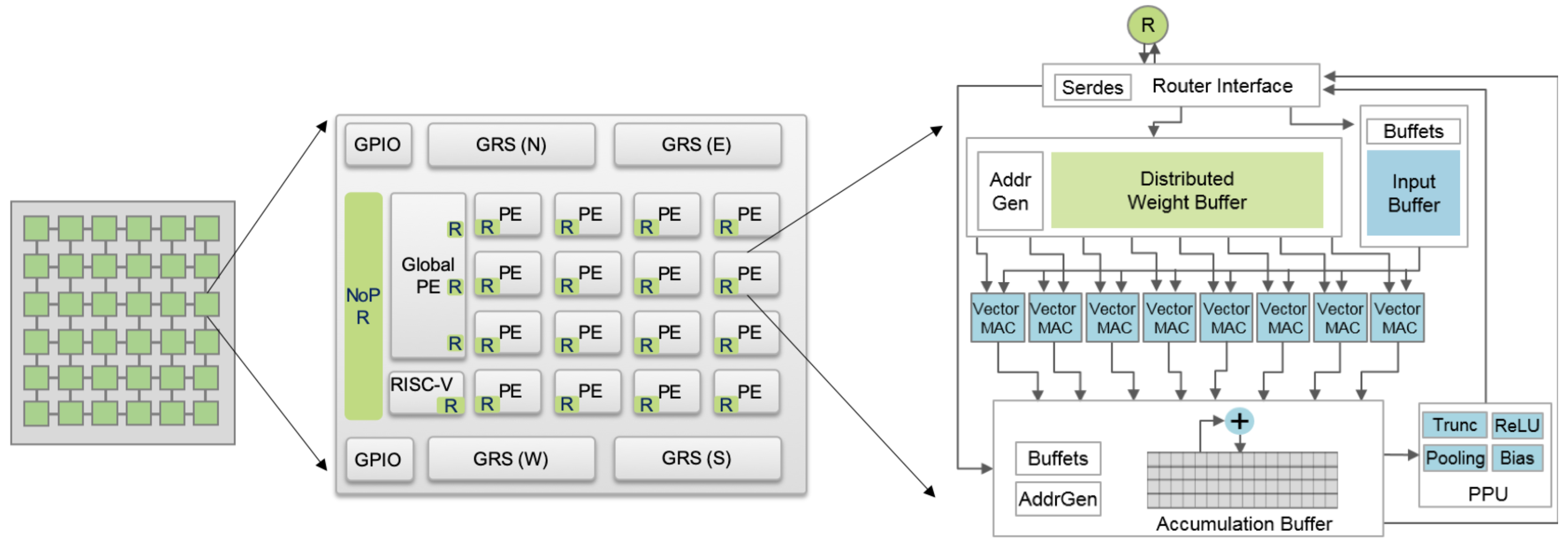}
    \caption{Simba architecture~\cite{shao_micro19} from left to right: package with 36 chiplets, chiplet, and processing element.}
    \label{fig:simba}
\end{figure}
Finally, as a representative chiplet-based DNN hardware accelerator, we report Simba~\cite{zimmer_jssc20,shao_micro19}. Simba is a scalable DNN accelerator consisting of 36 chiplets connected in a mesh network on a multi-chip module using ground-referenced signaling. Simba enables flexible scaling for efficient inference on a wide range of DNNs, from mobile to data center domains. The prototype achieves high area efficiency, energy efficiency, and peak performance for both one-chiplet and 36-chiplet systems. Simba architecture is shown in Figure~\ref{fig:simba}. It implements a tile-based architecture and adopts a hierarchical interconnect to efficiently connect different PEs. This hierarchical interconnect consists of a network-on-chip (NoC) that connects PEs on the same chiplet and a network-on-package (NoP) that connects chiplets together on the same package. Each Simba chiplet contains an array of PEs, a global PE, an NoP router, and a controller, all connected by a chiplet-level interconnect.
Table~\ref{tab:mcm_dnn_acc} presents a summary of the key characteristics of a representative subset of chiplet-based DNN accelerators that were reviewed earlier.
\begin{table}[b]
    \caption{Summary of Chiplet-based DNN Accelerators.}
    \label{tab:mcm_dnn_acc}
 \resizebox{0.75\columnwidth}{!}{%
    \begin{tabular}{lcccccc}
        \toprule
        & Simba \cite{shao_micro19} & Chimera \cite{prabhu2022chimera} & Arvon \cite{tang_jssc24} & Instinct \cite{smith_isscc24} & DOJO \cite{talpes_micro23} \\
        \midrule
        Technology & 16nm & 40nm & \makecell{14nm FPGA\\22nm DSP} & 6nm FinFET & 7nm\\
        Area & 6 mm$^{2}$* & 29.2 mm$^2$ & 32.3 mm$^2$ & - & 645 mm$^2$\\
        Power Efficiency &  9.1 TOPS/W** & 2.2 TOPS/W & 1.8 TFLOPS/W & 0.7 TFLOPS/W & 0.6 TFLOPS/W\\
        Throughput &4--128 TOPS & 0.92 TOPS & 4 TFLOPS & 383 TFLOPS & 362 TFLOPS\\
        Frequency &161 MHz--1.8 GHz & 200 MHz & 800 MHz & 1.7--2.1 GHz & 2 GHz \\
        Precisions & int8 & int8, fp16 & fp16 & multi & multi \\
        On-chip Memory & 752 KiB* & 2.5 MB\dag & - & 4 MB & 1.25 MB \\
        Chiplet Bandwidth & 100 GB/s & 1.9 Gbps & 7.68 Tb/s & 17.2 TB/s & 18 TB/s \\
        Interconnect & \makecell{Wired Mesh \\ (GRS)} & \makecell{Wired \\ App. specific\ddag} & Wired (EMIB) & \makecell{Infinity\\Fabric} & 2D mesh \\
        Applications & CNN Inference & \makecell{Inference \\ Training} & \makecell{NN, Comm.\\signal proc.} & \makecell{Inference\\Training} & \makecell{Training} \\
        \bottomrule
        \multicolumn{6}{l}{*One chiplet, **When operating at a minimum voltage of 0.42 V with a 161 MHz PE frequency}\\
        \multicolumn{6}{l}{\dag 2 MB RRAM, 0.5 MB SRAM, \ddag C2C links (77 pJ/bit, 1.9 Gbits/s)}
    \end{tabular}
  }
\end{table}

\subsection{Accelerators based on Quantum Computing and Photonic Computing}
Before concluding this Section, we would like to introduce some open challenges on two promising technologies to further speed up AI workloads: Quantum Computing and Photonic Computing.

There is a general agreement that Quantum computers will not replace conventional computing systems, but they will be used in combination with supercomputers to accelerate some hard-to-compute problems. Quantum computers will play the role of unconventional accelerators to outperform conventional supercomputers, thanks to the improved parallelism that enables the so-called \emph{quantum speedup}. 
Governments, supercomputing centers, and companies around the world have also started to investigate $How$/$When$/$Where$ quantum processing units (QPUs) could fit into HPC infrastructures to speed up some heavy tasks, such as DL workloads. Emerging trends and commercial solutions related to \emph{hybrid} quantum-classical supercomputers are described in ~\cite{HPCWIRE2022}.
To address this challenging trend, in October 2022, the EuroHPC Joint Undertaking initiative selected six supercomputing centers across the European Union to host quantum computers and simulators.
IBM Research was the first provider to offer a cloud-based QC service. IBM Qiskit ~\cite{QISKIT2023} is an open-source SDK based on a library of quantum gates/circuits: 
Remote users can develop quantum programs and execute them on quantum simulators and cloud-based quantum processors. 
Cloud providers have also jumped into the quantum race. 
As an example, Amazon Braket~\cite{BRAKET2023} is a QC service based on different types of quantum systems and simulators, including the quantum annealer from D-Wave. 
On this trend, there is a general agreement that GPUs will play a key role in hybrid quantum-classical computing systems. GPU company NVIDIA 
offers CuQuantum DGX hardware appliance integrating a software container on a full-stack quantum circuit simulator: The system uses NVIDIA’s A100 GPUs to accelerate quantum simulation workloads.

Recently, a survey on QC technologies appeared in ~\cite{GYONGYOSI201951}, while another survey on QC frameworks appeared in ~\cite{upama2022}. More specifically, there is a promising research trend on \emph{Quantum Machine Learning} ~\cite{Biamonte2017} which aims at developing quantum algorithms that outperform classical computing algorithms on ML tasks such as recommendation systems.
More in detail, classical DNNs inspired the development of \emph{Deep Quantum Learning} methods. The main advantage of these methods is that they do not require a large, general-purpose quantum computer. Quantum annealers, such as the D-Wave commercial solutions~\cite{D-WAVE2023}, are well-suited for implementing deep quantum learners. Quantum annealers are special-purpose quantum processors that are significantly easier to construct and scale up than general-purpose quantum computers. Following this research trend, Google proposed TensorFlow Quantum (TFQ) ~\cite{broughton2021}, an open-source quantum machine learning library for prototyping hybrid quantum-classical ML models.


The second challenging and promising research direction is represented by the use of Photonic Computing to further accelerate DL tasks. Photonic Computing relies on the computation of electromagnetic waves typically via non-linear modulation and interference effects. It was originally introduced in the 1980s to address optical pattern recognition and optical Fourier transform processing~\cite{ambs_2010}. Despite the potential advantages of processing parallelism and speed, optical computing has never translated into a widely adopted commercial technology. Only recently, due to the emergence of data-intensive computing tasks, such as AI, optical computing has seen a renewed interest.
There are two main advantages of optical computing, namely (i) the inherent speed of signal transmission, where light pulses can be transferred without the typical RC delays and IR drop of electrical interconnects, and (ii) the inherent parallelism, where multiple wavelengths, polarizations, and modes can be processed by the same hardware (e.g., waveguides, interferometers, etc.), without interfering with each other. These properties can provide strong benefits to data-intensive computing tasks such as DL. 
Photonic computing represents a promising platform for accelerating AI. For instance, it has been estimated that photonic MAC operations can show significant improvements over digital electronics in terms of energy efficiency ($>10^2$), speed ($>10^3$), and compute density ($>10^2$)~\cite{nahmias_2020}. However, there are still many challenges to developing an industrially feasible photonic system. The main challenge is the area/energy inefficiency of processing across the mixed optical/electronic domain. Optical-electrical conversion and vice versa result in considerable overhead in terms of area and power consumption. To bridge this gap, the trend is developing silicon photonic integrated circuits (PICs) with increasing robustness, manufacturability, and scalability. 
Photonic computing essentially operates in the analog domain, thus accuracy is deeply affected by accumulated noise and imprecision of optical devices, such as electro-optic and phase change modulators. These challenges, similar to those arising in analog IMC, might be mitigated by hardware-aware training and system-level calibration techniques. 
\section{Conclusions}
\label{sec:conclusion}

The Deep Learning ecosystem based on advanced computer architectures and memory technologies spans from edge and IoT computing solutions to high-performance servers, supercomputers, and up to large data centers for data analytics. In this context, the main objective of this survey is to provide an overview of the leading computing platforms utilized for accelerating the execution and enhancing the energy efficiency of Deep Learning applications.
Although GPUs have been crucial to boosting the deep learning revolution, especially for crunching in parallel a large amount of data, they are not the \textit{panacea} for all types of AI-applications. There are plenty of much smaller and customized AI accelerators, boosting their energy efficiency to make them suitable for mobile resource-constrained devices at a reasonable market price and without relying on sending data to the cloud. This survey reviews not only GPU-based solutions and Tensor Processor Units, but also ASIC- and FPGA-based accelerators, Neural Processing Units, and customized co-processors based on the open-hardware RISC-V architecture. To push further on the more advanced AI solutions, the survey also describes accelerators based on emerging computing paradigms and technologies, such as 3D-stacked processing in memory, emerging non-volatile memories, Multi-Chip Modules, chiplets, quantum-and photonic-based accelerating solutions.  


\ifdefined\arXiv
    \appendix
    \section{Appendix}
\label{sec:Appendix}
\subsection{Deep Learning Background: Concepts and Terminology}

Deep Learning \cite{Schmidhuber14, lecun2015} is a subset of 
ML methods that use artificial 
DNNs for automatically discover the representations needed for feature detection or classification from large data sets, by employing multiple layers of processing to extract progressively higher-level features. 
DNNs mimic human brain functionalities, in which neurons are interconnected with each other to receive information, process it, and pass it to other neurons. As shown in Figure~\ref{fig:perceptron}, in a way similar to the brain’s neuron, the simple model of a perceptron (artificial neuron) receives information from a set of inputs and applies a nonlinear function F (activation function) on a weighted (W) sum of the inputs (X) \cite{Rosenblatt1957}.
DNNs are composed of a number of layers of artificial neurons (hidden layers), organized between the input layer, which brings the initial data into the system, and the output layer, in which the desired predictions are obtained (see Figure~\ref{fig:neuralnetwork}).  In \textit{feed-forward networks}, the outputs of one layer become the inputs of the next layer in the model, while in \textit{recurrent networks}, the output of a neuron can be the input of neurons in the same or previous layers. The term “deep” in DNNs refers to the use of a large number of layers, which results in more accurate models that capture complex patterns and concepts.

\begin{figure} [b!]
  \centering
    \begin{subfigure}[c]{0.4\textwidth}
        \centering
        \includegraphics[width=\textwidth]{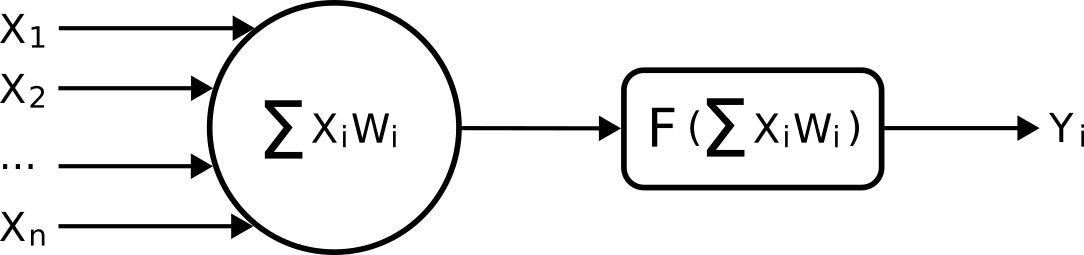}
         \caption{}
         \label{fig:perceptron}
    \end{subfigure}
    \qquad
    \begin{subfigure}[c]{0.4\textwidth}
         \centering
         \includegraphics[width=50mm]{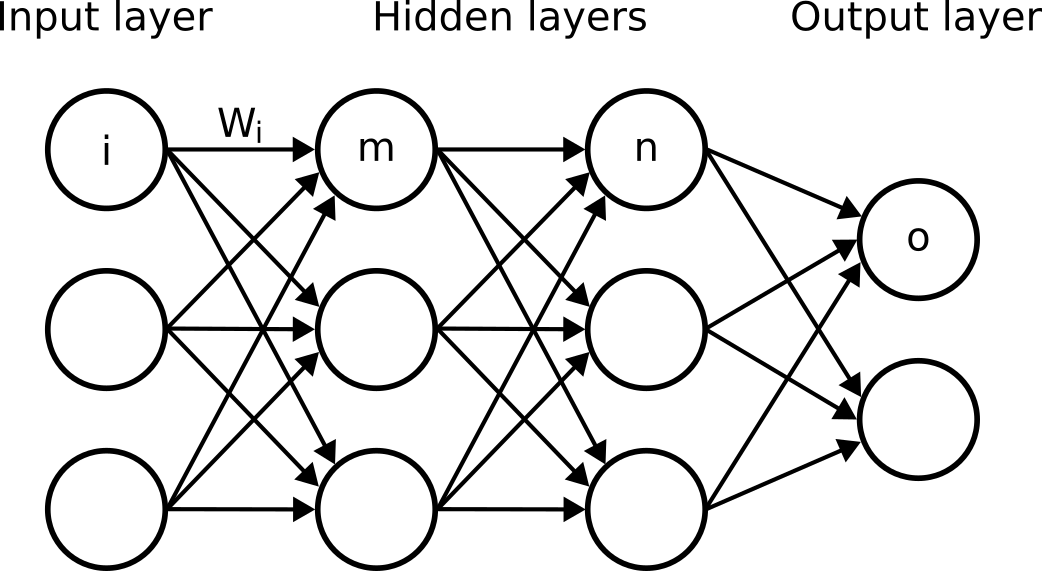}
         \caption{}
         \label{fig:neuralnetwork}
    \end{subfigure}
    \caption{Model of a perceptron (artificial neuron) (a) and of a multi-layer DNN (b).}
    \label{fig:perc_DNN}
\end{figure}
There are two phases in a DNN’s operation: training, and inference.
In the \emph{training} phase, the neural network model is fed on a curated data set so that it can “learn” everything it needs to about the type of data it will analyze. 
In the case of \emph{supervised} learning, a large set of examples and their corresponding labels indicating the correct classification are passed as input to the DNN. 
A forward pass is executed, and the error against the correct labels is measured. 
Then, the error is used in the DNN’s backward pass to update the weights. 
This loop is performed repeatedly until the DNN model achieves the desired accuracy. 
In \emph{unsupervised} learning, the DNN uses unlabeled data to create an encoded self-organization of weights and activations that captures patterns as probability densities.
With \emph{semi-supervised} learning, during training a small amount of labeled data is combined with a large amount of unlabeled data.
In the \emph{inference} phase, the trained DNN model is used to make predictions on unseen data.
When it comes to deployment, the trained model is often modified and simplified to meet real-world power and performance requirements.
The two phases present different computational characteristics.
On the one hand, the training phase of a model is computationally expensive but usually performed only once. On the other hand, the trained model is used for predictions on multiple input data, often under strict latency and/or energy constraints.

Three general types of DNN are mostly used today: Multi-Layer Perceptrons (MLPs), Convolutional Neural Networks (CNNs), and  Recurrent Neural Networks (RNNs).
MLPs \cite{Rosenblatt1957} are feed-forward ANNs composed of a series of fully connected layers, where each layer is a set of nonlinear functions of a weighted sum of all outputs of the previous one. On the contrary, in a CNN \cite{Lecun1998}, a convolutional layer extracts the simple features from the inputs by executing convolution operations. Each layer is a set of nonlinear functions of weighted sums of different subsets of outputs from the previous layer, with each subset sharing the same weights. Each convolutional layer in the model can capture a different high-level representation of input data, allowing the system to automatically extract the features of the inputs to complete a specific task, e.g., image classification, face authentication, and image semantic segmentation. Finally, RNNs \cite{Schmidhuber14} address the time-series problem of sequential input data. Each RNN layer is a collection of nonlinear functions of weighted sums of the outputs of the previous layer and the previous state, calculated when processing the previous samples and stored in the RNN’s internal memory. RNN models are widely used in Natural Language Processing (NLP) for natural language modeling, word embedding, and machine translation.

On the one hand, DNNs such as AlexNet\cite{Krizhevsky2012} and the more recent GoogLeNet\cite{Szegedy2015} are composed of tens of layers, with millions of weights to be trained and used in every prediction, requiring tens to hundreds of megabytes (or even gigabytes) of memory for their storage.
The calculation of the weighted sums requires a large number of data movements between the different levels of the memory hierarchy and the processing units, often posing a challenge to the available energy, memory bandwidth, and memory storage of the computing architecture.
On the other hand, Tiny machine learning (TinyML) DNN models \cite{Warden2019} have been investigated to run on small, battery-operated devices like microcontrollers, trading off prediction accuracy with respect to low-latency, low-power, and low-bandwidth model inference of sensor data on edge devices.

Besides DNNs, the other major category of DL algorithms is that of Transformer-based models~\cite{vaswaniAttentionAllYou2017}, which have recently captured great attention.
Transformers were originally proposed for NLP~\cite{vaswaniAttentionAllYou2017}, and are designed to recognize long-distance dependencies between data by means of \emph{attention} layers.
In attention layers, the weights used to linearly transform the input data are computed dynamically based on the input data itself.
Transformer models are flexible (e.g., they can also be used for vision tasks~\cite{dosovitskiyImageWorth16x162021}) and, most importantly, empirical scaling laws~\cite{kaplanScalingLawsNeural2020} govern their expressiveness: larger transformers trained for more time on larger datasets deliver better performance in consistent and predictable way capabilities.
This makes it possible to construct and pre-train larger and larger models (e.g., GPT-4~\cite{openaiGPT4TechnicalReport2023}, PaLM~\cite{anilPaLMTechnicalReport2023}, LLaMA~\cite{touvronLLaMAOpenEfficient2023}) with hundreds of billions of trillions of parameters, which can be then used as \emph{Foundation Models}~\cite{bommasaniOpportunitiesRisksFoundation2022} to be fine-tuned for more focused applications.

\subsection{Technology for GPU and TPU Architectures}
In this sub-section, we review the basic features of NVIDIA GPU architectures to boost the performance of HPC and Deep Learning applications.
The technological development of the last ten years significantly increased the computing power of GPU devices, which, due to their highly parallel nature, are incidentally very well suited to accelerate neural network training algorithms. 

The hardware architecture of a GPU is based on a multicore design of processing elements called {\em Streaming Multiprocessors} (SM). 
Each SM, in turn, includes a number of compute units, called CUDA-cores in NVIDIA jargon, to execute at each clock-cycle multiple warps, i.e. groups of 32 operations called CUDA-threads processed by the {\em Single Instruction Multiple Thread} (SIMT) fashion. SIMT execution enables different threads of a group to take different branches (with a performance penalty).
By varying CPU threads, context switches among active CUDA threads are very fast. 
Typically one CUDA-thread processes one element of the target data set. This helps to exploit the available parallelism of the algorithm and to hide the latency by swapping among threads waiting for data coming from memory and threads ready to run.
This structure remained stable across generations, with several enhancements implemented in the most recent architectures making available more registers addressable to each CUDA thread. 
Considering each generation of NVIDIA architecture, some minor differences occurred. 
The C2050 and C2070 boards based on the {\em Fermi} processor architecture differ in the amount of available global memory. Both cards have a peak performance of $\approx 1$ TFLOPS in single-precision (SP), and $\approx 500$ GFLOPS in double-precision (DP), and the peak memory bandwidth is $144$ GB/s.

The K20, K40, and K80 are boards based on the {\em Kepler} architecture. The K40 processor has more global memory than the K20 and slightly improves memory bandwidth and floating-point throughput, while the K80 has two enhanced {\em Kepler} GPUs with more registers and shared memory than K20/K40 and extended GPUBoost features. On the {\em Kepler} K20 and K40, the peak SP (DP) performance is $\approx 5$ TFLOPS ($\approx 1.5$ TFLOPS), while on the K80 the aggregate performance of the two GPUs delivers a peak SP (DP) of $\approx 5.6$ TFLOPS ($\approx 1.9$ TFLOPS). The peak memory bandwidth is $250$ and $288$ GB/s respectively for the K20X and the K40 while on the K80 the aggregate peak is $480$ GB/s.

The P100 board is based on the {\em Pascal} architecture, engineered to tackle memory challenges using stacked memory, a technology that enables multiple layers of DRAM components to be integrated vertically on the package along with the GPU. The P100 is the first GPU accelerator to use High Bandwidth Memory 2 (HBM2) to provide greater bandwidth, more than twice the capacity, and higher energy efficiency, compared to off-package GDDR5 used in previous generations. The SXM-2 version of P100 board also integrates the NVLinks, NVIDIA’s new high-speed interconnect technology for GPU-accelerated computing significantly increasing performance for both GPU-to-GPU communications and for GPU access to system memory. The P100 delivers a peak performance of $\approx 10.5$ TFLOPS SP and $\approx 5.3$ in DP, while the peak memory bandwidth has been increased to $732$ GB/s.

The {\em Volta} architecture has been developed and engineered for the convergence of HPC and AI. Key compute features of Tesla V100 include a new SM Architecture Optimized for Deep Learning, integrating Tensor Cores designed specifically for deep learning. Also, the Tesla V100 board integrates a second-generation NVLink supporting up to 6 links at 25 GB/s for a total of 300 GB/s, and1 6GB of HBM2 memory subsystem delivering 900 GB/sec peak memory bandwidth provides 1.5x delivered memory bandwidth versus Pascal GP100. {\em Volta} increases the computing throughput to 7.5 TFLOPS DP, and the memory bandwidth to 900 GB/s, respectively a factor 1.4X and 1.2X w.r.t. the Pascal architecture. 

The {\em Ampere} architecture adds a powerful new generation of Tensor Core that boosts throughput over V100 for Deep Learning applications running 10x faster. The peak performance in DP has been increased to 9.7 TFLOPS, and to 19.5 TFLOPS using Tensor Core or single precision FP32 operations. The A100 40GB of high-speed HBM2 memory with a peak bandwidth of 1555 GB/sec, corresponding to a 73\% increase compared to the Tesla V100. It also supports a third-generation of NVIDIA NVLink with a data rate of 50 Gbit/sec per signal pair, nearly doubling the 25.78 Gbits/sec rate in V100.

The {\em Hopper} is the latest architecture developed by NVidia providing a new generation of streaming multiprocessors with several new features. Tensor Cores are up to 6x faster chip-to-chip compared to A100, the memory subsystem is based on HBM3 modules providing  nearly a 2x bandwidth increase over the previous generation, and integrate a fourth-generation of NVlinks providing a 3x bandwidth increase. The peak performance is boosted up to 24 TFLOPS in DP, and 48 TFLOPS using FP64 tensor core and FP32 operations. The H100 SXM5 GPU raises the bar considerably by supporting 80 GB (five stacks) of fast HBM3 memory, delivering over 3 TB/sec of memory bandwidth, effectively a 2x increase over the memory bandwidth of the A100 that was launched just two years ago. The PCIe H100 provides 80 GB of fast HBM2e with over 2 TB/sec of memory bandwidth. The H100 also introduces DPX instructions to accelerate the performance of Dynamic Programming algorithms. These new instructions provide support for advanced fused operands for the inner loop of many dynamic programming algorithms. This leads to dramatically faster times-to-solution in disease diagnosis, logistics routing optimizations, and even graph analytics. For a more complete description, we can refer to \cite{fermi,kepler,pascal,volta,Ampere,h100}, while the work in \autoref{gpu-evolution} summarizes just a few relevant parameters of NVIDIA GPU architectures.
\begin{table}
\caption{Summary of hardware features of NVIDIA GPU architectures.}
\centering
\resizebox{\textwidth}{!}{
\begin{tabular}{lrrrrlrrrr}
\toprule
Architecture                   &   Fermi         & Kepler& Kepler   & Kepler &                         & Pascal & Volta & Ampere &  Hopper \\
GPU                            &   GF100         & GK110 & GK110B   & GK210 &\hspace{-1em} $\times$ 2 & P100 & V100 & A100 & H100 \\
\midrule
Year                           &  2011           & 2012  & 2013     & 2014  &                         & 2016 & 2017 & 2021 &  2022 \\
\midrule
\#SMs                          &   16            & 14    & 15       & 13    &\hspace{-1em} $\times$ 2 & 56   & 80   &  108 &   132 \\
\#CUDA-cores                   &   448           & 2688  & 2880     & 2496  &\hspace{-1em} $\times$ 2 & 3584 & 5120 & 6912 & 16896 \\ 
Base clock (MHz)               &   1.15          & 735   & 745      & 562   &                         & 1328 & 1370 & 1700 &  1600 \\
Base DP (GFLOPS)               &   515           & 1310  & 1430     & 935   &\hspace{-1em} $\times$ 2 & 4755 & 7000 & 9700 & 30000 \\
\midrule     
Total available memory (GB)    &   3             & 6     & 12       & 12    &\hspace{-1em} $\times$ 2 & 16   &   16 &   40 &    80 \\
Memory bus width (bit)         &   384           & 384   & 384      & 384   &\hspace{-1em} $\times$ 2 & 4096 & 4096 & 5120 &  5120 \\
Peak mem. BW  (GB/s)           &   144           & 250   & 288      & 240   &\hspace{-1em} $\times$ 2 & 732  &  900 & 1555 &  3072 \\
\bottomrule
\end{tabular}
}
\label{gpu-evolution}
\end{table}

The number of GPU modules available within the DGX~\cite{dgx} line of server and workstation platforms varies from 4 to 16 Tesla daughter cards integrated into the system using a version of the high-bandwidth SMX\cite{smx} socket solution.
The DGX-1 server, the first of DGX line, was announced in 2016, and it was first based on 8 Pascal cards, after upgraded to Volta, interconneced by an NVLink mesh network. The Pascal-based DGX-1 delivered 170 TFLOPS using FP16 half-precision processing, while the Volta-based upgrade increased this to 960 TFLOPS using FP16 tensor computing.
The DGX-2, the successor of DGX-1, was announced in 2018; it is based on 16 V100 32 GB GPU cards in a single unit interconnected by a NVSwitch~\cite{nvlink} for high-bandwidth GPU-to-GPU communications, and delivers nearly 2 PFLOPS using FP16 tensor processing, ans assemble a total of 512~GB of HBM2 memory.
The DGX Station is a workstation designed as a deskside AI system that can operate completely independent without the typical infrastructure of a datacenter. The DGX Station is a tower chassis, and the first available was including four Testa V100 accelerators each with 16~GB of HBM2 memory, delivering an aggregate computing performance of nearly 500 TFops using FP16 tensor computing.
The Ampere version of the DGX Station includes four A100 accelerators configured with either 40 or 80 GB of memory each, resulting either in 160~GB or 320~GB variants, and a peak FP16-tensor computing performance of approximately 1 PFLOPS.
The DGX A100 server is the 3rd generation of DGX servers announced in 2002. It includes 8 A100 accelerators, and it is the first DGX server replacing the Intel Xeon CPUs with the AMD EPYC CPUs, delivering a peak FP16-tensor computing performance of approximately 2.5 PFLOPS. 
The DGX H100 Server has been announced in 2022, and it is the 4th generation of DGX servers. It includes 8 Hopper H100 cards delivering a total of 16 PFLOPS of FP16-tensor AI computing, and assembling a total of 640 GB of HBM3 memory.  
The DGX SuperPod is a high-performance turnkey supercomputer solution based on DGX hardware, combining high-performance DGX compute nodes with fast storage and high bandwidth networking, that can be used as building-block to assemble large supercomputer systems. 
The Selene Supercomputer, installed at the Argonne National Laboratory, is one example of a DGX SuperPod-based system, built from 280 DGX A100 nodes. 
The new version of SuperPod based on H100 DGX can scale up to 32 nodes, for a total of 256 H100 GPUs and 64 x86 CPUs. This gives the complete SuperPod a total 20TB of HBM3 memory, 70.4 TB/s of bisection bandwidth, and up to 1 EFlop of FP8 and 500 PFLOPS of FP16 tensor AI computing.
The Eos\cite{eos} supercomputer announced in March 2022, designed, built, and operated by NVIDIA, is based on 18 H100 SuperPods, for a total of 576 DGX H100 systems. 
This allows Eos to deliver approximately 18 EFLOPS of FP8 and 9 EFLOPS of FP16 computing, making Eos the fastest AI supercomputer in the world.
Table \autoref{gpu-platforms} summarizes the computing performance of a few DGX systems. We report the peak computing performance using tensor FP16 operations relevant for AI applications, and the standard FP32 and FP64 relevant for many scientific applications.

\begin{table}
\caption{Performance in TFLOPS of DGX based platforms; for H100 platforms, sparsity features are used.}
\begin{tabular}{lrrrr}
\toprule
Platform          & \#GPUs    & FP16 Tensor  & F32     &  FP64  \\
\midrule
DGX1-P100         &   8x P100 & --           & 85      &     42 \\
DGX1-V100         &   8x V100 & 1000         & 124     &     62 \\
DGX2              &  16x V100 & 2000         & 248     &    124 \\
DGX-A100 Server   &   8x A100 & 2496         & 154     &     77 \\
DGX-H100 Server   &   8x H100 & 16000        & 544     &    272 \\
\midrule
Supercomputer     &                          &         &        \\
\midrule
Selene            &2240x A100 &  698880      &  43120  &  21560 \\
Eos               &4608x H100 & 9216000      & 313344  & 156672 \\
\bottomrule
\end{tabular}
\label{gpu-platforms}
\end{table}
The TPU presented in TPU~\cite{Jouppi_2017,jouppiMotivationEvaluationFirst2018} is centered on a large (256$\times$256) systolic array operating on signed or unsigned 8-bit integers and targeting exclusively data center inference applications; this is coupled with a large amount of on-chip SRAM for activations (24 MiB) and a high-bandwidth (30 GiB/s) dedicated path to off-chip L3 DRAM for weights.
The next design iterations (TPUv2, TPUv3)~\cite{jouppiDomainspecificSupercomputerTraining2020} forced to move from an inference-oriented design to a more general engine tuned for both inference and training, employing the 16-bit BF16 floating-point format, more cores (2 per chip) using each one or two 4$\times$ smaller arrays than TPUv1 (128$\times$128, to reduce under-usage inefficiencies).
TPUv2/v3 also introduced high-bandwidth memory support, which results in more than 20$\times$ increase in the available off-chip memory bandwidth.
Conversely, Goya \cite{medina2019hotchips} relies on PCIe 4.0 to interface to a host processor and exploits a design that uses a heterogenous approach comprising of a large General Matrix Multiply (GMM) engine, TPUs, and a large shared DDR4 memory pool. Each TPU also incorporates its own local memory that can be either hardware-managed or fully software-managed, allowing the compiler to optimize the residency of data and reduce movement. Each of the individual TPUs is a VLIW design that has been optimized for AI applications. The TPU supports mixed-precision operations including 8-bit, 16-bit, and 32-bit SIMD vector operations for both integer and floating-point. Gaudi has an enhanced version of the TPUs and uses HBM global memories rather than the DDR used in Goya, increasing the support towards bfloat16 data types and by including more operations and functionalities dedicated to training operations. 
As a counterpart of smaller TPUs, NVIDIA TensorCores are small units, designed to perform a 4$\times$4$\times$4 FP16 GEMM operation per cycle in Volta (doubled in Ampere and quadrupled in Hopper, adding also support for other data types).
Performance is then obtained by parallelization: each Streaming Multiprocessor includes eight TensorCores controlled by 32 threads; and, depending on the specific chip, GPUs can contain tens of Streaming Multiprocessors.

\subsection{FPGA Technology}
FPGAs are semiconductor devices that provide a unique combination of flexibility and performance thanks to their fundamental building blocks, known as Configurable Logic Blocks (CLBs) or simply Logic Elements (LEs). They consist of look-up tables (LUTs) and flip-flops that can be used to implement arbitrary combinational and sequential bit-level operations, on the basis of user-defined tasks.
Programmable interconnects provide the necessary routing resources to establish connections between different elements within the device and to facilitate the seamless flow of data and control signals. 
FPGAs provide various types of memory resources that can be utilized for different purposes:
\begin{itemize}
    \item Block RAM (BRAM): specialized on-chip memory resource that offers dedicated storage for data or program code. BRAMs are characterized by their dual-port or true dual-port design, providing high bandwidth and low latency. Organized into fixed-sized blocks, BRAMs serve a variety of purposes including data buffering, cache memory, FIFO implementation, and storage for coefficients or tables utilized in digital signal processing applications. BRAMs play a crucial role in optimizing data access and manipulation within FPGA designs.
    \item Distributed RAM: memory elements that are distributed across the logic fabric of an FPGA. Unlike BRAM, which is dedicated memory, distributed RAM is implemented using the LUTs present in the CLBs. These LUT-based memories offer smaller storage capacity compared to BRAM but are more flexible and can be used for smaller data sets or temporary storage within the FPGA.
    \item Configuration Memory: a specialized form of memory dedicated to storing the configuration data necessary to define the desired behavior of the FPGA. This memory contains the bitstream or configuration file, which is loaded into the FPGA upon startup. Configuration memory can be implemented using diverse technologies, such as SRAM-based or flash-based configurations, offering flexibility in how the FPGA is programmed and initialized.
\end{itemize}
Moreover, in order to meet the demands of emerging technologies and applications, FPGAs provide the designers with specialized macros, such as Digital Signal Processors (DSPs) and embedded multipliers, that can be exploited to enhance processing capabilities, improve power efficiency, and increase the flexibility of hardware accelerators for DL. The latter exploit FPGAs mostly to accelerate inference, while training is delegated to GPUs: this reflects the differences between the two phases, as training is only executed once and requires high throughput, while for inference, especially on edge devices, latency, and power consumption become critical \cite{guo2019dl, blaiech2019survey}. 
FPGAs are also often used as a prototyping platform to explore different architectures before committing to ASIC manufacturing \cite{Spagnolo2022_3}.

\subsection{EDA Frameworks}
Implementing hardware accelerators for ML algorithms, particularly DNNs, is a complex task that is rarely addressed through manual coding in low-level  Hardware Description Languages (HDL).
When Register Transfer Level (RTL) design is required to achieve high performance, templated components may be used \cite{Suda2016}.
Instead, there are several electronic design automation (EDA) tools that bridge the gap between ML models and FPGAs/ASICs, allowing researchers to focus on developing the algorithms at a high level of abstraction \cite{venieris2018toolflows}.

Vitis AI, Xilinx's development environment for AI inference \cite{VitisAI}, supports models developed in major frameworks such as PyTorch \cite{torch}, TensorFlow \cite{TF} and Caffe \cite{Caffe}, and maps them on deep learning processor unit (DPU) cores present on modern Xilinx boards alongside the standard FPGA logic.
The  work in \cite{Ajili2022} describes the implementation of DeepSense, a framework that includes CNN and RNN, with a focus on the choice of parameters to define DPUs used by Vitis AI; \cite{Vandendriessche2022} performs a parametric study of the DPU architecture used by Vitis AI and examines the tradeoffs between the resources used and the clock frequency, as well as their impact on power consumption; \cite{Wang2021} compares the FPGA implementation of YOLOv3 provided by Vitis AI with its GPU counterpart, showing higher throughput and lower power consumption; \cite{Ushiroyama2022} evaluates the implementation of three different CNNs in terms of precision, power consumption, throughput, and design man-hours, and compares these figures with their GPU counterparts.

High-Level Synthesis (HLS) plays a crucial role to automate the design of ML accelerators.
HLS tools such as Vitis HLS \cite{vitishls}, Bambu \cite{Bambu}, Intel HLS Compiler \cite{IntelHLS2022}, Catapult \cite{CatapultHLS2022}, Stratus HLS \cite{StratusHLS2022}, or LegUp \cite{LegUpHLS2013} provide users with a high level of abstraction where they can describe the desired functionality with a software programming language (C/C++/SystemC) and automatically obtain a corresponding high-performance HDL implementation. 
HLS thus boosts the productivity of hardware designers, who can benefit from faster design changes and functional verification.
In fact, HLS allows generating accelerators for different platforms (e.g., larger or smaller FPGAs) without altering the C/C++ source code apart from a few design directives. This makes it possible to explore the design space and find the best implementation much faster than with HDL design.
Note that code must be written with hardware knowledge in mind in order to meet given performance and resource usage results. Arbitrary software code, written for a CPU target, could achieve very low performance since it typically does not expose enough parallelism to exploit the spatial concurrency available on FPGA or ASIC.

There are also academic open-source HLS tools available, such as Bambu \cite{Bambu}. Bambu aids designers in the high-level synthesis of complex applications. It supports various C/C++ constructs and follows a software compilation-like flow. The tool consists of three phases: front-end, middle-end, and back-end. In the front end, the input code is parsed and translated into an intermediate representation. The middle-end performs target-independent analyses and optimizations. Lastly, the back-end synthesizes Verilog/VHDL code for simulation, logic synthesis, and implementation using external tools. Bambu offers a command-line interface and is particularly useful for designers seeking assistance in HLS and optimizing hardware designs efficiently.

In order to explore the acceleration of DNN inference on FPGAs, several frameworks and packages have been developed based on HLS.
They can be divided into two categories: tools based on libraries of HLS templates, such as FINN \cite{FINN} and hls4ml \cite{hls4ml}, and tools that use a compiler-based approach, such as SODA \cite{sodaMICRO} and ScaleHLS \cite{scalehls2022dac}.
In \cite{Machura2022}, a comparison between a custom implementation of two DNNs written in SystemVerilog and an implementation using the Xilinx tools FINN and Vitis AI is presented; a comparison between FINN and Vitis AI is reported in \cite{Hamanaka2023}, where a ResNet model is implemented using a widely used set of configurations of FINN and Vitis AI.
Both FINN and hls4ml use Vitis HLS as a backend; they parse a model exported from high-level ML frameworks and replace operators with C/C++ functions taken from a library of templates that already contains Vitis optimization directives.
The HLS tool processes the C/C++ code and produces a corresponding accelerator design.
The library of templates is necessarily tied to a specific HLS tool, and it requires expert HLS developers to implement in advance the best version of all necessary ML operators for a pre-determined backend tool.
On the other hand, SODA and ScaleHLS use a compiler infrastructure (MLIR, the Multi-Level Intermediate Representation from the LLVM project \cite{lattner2021mlir}) to progressively translate the input model through representations at different levels of abstraction, until they can be passed to the HLS tool as a C++ representation or an LLVM IR.
This second approach exploits the existing MLIR infrastructure for machine learning, without requiring to create and maintain a library of operators.
A hybrid RTL–HLS approach has been proposed in \cite{Guan2017} to improve performance and development time for various DL algorithms.

\subsection{Accelerating Arithmetic Data-paths}
Performance achievable with ASIC accelerators for the inference of DL models is mainly dependent on the structure of the arithmetic data path. At its core, DL systems perform several finite impulse response operations over a large set of data. Hence the accelerator can be optimized by exploiting techniques used for the efficient implementation of the underlying arithmetic operations. As shown in Figure ~\ref{fig:ASIC_datapath_taxonomy}, three main types of optimization can be performed on the arithmetic data path.
The first optimization approach takes into account that convolution is one of the main operations performed in a DL system. As demonstrated in \cite{Cheng_2004, Tsao_2012}, mono-dimensional convolutions can be efficiently computed by performing a reduced number of multiplications, thus improving the trade-off between the throughput of the circuit and the number of hardware resources needed for its implementation. This technique has been further developed in \cite{Wang_2018, Cheng_2020, Wang_2022}, where it is applied to multi-dimensional convolutions.

The multiplication itself can be implemented with optimized circuits. In \cite{MinJou_1999, Petra_2011, Frustaci2020} the area and power dissipation of the multiplier circuit is reduced by discarding part of the partial products used to compute the result. These circuits trade off precision and circuit complexity to improve speed and power consumption. This approach is often referred to as \emph{approximate computing paradigm}, providing a way to approximate the design at the cost of an acceptable accuracy loss. Approximate computing techniques proposed in \cite{Kulkarni_2011, Zervakis_2016} provide a reduced complexity multiplier by modifying the way the partial products are computed. In \cite{Zacharelos_2022}, a recursive approach is proposed, in which the multiplier is decomposed into small approximate units. In the approach proposed in \cite{Esposito_2017}, the approximation is implemented in the way the partial products are summed.
Finally, the approximate computing paradigm can also be implemented in the 4-2 compressors \cite{Ahmadinejad_2019, Yang_2015, Ha_2018, Strollo_2020, Park_2021, Kong_2021} that represent the atomic blocks used for the compression of the partial products.

Different from the previous works, the segmentation method aims at reducing the bit-width of the multiplicands. The approaches in \cite{Hashemi_2015, Vahdat_2019} describe a dynamic segmentation method in which the segment is selected starting from the leading one of the multiplicand binary representation. On the contrary, the paper in \cite{Narayanamoorthy_2015} proposes a static segmentation method, which reduces the complexity of the selection mechanism by choosing between two segments with a fixed number of bits. The paper in \cite{Strollo_2022} improves the accuracy of the static segmentation method multipliers by reducing the maximum approximation error, whereas in \cite{Li_2021} the authors propose a hybrid approach in which a static stage is cascaded to a dynamic one.

\begin{figure}[!t]
    \centering
    {
        \tiny
        \begin{forest}
            rounded/.style={ellipse,draw},
            squared/.style={rectangle,draw},
            qtree,
            [{\textbf{Arithmetic Data-path}}
              [{\textit{Convolution Optimization}}
                    [{
					Cheng \cite{Cheng_2004}\\
					Tsao \cite{Tsao_2012}\\
					Wang \cite{Wang_2018}\\
					Cheng \cite{Cheng_2020}\\
					Wang \cite{Wang_2022}\\
                    }]
              ],
              [{\textit{Approximate Computing}}
			    [{\textit{Multiplication}}
				  [{
				    Jou \cite{MinJou_1999}, Petra \cite{Petra_2011}\\
					Kulkarni \cite{Kulkarni_2011}, Zervakis \cite{Zervakis_2016}\\
					Zacharelos \cite{Zacharelos_2022}, Esposito \cite{Esposito_2017}\\
				  }]
				],
				[{\textit{4-2 compressors}}
				  [{
				    Ahmadinejad \cite{Ahmadinejad_2019}, Yang \cite{Yang_2015}\\
					Ha \cite{Ha_2018}, Strollo \cite{Strollo_2020}\\
					Park \cite{Park_2021}, Kong \cite{Kong_2021}\\
				  }]
				]
              ],
              [{\textit{Segmentation Methods}}
			    [{
				  Hashemi \cite{Hashemi_2015}\\
				  Vahdat \cite{Vahdat_2019}\\
				  Narayanamoorthy \cite{Narayanamoorthy_2015}\\
				  Strollo \cite{Strollo_2022}\\
				  Li \cite{Li_2021}\\
				}]
              ]
            ]
        \end{forest}
        }
    \caption{Taxonomy of the data-path architectures.}
    \label{fig:ASIC_datapath_taxonomy}
\end{figure}



\subsection{Sparse Matrices}
The most famous definition of \emph{sparse matrix} is attributed to James Wilkinson and dates back to more than 50 years ago~\cite{davis2007}: any matrix with enough zeros that it pays to take advantage of them.
A more recent and quantitative definition by Filippone et al.~\cite{filippone2017sparse} states that a matrix is \emph{sparse} if its number of non-zero coefficients   
is $O(n)$, where $n$ is the number of rows (columns) of the matrix.

Sparse matrices are usually stored in a compressed format to avoid redundant storage as well as a lot of useless calculations. 
That is, the \textit{storage format} attempts to take advantage of the zeros by avoiding their explicit storage. 
The counterpart is that the traditional simple mapping between the index pair of each matrix coefficient and the position of the coefficient in memory is destroyed. 
Therefore, all sparse matrix storage formats are devised around rebuilding this mapping by using some auxiliary index information. 
This rebuilding has a non-negligible cost and impact on the matrix operations to be performed. Therefore, 
the performance of sparse matrix computations depends on the selected storage format.

Widely used sparse matrix storage formats include COOrdinate (COO), Compressed Sparse Rows (CSR), and Compressed Sparse Columns (CSC)~\cite{filippone2017sparse}.  
CSR is perhaps the most popular sparse matrix representation: It compresses the sparse matrix into three different arrays. 
The first array represents the non-zero values, the second contains the column indexes, and the third marks the boundaries of each row in the matrix. The above formats can be considered general-purpose, meaning that they can be used on most hardware with little or no changes. However, hardware-oriented formats become attractive when moving to special computing architectures such as accelerators. 
Many storage formats, such as ELLPACK, are specifically developed for vector machines to introduce a certain degree of regularity 
in the data structure to enable the efficient exploitation of vector instructions. 

A factor that can drive the choice of the storage format is the \textit{sparsity pattern} of the matrix that is, the pattern of non-zero entries contained in the matrix. 
Common sparsity patterns include unstructured (where nonzeros are randomly and irregularly scattered), 
diagonal (where nonzeros are restricted to a small number of matrix diagonals), 
and block sparse (either coarse-grain or fine-grain). 
Each of these sparsity patterns is best addressed using different formats. 
For instance, the diagonal format (DIA) is an appropriate representation for diagonal matrices. 

DNN models are composed of large, dense matrices which are typically used in matrix multiplication and convolutions.  
In the last years, state-of-the-art DL models have dramatically increased in size, with hundreds of billions of parameters (e.g., large language models as GPT-3 require  
175B parameters~\cite{brown:2020}) and trillions of compute operations per input sample. 
In order to reduce DNN model sizes and computation requirements (including the energy footprint), 
\textit{pruning} (i.e., setting to zero) of DNN weights has emerged as a particularly effective and promising technique. 
Pruning entails identifying unnecessary redundancy in DNN-trained model weights and 
zero out these nonessential weights~\cite {han:nips2015, srinivas:2015}, thus allowing to discard of zero values from storage and computations. 
Therefore, pruning induces sparsity in the DL model, in which a large proportion (typically 
between 50\%\cite{park:arxiv2016} to 90\%~\cite{hao:arxiv2015})  
of the weights are zero. 
Pruning methods allow keeping model accuracy with little 
loss in model quality, thus achieving the same expressive power as dense model counterparts, 
while leading to models that are more efficient in terms of computing and storage resources demand. 

The second factor that induces sparsity in DNN models is the ReLU (rectified linear unit) operator, which is frequently used as an activation function. 
Indeed, ReLU resets all the negative values in the matrices of the 
activations\footnote{The activations are the output values of an individual layer that are passed as inputs to the next layer.} to zero. 

Because of network pruning and zero-valued activations, sparsity has become an active area of research in DNN models. 
These two techniques allow to reduce both the memory size and the memory accesses, the latter thanks to the removal of useless operations (i.e., multiply by zero), 
which also save processing power and energy consumption. 
As regards the memory size, the number of non-zero entries in the resulting sparse matrices can be reduced to 20-80\% and 50-70\% for weights and activations, 
respectively~\cite{han:nips2015,SCNN:isca2017}. 
Sparse matrices can thus be stored using a compressed representation, thus leading to at least 2-3x memory size reduction. 
However, the main disadvantage of a sparse matrix is that the indexes become relative, which adds extra levels of indirection that 
add complexity and need to be carefully managed to avoid inefficiency.

As regards the sparsity pattern of DNN models, it can range from unstructured as a result of fine-grain pruning, which maintains model accuracy, 
to structured when coarse-grain pruning is applied to improve execution efficiency at the cost of downgrading the model accuracy~\cite{han:nips2015, wen:nisp2016}. 
Randomly distributed nonzeros can lead to irregular memory accesses, that are unfriendly
on commodity architectures, e.g., GPU, as well as to irregular computations that introduce conditional branches to utilize the sparsity. 
The latter are hardly applicable for accelerators, which are designed for fine-grained data or thread parallelism rather than flexible data path control. 
On the other hand, hardware-friendly structured sparsity can efficiently accelerate the DNN evaluation at the cost of model accuracy degradation. 

Moreover, sparsity is becoming ubiquitous in modern deep learning workloads (i.e., not only because of the application of 
compression techniques such as network pruning and zero-valued activations) due to the application of deep learning to graphs for 
modeling relations (in social networks, proteins, etc.) using highly sparse matrices, such as in Graph Neural Networks (GNNs). 

The key computational kernel within most DL workloads is general matrix-matrix multiplications (GEMM)~\cite{gao2023acm}. 
It appears frequently during both the forward pass (inference and training) and backward pass (training); for instance, experiments reported in~\cite{sigma:2020} show 
that GEMM comprises around 70\% of the total compute cycles during training for Transformer  and Google Neural Machine Translation 
workloads. 
Therefore, GEMM represents a primary target for hardware acceleration in order to speed up training and inference. 
However, GEMM in DL is characterized by sparsity of matrices, which arises from pruning as explained above, 
and non-square matrix dimensions, which arise from mini-batches and weight factorization~\cite{park-facebook:arxiv2018}.
A popular computational kernel for convolutional neural networks (CNNs) is a sparse vector-vector dot product. 
Sparse-dense matrix multiplication (SpMM)  and sampled-dense matrix multiplication (SDDMM)  are two of the most generic kernels in GNNs.

Spatial-architecture-based hardware accelerators that exploit sparsity have different architectures 
that allow to adapt the computation to sparse matrices. 
In the following, we review their main features. 

\subsection{Emerging 3D-stacked Processing-in-memory Technologies}
3D integration technologies \cite{Kada2015} enable to stack as many as 16 or more 2D integrated circuits and interconnect them vertically using, for instance, through-silicon vias (TSVs), micro bumps, or Cu-Cu connections. 
In this way, a 3D circuit behaves as a single device achieving a smaller area footprint than conventional 2D circuits, while reducing power and latency in data transfer. 
In general, 3D integration is a term that includes such technologies as 3D wafer-level packaging (3DWLP) \cite{Soussan2008}, 2.5D and 3D interposer-based integration \cite{Lau2011}, 3D stacked ICs (3D-SICs), 3D heterogeneous integration, and 3D systems integration, as well as true monolithic 3D ICs \cite{Knechtel2017, Mathur2021}.

These technologies have been employed in the development of new memory devices, which stack layers of conventional 2D DRAM or other memory types (for instance, Non-Volatile Memory (NVM) based on ReRAM \cite{Clarke2015}) together with one or more optional layers of logic circuits.
These logic layers are often implemented with different process technology and can include buffer circuitry, test logic, and processing elements.
Compared to 2D memories, 3D stacking increases memory capacity and bandwidth, reduces access latency due to the shorter on-chip wiring interconnection and the use of wider buses, and potentially improves the performance and power efficiency of the system. 
In fact, 3D stacking of DRAM memory provides an order of magnitude higher bandwidth and up to $5\times$ better energy efficiency than conventional 2D solutions, making the technology an excellent option for meeting the requirements in terms of high throughput and low energy of DNN accelerators \cite{Hassanpour2022}.

Two main 3D stacked memory standards have been recently proposed: the Hybrid Memory Cube (HMC) and the High Bandwidth Memory (HBM). 
3D stacked processing-in-memory accelerator proposals modify the architecture of the 3D memory block inserting processing logic near the memory elements.
Two approaches can be commonly found. 
In the first approach, the computing logic is embedded into the logic die (logic die-level processing-in-memory); 
in the second approach, the processing logic is integrated into each DRAM die at the level of memory banks, after the column decoder and selector blocks (bank-level processing-in-memory).
We present in the following subsections the main characteristics of the two 3D stacked memory standards and overview the existing accelerators adopting them.

\subsubsection{Hybrid Memory Cube}
The Hybrid Memory Cube is a single package containing four to eight DRAM die and one logic die, all stacked together using thousands of TSVs, achieving a much desired high memory bandwidth \cite{Jeddeloh2012,HMC2014}.
As shown in Figure~\ref{fig:HMC}, in a HMC, memory is organized vertically, and portions of each memory die are combined with the corresponding portions of the other memory dies and the logic die.
This creates a 2D grid of vertical partitions, referred as vaults \cite{Pawlowski2011}.
Each vault is functionally and operationally independent and includes in the logic layer a memory controller that manages all memory reference operations within that vault, as well as determining timing requirements and dealing with refresh operations, eliminating these functions from the host memory controller.
The independence of each vault allows to exploit memory level parallelism, as multiple partitions in the DRAM dies can be accessed simultaneously.

Commands and data are transmitted from and to the host across external I/O links consisting of up to four serial links, each with a default of 16 input lanes and 16 output lanes for full duplex operation (HMC2 specifications \cite{HMC2014}).
All in-band communication across a link is packetized.
According to specifications, up to 320 GB/s effective bandwidth can be achieved by considering 30 Gb/s SerDes I/O interfaces, with a storage capacity, depending on the number of stacked layers, of 4GB and 8GB \cite{Micron2018, HMC2014}.
\begin{figure}[t]
  \centering
    \begin{subfigure}[c]{0.3\textwidth}
        \centering
        \includegraphics[width=\textwidth]{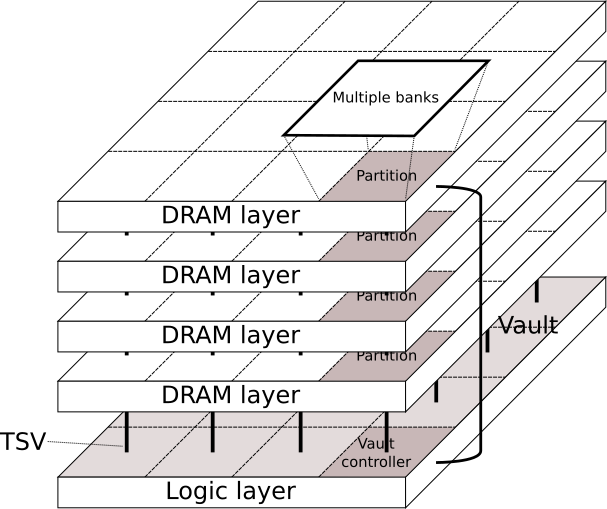}
         \caption{}
         \label{fig:HMC}
    \end{subfigure}
    \qquad
    \qquad
    \begin{subfigure}[c]{0.5\textwidth}
         \centering
         \includegraphics[width=\textwidth]{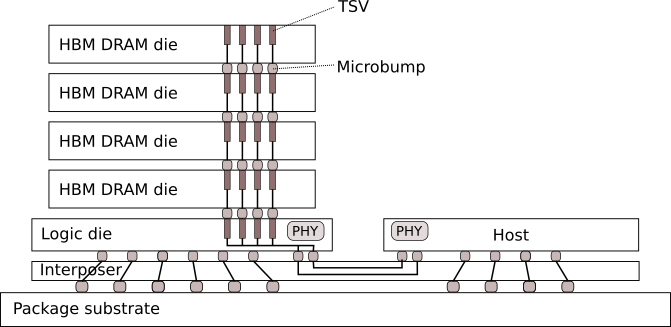}
         \caption{}
         \label{fig:HBM}
    \end{subfigure}
    \caption{(a) High-level architecture of the Hybrid Memory Cube. (b) Cut through image of a computing system with HBM.}
    \label{fig:none}
\end{figure}
\subsubsection{High Bandwidth Memory}
The High Bandwidth Memory is a high-speed computer memory interface for 3D-stacked synchronous dynamic random-access memory (SDRAM) \cite{Joonyoung2014, Sohn2017}.
Each memory module is composed by stacking up to eight DRAM dies and an optional base die including buffer circuitry and test logic.
Dies are vertically interconnected by TSVs and microbumps, in a way similar to the HMC.
As shown in Figure~\ref{fig:HBM}, the memory stack is often connected to the memory controller on the host (e.g., GPU or CPU) through purpose-built silicon chip, called interposer \cite{Lau2011}, which is effectively a miniature PCB that goes inside the package and decreases the memory paths by allowing the host and the memory to be physically close.
However, as semiconductor device fabrication is significantly more expensive than printed circuit board manufacture, this adds cost to the final product.
Alternatively, the memory die can be stacked directly on the host processor chip.

The HBM DRAM is tightly coupled to the host computer through a distributed interface, which is divided into independent channels.
The HBM DRAM uses a wide-interface architecture to achieve high-speed, low-power operation. 
Each channel interface maintains a 128-bit (HMB2) or 64-bit (HMB3) data bus operating at double data rate (DDR).
The latest version (2022) of the HBM (HBM3) supports up to 16 channels of 64 bits, with a total number of data pins equal to 1024, and with an overall package bandwidth of 600 GB/s \cite{JEDEC2023, Prickett2022}. 
Depending on the producer, the HBM stack consists of 8 or 12 16Gb DRAMs, with a total maximum memory capacity of 24 GB \cite{Robinson2022}.

\subsubsection{3D stacked PIM: Accelerating applications loosely related to DNNs}
In addition to the DNN accelerators described in Section 5.2, we briefly overview 3D stacked accelerators for applications loosely related to DNNs. 
The use of PEs in the logic layer of an HMC is discussed in \cite{Oliveira2017}, to support the simulation of large networks on neurons.
The proposed Neuron In-Memory (NIM) architecture is composed of 2,048 functional units, operating on integer and floating-point data, and a small register file with 8 × 16 registers of 32 bits each per vault.
Fast vector elements operation is also supported.
When compared with traditional multi-core environments, NIM provides overall system acceleration and reduces overall energy consumption, taking advantages of the broad bandwidth available in 3D-stacked memory devices.

Millipede \cite{Nitin2018} is an NDP architecture for Big data Machine Learning Analytics (BMLA) that implements its processors in the logic layer of 3D-stacked memories.
These processors have a local memory, register file, pipeline, cache, and prefetch buffers.

The authors in \cite{Mathur2021} explore the design trade-offs and thermal implications of  3D stacking in different configurations layers of SRAM buffers and systolic accelerators composed of MAC elements, while targeting Deep learning applications. 
The main memory (DRAM) is however not necessarily stacked with the rest of the system.
Their simulations show that stacking PE array on top of the SRAM stack in a logic-over-memory
fashion can not only achieve low energy but also mitigate the thermal impact of 3-D.

iPIM \cite{Gu2020} uses a near-bank architecture for image processing. 
The control and the execution are decoupled to obtain a higher bank-level bandwidth and maximize the parallel execution of processing engines on the memory dies.
Each vault contains a control core in the logic die, while the execution logic is placed in the memory die of each vault.
Each control core manages intra/inter-vault data communication and executes instruction decoding with the support of the single-instruction-multiple-bank (SIMB) instruction set architecture (ISA).

Neurosensor \cite{Amir2018} is a 3D CMOS image sensor system with an integrated convolutional neural network computation. 
The image sensor, read-out circuits, memory, and neural computation logic layers are integrated in a single stack.
The DNN computation platform is an adaptation from Neurocube \cite{Kim2016}, and consists of a global controller, processing elements, a 2D mesh NoC connecting the PEs, and a programmable neurosequence generator for DRAM.
The DNN computation is split between the sensor and the host, and the optimal task distribution depends on the processing capabilities of the sensor, the available amount of in-sensor memory for storing the synaptic weights, and the available bandwidth between the sensor and the host.
\subsubsection{3D stacked PIM: Some considerations}
First, the amount of processing elements that can be integrated into 3D stacked memories is limited by the size of the package. 
Moreover, the overall power dissipation of these elements is limited by thermal issues of 3D stacking, as an increase in the operation temperature would result in performance degradation from overheating \cite{Heyman2022}.
Second, the stacking of multiple IC layers has a high manufacturing complexity, which leads to lower yield and problematic testability.
Therefore, in order to support the adoption of this technology, proper cooling methods and better manufacturing solutions are required.

Apart from the above-mentioned technological challenges, embedding processing elements into the memory and moving the computation closer to it requires to rethink of the optimization of the system design in order to take into account the  proximity of the processing logic to the main memory.
Depending on the use case, this might involve the redesign of the on-chip buffers in the logic die, to support the lower latency and energy cost of the accesses to main memory, as well as the use of new approaches for representing, partitioning, and mapping the dataflow of the application in order to exploit the highly parallel system supported by the availability of multiple channels \cite{Hassanpour2022}.

\subsection{RRAM and PCM Technologies for IMC}
Besides conventional CMOS designs, emerging non-volatile memories such as the RRAM and the PCM have been recently explored for integration in stand-alone DNN accelerators. The RRAM device structure (see Figure~\ref{fig:npu2}a) is a metal-insulator-metal (MIM) structure that consists of a top electrode (TE), a bottom electrode (BE), and a metal-oxide layer sandwiched between them. By applying a proper voltage across the electrodes and setting the maximum current flowing in the MIM stack (through a series transistor), an RRAM cell can modulate the shape of a conductive filament created in the metal-oxide layer. In PCM the active material is a chalcogenide phase change material, which can remain in either crystalline or amorphous states for long periods of time at moderately high temperatures. Starting from the amorphous state, the application of voltage pulses with relatively low amplitude causes the crystallization induced by Joule heating, whereas the application of pulses at higher amplitudes can lead to local melting and consequent amorphization. A typical PCM cell has a mushroom shape shown in Figure~\ref{fig:npu2}a, where the pillar-like bottom electrode confines heat and current, thus resulting in a hemispherical shape of the molten material. In both technologies, their resistance state can be tuned not only as a digital memory but also as a continuous analog memory with multiple states to perform in-memory computing \cite{ielmini2018nature}. This characteristic allows efficient matrix-vector multiplication when RRAM and PCM are arranged in crossbar structures (see Figure~\ref{fig:npu2}b).

\begin{figure}[!t]
 \centering
 \includegraphics[width=0.5\columnwidth]{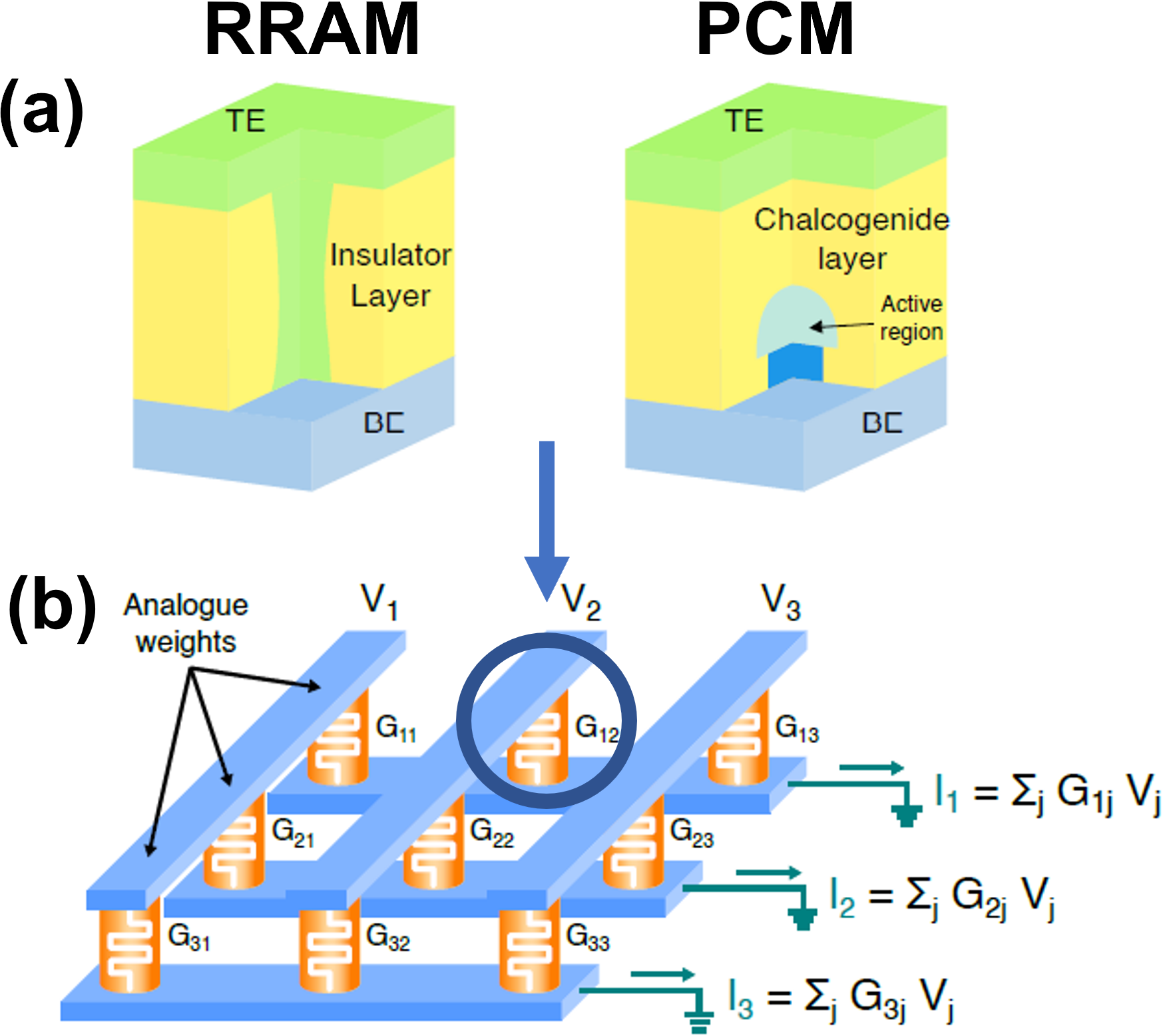}
 \caption{(a) RRAM and PCM devices structure and (b) their arrangement in a crossbar structure for matrix-vector multiplication. Adapted with permission from \cite{lepri2023jeds} under Creative Commons License (CC-BY-4.0).}
 \label{fig:npu2}
\end{figure}
\subsection{Alternative Integration Technologies}
The semiconductor industry has grown significantly as a result of increased integration complexity, resulting in improved performance and cost-effectiveness of transistors. 
Unfortunately, the trend of increasing the number of transistors per die is slowing down, leading to a power-efficiency-driven design era known as \emph{dark silicon}~\cite{esmaeilzadeh_isca11}. While the number of transistors per die continues to increase, many foundries are struggling to achieve the targeted area scaling per transistor, and new process technologies are expected to slow down. The cost per transistor may no longer hold, resulting in yield challenges and additional wafer costs. Circuit designers and computer architects can no longer rely on the free availability of additional transistors and integration opportunities with each new process node, and non-recurring engineering costs have also increased due to fabrication and system complexity challenges~\cite{shalf_ptrs19}.

Alternative integration technologies can provide cost reductions and increase the number of transistors per circuit. These technologies include die-level integration such as 3D die stacking with connections through micro-bumps or Through-Silicon Vias (TSVs)~\cite{hu_micro18}, or through interposer-based 2.5D integration~\cite{zhang_apr15}. By partitioning a monolithic SoC across multiple small dies, namely \emph{chiplets}, (see Figure~\ref{fig:monolitic_vs_25-3d}), yield per die can be improved and metal layer count can be reduced, which can lead to a lower total IC cost~\cite{stow_iccad16}. In fact, larger chips cost more due to two main factors: geometry and manufacturing defects. Fewer larger chips can fit in a wafer, while defects in larger chips waste more silicon than defects in smaller chips~\cite{ajaykumar_micro15}. Smaller chips can be packed more tightly, resulting in more chips that work. In general, making smaller chips results in a higher yield of functioning chips (see Figure~\ref{fig:yield}).

\begin{figure}[t]
  \centering
    \begin{subfigure}[b]{0.45\textwidth}
        \centering
        \includegraphics[width=\textwidth]{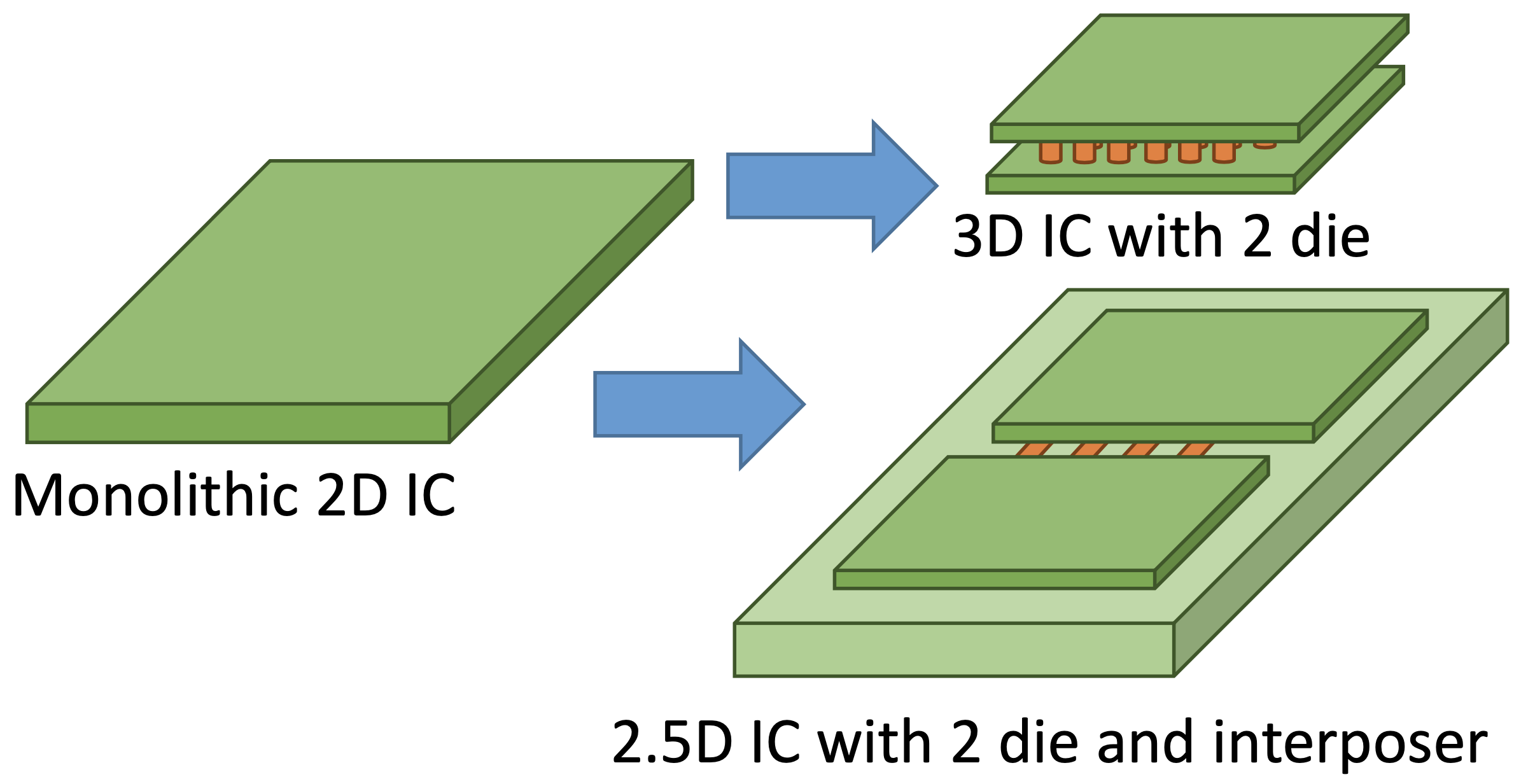}
         \caption{}
         \label{fig:monolitic_vs_25-3d}
    \end{subfigure}
    \hfill
    \begin{subfigure}[b]{0.45\textwidth}
         \centering
         \includegraphics[width=\textwidth]{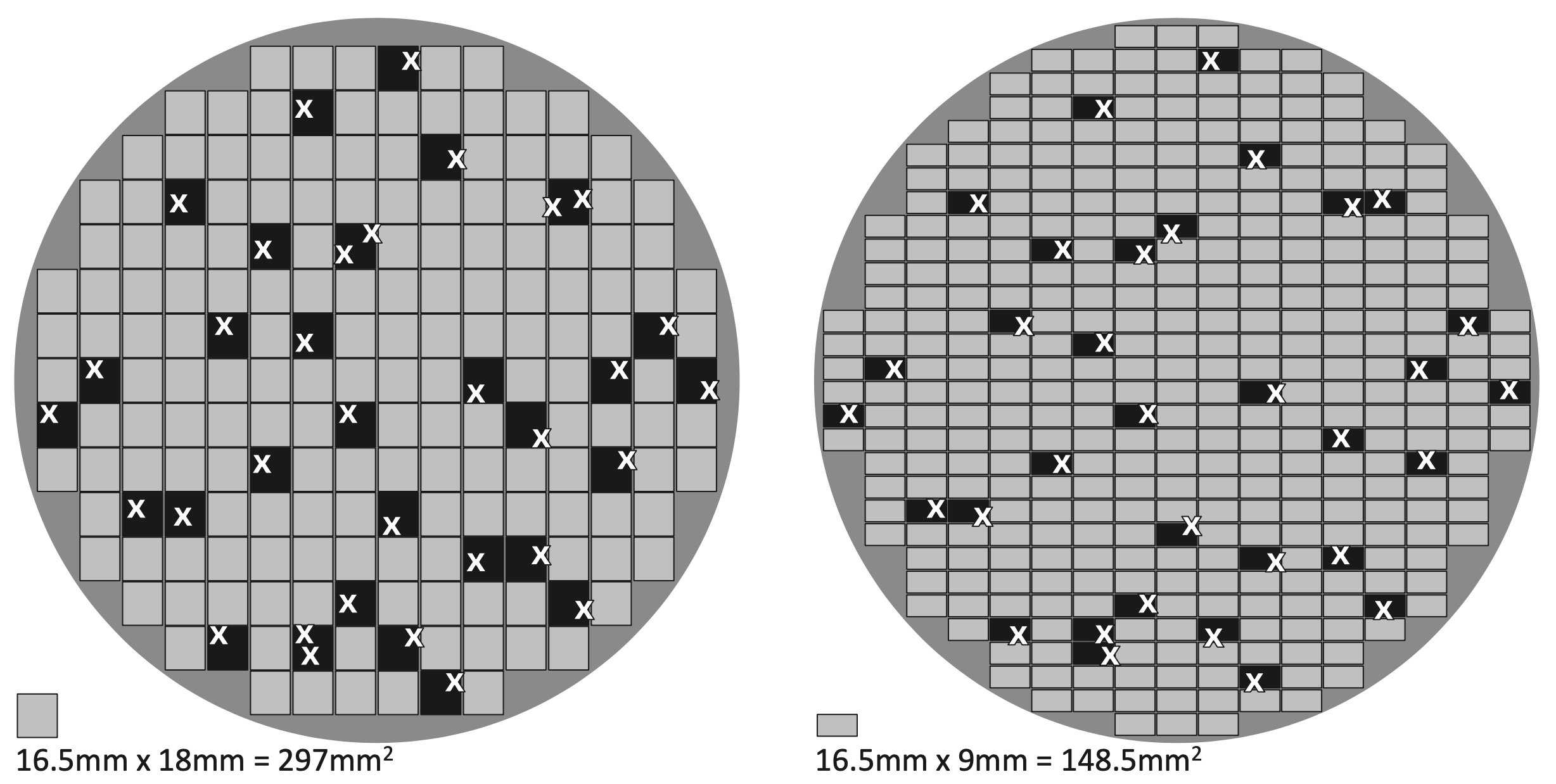}
         \caption{}
         \label{fig:yield}
    \end{subfigure}
    \caption{Die-level integration through TSV-based 3D and interposer-based 2.5D technologies~\ref{fig:monolitic_vs_25-3d}~\cite{stow_iccad16} and overall number of chips and impact on yield of an example defect distribution for two different chip sizes~\cite{ajaykumar_micro15}}
    \label{fig:technologies_yield}
\end{figure}

Die-level integration provides innovative integration strategies, like heterogeneous process integration between dies that can improve performance and reduce costs~\cite{zhang_apr15}. Additionally, this technology can be exploited for the reuse of intellectual property to configure SoCs with different die combinations and reduce non-recurring overheads.

In multichip-module (MCM) silicon interposer-based integration, the interposer uses micro-bumps to connect the smaller chips, which have a higher density than traditional C4 bumps. The impedance across the interposer is the same as conventional on-chip interconnects. The only downside is the additional cost of the interposer. Vertical 3D chip stacking involves combining multiple chips with through-silicon vias (TSVs) for vertical interconnects. This technique has the potential to offer the highest bandwidth but it requires significant cost and overall process complexity as each die must be thinned and processed for TSVs. Overall, as 3D stacking is more expensive and complex, while also potentially causing thermal issues, we focus on MCM silicon interposer-based design in the following. 

\subsubsection{MCM Silicon Interposer-based Design}
In 2.5D integration technology, an interposer is a substrate that connects multiple dies (chiplets) together. There are two types of interposers: passive interposers and active interposers~\cite{stow_iccad17}. Passive interposers are simple substrates that connect multiple dies together without adding any active components. They mainly provide electrical connections, signal routing, and thermal management between the dies. On the other hand, active interposers contain active components such as transistors, capacitors, and inductors, in addition to the electrical connections and signal routing provided by passive interposers. Active interposers can perform some processing and signal conditioning functions between the dies. 

\subsubsection{General Purpose Chiplet-based Architectures}
Chiplet-based designs are being utilized across a wide range of platforms, tailored to support various application contexts. The challenges of creating integrated SoCs for aerospace platforms in advanced semiconductor nodes are reported in~\cite{mounce_ac16} in which authors highlight the possibility of creating heterogeneous mixtures of chiplets, including different embodiments of processors, ultradense memory servers, field-programmable gate array clusters, and configurable analog and radiofrequency functional blocks. Further, some of the features necessary to support scalability and heterogeneity with multi-domain, hybrid architectures involving a mixture of semiconductor technologies and advanced packaging approaches are also outlined. In~\cite{vijayaraghavan_hpca17} a chiplet-based computing system for climate prediction is presented. It integrates a high-throughput and energy-efficient GPU chiplet, high-performance multi-core CPU chiplet, and large-capacity 3D memory. The system can achieve a bandwidth of 3~TB/s and power consumption of 160~W at 1~GHz.

GPU platforms are also benefiting from chiplet-based integration. In~\cite{arunkumar_isca17} a single-chip multi-core GPU is broken down into multiple GPU chiplets to improve both performance and energy efficiency by increasing hardware resource utilization for both the GPU and DRAM chiplets, while also mitigating the dark silicon effect. Additionally, breaking the larger GPU into multiple smaller chiplets has resulted in improved wafer yield.

The design and implementation of a dual-chiplet Chip-on-Wafer-on-Substrate are presented in~\cite{lin_vlsi19}, where each chiplet has four ARM Cortex-A72 processors operating at 4~GHz. The on-die interconnect mesh bus operates above 4~GHz at a 2~mm distance and the inter-chiplet connection features a scalable, power-efficient, high-bandwidth interface achieving 8~Gb/s/pin and 320~GB/s bandwidth. 

The above work uses 2.5D integration technology based on a passive interposer. In~\cite{vivet_jssc21}, the authors observe that current passive interposer solutions still lack flexibility and efficiency when it comes to long-distance communication, smooth integration of chiplets with incompatible interfaces, and easy integration of less-scalable analog functions, such as power management and system input/output signals (IOs). Thus, they present a CMOS Active Interposer that integrates power management and distributed interconnects, enabling a scalable cache-coherent memory hierarchy. The proposed platform integrates six chiplets onto the active interposer, offering a total of 96 cores.

The exploitation of active interposers as a way to address energy-efficiency and computing density issues in high-performance computing (HPC) for Exascale architectures is discussed in~\cite{martinez_vlsi20}. The authors suggest that the integration of chiplets, active interposers, and FPGA can lead to dense, efficient, and modular computing nodes. They detail the ExaNoDe multi-chip-module which combines various components and demonstrate that multi-level integration allows for tight integration of hardware accelerators in a heterogeneous HPC compute node.

\fi

\begin{acks}
This work has been supported by the Spoke 1 on
\emph{Future HPC} of the Italian Research Center on High-Performance Computing, Big Data and Quantum Computing (ICSC)  funded by MUR Mission 4 - Next Generation EU.  
\end{acks}


\newcommand{\noDOI}{}

\ifdefined\noDOI
    \bibliographystyle{acm} 
\else
    \bibliographystyle{ACM-Reference-Format} 
\fi

\bibliography{biblio/biblio,biblio/rebuttal}

\begin{thebibliography}{100}

\bibitem{TF}
{\sc Abadi, M., Agarwal, A., Barham, P., Brevdo, E., Chen, Z., Citro, C., Corrado, G.~S., Davis, A., Dean, J., Devin, M., Ghemawat, S., Goodfellow, I., Harp, A., Irving, G., Isard, M., Jia, Y., Jozefowicz, R., Kaiser, L., Kudlur, M., Levenberg, J., Man\'{e}, D., Monga, R., Moore, S., Murray, D., Olah, C., Schuster, M., Shlens, J., Steiner, B., Sutskever, I., Talwar, K., Tucker, P., Vanhoucke, V., Vasudevan, V., Vi\'{e}gas, F., Vinyals, O., Warden, P., Wattenberg, M., Wicke, M., Yu, Y., and Zheng, X.}
\newblock { TensorFlow: Large-Scale Machine Learning on Heterogeneous Systems}, 2015.
\newblock Software available from tensorflow.org.

\bibitem{agrawal20219}
{\sc Agrawal, A., Lee, S.~K., Silberman, J., Ziegler, M., Kang, M., Venkataramani, S., Cao, N., Fleischer, B., Guillorn, M., Cohen, M., et~al.}
\newblock {9.1 A 7nm 4-core AI chip with 25.6 TFLOPS hybrid FP8 training, 102.4 TOPS INT4 inference and workload-aware throttling}.
\newblock In {\em 2021 IEEE International Solid-State Circuits Conference (ISSCC)\/} (Feb. 2021), vol.~64, IEEE, pp.~144--146.

\bibitem{Ahmadinejad_2019}
{\sc Ahmadinejad, M., Moaiyeri, M.~H., and Sabetzadeh, F.}
\newblock {Energy and area efficient imprecise compressors for approximate multiplication at nanoscale}.
\newblock {\em AEU - International Journal of Electronics and Communications 110\/} (2019), 152859.

\bibitem{Aimar2019}
{\sc Aimar, A., Mostafa, H., Calabrese, E., Rios-Navarro, A., Tapiador-Morales, R., Lungu, I.-A., Milde, M.~B., Corradi, F., Linares-Barranco, A., Liu, S.-C., and Delbruck, T.}
\newblock {NullHop: A Flexible Convolutional Neural Network Accelerator Based on Sparse Representations of Feature Maps}.
\newblock {\em IEEE Transactions on Neural Networks and Learning Systems 30}, 3 (2019), 644--656.

\bibitem{akopyan2015tcad}
{\sc Akopyan, F., Sawada, J., Cassidy, A., Alvarez-Icaza, R., Arthur, J., Merolla, P., Imam, N., Nakamura, Y., Datta, P., Nam, G.-J., Taba, B., Beakes, M., Brezzo, B., Kuang, J.~B., Manohar, R., Risk, W.~P., Jackson, B., and Modha, D.~S.}
\newblock {TrueNorth: Design and Tool Flow of a 65 mW 1 Million Neuron Programmable Neurosynaptic Chip}.
\newblock {\em IEEE Transactions on Computer-Aided Design of Integrated Circuits and Systems 34}, 10 (2015), 1537--1557.

\bibitem{cnvlutin:isca2016}
{\sc Albericio, J., Judd, P., Hetherington, T., Aamodt, T., Jerger, N.~E., and Moshovos, A.}
\newblock {Cnvlutin: Ineffectual-Neuron-Free Deep Neural Network Computing}.
\newblock In {\em Proceedings of the 43rd International Symposium on Computer Architecture\/} (2016), ISCA'16.

\bibitem{Alves2016}
{\sc Alves, M. A.~Z., Diener, M., Santos, P.~C., and Carro, L.}
\newblock {Large vector extensions inside the HMC}.
\newblock In {\em 2016 Design, Automation \& Test in Europe Conference \& Exhibition (DATE)\/} (2016), pp.~1249--1254.

\bibitem{ambs_2010}
{\sc Ambs, P.}
\newblock {Optical Computing: A 60-Year Adventure}.
\newblock {\em Advances in Optical Technologies 2010}, 372652 (May 2010).

\bibitem{mi250x}
{\sc AMD}.
\newblock {AMD Instinct MI200 series accelerator}, Jan 2021.

\bibitem{VitisAI}
{\sc AMD-Xilinx}.
\newblock {VitisAI develop environment}, 2023.

\bibitem{Amir2018}
{\sc Amir, M.~F., Ko, J.~H., Na, T., Kim, D., and Mukhopadhyay, S.}
\newblock {3-D Stacked Image Sensor With Deep Neural Network Computation}.
\newblock {\em IEEE Sensors Journal 18}, 10 (2018), 4187--4199.

\bibitem{h100}
{\sc Andersch, M., Palmer, G., Krashinsky, R., Stam, N., Mehta, V., Brito, G., and Ramaswamy, S.}
\newblock {NVIDIA Hopper Architecture In-Depth}, Mar 2022.

\bibitem{anilPaLMTechnicalReport2023}
{\sc Anil, R., Dai, A.~M., Firat, O., Johnson, M., Lepikhin, D., Passos, A., Shakeri, S., Taropa, E., Bailey, P., Chen, Z., Chu, E., Clark, J.~H., Shafey, L.~E., Huang, Y., {Meier-Hellstern}, K., Mishra, G., Moreira, E., Omernick, M., Robinson, K., Ruder, S., Tay, Y., Xiao, K., Xu, Y., Zhang, Y., Abrego, G.~H., Ahn, J., Austin, J., Barham, P., Botha, J., Bradbury, J., Brahma, S., Brooks, K., Catasta, M., Cheng, Y., Cherry, C., {Choquette-Choo}, C.~A., Chowdhery, A., Crepy, C., Dave, S., Dehghani, M., Dev, S., Devlin, J., D{\'i}az, M., Du, N., Dyer, E., Feinberg, V., Feng, F., Fienber, V., Freitag, M., Garcia, X., Gehrmann, S., Gonzalez, L., {Gur-Ari}, G., Hand, S., Hashemi, H., Hou, L., Howland, J., Hu, A., Hui, J., Hurwitz, J., Isard, M., Ittycheriah, A., Jagielski, M., Jia, W., Kenealy, K., Krikun, M., Kudugunta, S., Lan, C., Lee, K., Lee, B., Li, E., Li, M., Li, W., Li, Y., Li, J., Lim, H., Lin, H., Liu, Z., Liu, F., Maggioni, M., Mahendru, A., Maynez, J., Misra, V., Moussalem, M., Nado, Z., Nham, J., Ni,
  E., Nystrom, A., Parrish, A., Pellat, M., Polacek, M., Polozov, A., Pope, R., Qiao, S., Reif, E., Richter, B., Riley, P., Ros, A.~C., Roy, A., Saeta, B., Samuel, R., Shelby, R., Slone, A., Smilkov, D., So, D.~R., Sohn, D., Tokumine, S., Valter, D., Vasudevan, V., Vodrahalli, K., Wang, X., Wang, P., Wang, Z., Wang, T., Wieting, J., Wu, Y., Xu, K., Xu, Y., Xue, L., Yin, P., Yu, J., Zhang, Q., Zheng, S., Zheng, C., Zhou, W., Zhou, D., Petrov, S., and Wu, Y.}
\newblock {{PaLM}} 2 {{Technical Report}}, May 2023.

\bibitem{arunkumar_isca17}
{\sc Arunkumar, A., Bolotin, E., Cho, B., Milic, U., Ebrahimi, E., Villa, O., Jaleel, A., Wu, C.-J., and Nellans, D.}
\newblock {MCM-GPU: Multi-chip-module GPUs for continued performance scalability}.
\newblock In {\em 2017 ACM/IEEE 44th Annual International Symposium on Computer Architecture (ISCA)\/} (2017), pp.~320--332.

\bibitem{assirArrowRISCVVector2021}
{\sc Assir, I.~A., Iskandarani, M.~E., Sandid, H. R.~A., and Saghir, M. A.~R.}
\newblock {Arrow: A RISC-V Vector Accelerator for Machine Learning Inference}.
\newblock {\em CoRR abs/2107.07169}, arXiv:2107.07169 (July 2021).

\bibitem{Bengio_2009}
{\sc Bengio, Y.}
\newblock {Learning Deep Architectures for AI}.
\newblock {\em Foundations and Trends in Machine Learning 2}, 1 (2009), 1--127.

\bibitem{bertacciniMiniFloatNNExSdotpISA2022}
{\sc Bertaccini, L., Paulin, G., Fischer, T., Mach, S., and Benini, L.}
\newblock {MiniFloat-NN and ExSdotp: An ISA Extension and a Modular Open Hardware Unit for Low-Precision Training on RISC-V Cores}.
\newblock In {\em 2022 {{IEEE}} 29th {{Symposium}} on {{Computer Arithmetic}} ({{ARITH}})\/} (Sept. 2022), pp.~1--8.

\bibitem{Biamonte2017}
{\sc Biamonte, J., Wittek, P., Pancotti, N., Rebentrost, P., Wiebe, N., and Lloyd, S.}
\newblock {Quantum machine learning}.
\newblock {\em Nature 549}, 7671 (sep 2017), 195--202.

\bibitem{blaiech2019survey}
{\sc Blaiech, A.~G., Khalifa, K.~B., Valderrama, C., Fernandes, M.~A., and Bedoui, M.~H.}
\newblock {A survey and taxonomy of FPGA-based deep learning accelerators}.
\newblock {\em Journal of Systems Architecture 98\/} (2019), 331--345.

\bibitem{FINN}
{\sc Blott, M., Preu{\ss}er, T.~B., Fraser, N.~J., Gambardella, G., O'brien, K., Umuroglu, Y., et~al.}
\newblock {FINN-R: An end-to-end deep-learning framework for fast exploration of quantized neural networks}.
\newblock {\em ACM Transactions on Reconfigurable Technology and Systems 11}, 3 (2018), 1--23.

\bibitem{sodaMICRO}
{\sc Bohm~Agostini, N., Curzel, S., Zhang, J.~J., Limaye, A., Tan, C., Amatya, V., Minutoli, M., Castellana, V.~G., Manzano, J., Brooks, D., Wei, G.-Y., and Tumeo, A.}
\newblock {Bridging Python to Silicon: The SODA Toolchain}.
\newblock {\em IEEE Micro 42}, 5 (2022), 78--88.

\bibitem{bommasaniOpportunitiesRisksFoundation2022}
{\sc Bommasani, R., Hudson, D.~A., Adeli, E., Altman, R., Arora, S., {von Arx}, S., Bernstein, M.~S., Bohg, J., Bosselut, A., Brunskill, E., Brynjolfsson, E., Buch, S., Card, D., Castellon, R., Chatterji, N., Chen, A., Creel, K., Davis, J.~Q., Demszky, D., Donahue, C., Doumbouya, M., Durmus, E., Ermon, S., Etchemendy, J., Ethayarajh, K., {Fei-Fei}, L., Finn, C., Gale, T., Gillespie, L., Goel, K., Goodman, N., Grossman, S., Guha, N., Hashimoto, T., Henderson, P., Hewitt, J., Ho, D.~E., Hong, J., Hsu, K., Huang, J., Icard, T., Jain, S., Jurafsky, D., Kalluri, P., Karamcheti, S., Keeling, G., Khani, F., Khattab, O., Koh, P.~W., Krass, M., Krishna, R., Kuditipudi, R., Kumar, A., Ladhak, F., Lee, M., Lee, T., Leskovec, J., Levent, I., Li, X.~L., Li, X., Ma, T., Malik, A., Manning, C.~D., Mirchandani, S., Mitchell, E., Munyikwa, Z., Nair, S., Narayan, A., Narayanan, D., Newman, B., Nie, A., Niebles, J.~C., Nilforoshan, H., Nyarko, J., Ogut, G., Orr, L., Papadimitriou, I., Park, J.~S., Piech, C., Portelance, E.,
  Potts, C., Raghunathan, A., Reich, R., Ren, H., Rong, F., Roohani, Y., Ruiz, C., Ryan, J., R{\'e}, C., Sadigh, D., Sagawa, S., Santhanam, K., Shih, A., Srinivasan, K., Tamkin, A., Taori, R., Thomas, A.~W., Tram{\`e}r, F., Wang, R.~E., Wang, W., Wu, B., Wu, J., Wu, Y., Xie, S.~M., Yasunaga, M., You, J., Zaharia, M., Zhang, M., Zhang, T., Zhang, X., Zhang, Y., Zheng, L., Zhou, K., and Liang, P.}
\newblock On the {{Opportunities}} and {{Risks}} of {{Foundation Models}}, July 2022.

\bibitem{BRAKET2023}
{Quantum Computing Service - Amazon Braket - AWS}, 2023.

\bibitem{broughton2021}
{\sc Broughton, M., Verdon, G., McCourt, T., Martinez, A.~J., Yoo, J.~H., Isakov, S.~V., Massey, P., Halavati, R., Niu, M.~Y., Zlokapa, A., Peters, E., Lockwood, O., Skolik, A., Jerbi, S., Dunjko, V., Leib, M., Streif, M., Dollen, D.~V., Chen, H., Cao, S., Wiersema, R., Huang, H.-Y., McClean, J.~R., Babbush, R., Boixo, S., Bacon, D., Ho, A.~K., Neven, H., and Mohseni, M.}
\newblock {TensorFlow Quantum: A Software Framework for Quantum Machine Learning}, 2021.

\bibitem{brown:2020}
{\sc Brown, T.~B., Mann, B., Ryder, N., Subbiah, M., Kaplan, J., Dhariwal, P., Neelakantan, A., Shyam, P., Sastry, G., Askell, A., Agarwal, S., Herbert{-}Voss, A., Krueger, G., Henighan, T., Child, R., Ramesh, A., Ziegler, D.~M., Wu, J., Winter, C., Hesse, C., Chen, M., Sigler, E., Litwin, M., Gray, S., Chess, B., Clark, J., Berner, C., McCandlish, S., Radford, A., Sutskever, I., and Amodei, D.}
\newblock {Language Models are Few-Shot Learners}.
\newblock {\em CoRR abs/2005.14165\/} (2020).

\bibitem{bruschiEndtoEndDNNInference2022}
{\sc Bruschi, N., Tagliavini, G., Garofalo, A., Conti, F., Boybat, I., Benini, L., and Rossi, D.}
\newblock End-to-end {DNN} inference on a massively parallel analog in memory computing architecture.
\newblock In {\em Design, Automation {\&} Test in Europe Conference {\&} Exhibition, {DATE}\/} (2023), {IEEE}, pp.~1--6.

\bibitem{StratusHLS2022}
{\sc {Cadence}}.
\newblock {\em {Stratus High-Level Synthesis}}, 2022.

\bibitem{LegUpHLS2013}
{\sc Canis, A., Choi, J., Aldham, M., Zhang, V., Kammoona, A., Czajkowski, T.~S., Brown, S.~D., and Anderson, J.~H.}
\newblock {LegUp: An open-source high-level synthesis tool for FPGA-based processor/accelerator systems}.
\newblock {\em {ACM} Trans. Embed. Comput. Syst. 13}, 2 (2013), 24:1--24:27.

\bibitem{cao2022tcomp}
{\sc Cao, W., Zhao, Y., Boloor, A., Han, Y., Zhang, X., and Jiang, L.}
\newblock {Neural-PIM: Efficient Processing-In-Memory With Neural Approximation of Peripherals}.
\newblock {\em IEEE Transactions on Computers 71}, 9 (2022), 2142--2155.

\bibitem{amber2022vlsi}
{\sc Carsello, A., Feng, K., Kong, T., Koul, K., Liu, Q., Melchert, J., Nyengele, G., Strange, M., Zhang, K., Nayak, A., Setter, J., Thomas, J., Sreedhar, K., Chen, P.-H., Bhagdikar, N., Myers, Z., D’Agostino, B., Joshi, P., Richardson, S., Bahr, R., Torng, C., Horowitz, M., and Raina, P.}
\newblock {Amber: A 367 GOPS, 538 GOPS/W 16nm SoC with a Coarse-Grained Reconfigurable Array for Flexible Acceleration of Dense Linear Algebra}.
\newblock In {\em {2022 IEEE Symposium on VLSI Technology and Circuits (VLSI Technology and Circuits)}\/} (2022), pp.~70--71.

\bibitem{cassidy2013ijcnn}
{\sc Cassidy, A.~S., Merolla, P., Arthur, J.~V., Esser, S.~K., Jackson, B., Alvarez-Icaza, R., Datta, P., Sawada, J., Wong, T.~M., Feldman, V., Amir, A., Rubin, D. B.-D., Akopyan, F., McQuinn, E., Risk, W.~P., and Modha, D.~S.}
\newblock {Cognitive computing building block: A versatile and efficient digital neuron model for neurosynaptic cores}.
\newblock In {\em {The 2013 International Joint Conference on Neural Networks (IJCNN)}\/} (2013), pp.~1--10.

\bibitem{cavalcanteAra1GHzScalable2020}
{\sc Cavalcante, M., Schuiki, F., Zaruba, F., Schaffner, M., and Benini, L.}
\newblock {Ara: A 1-GHz+ Scalable and Energy-Efficient RISC-V Vector Processor With Multiprecision Floating-Point Support in 22-Nm FD-SOI}.
\newblock {\em IEEE Transactions on Very Large Scale Integration (VLSI) Systems 28}, 2 (Feb. 2020), 530--543.

\bibitem{cavalcanteSpatzCompactVector2022}
{\sc Cavalcante, M., W{\"u}thrich, D., Perotti, M., Riedel, S., and Benini, L.}
\newblock {Spatz: A Compact Vector Processing Unit for High-Performance and Energy-Efficient Shared-L1 Clusters}.
\newblock In {\em 41st {{IEEE}}/{{ACM International Conference}} on {{Computer-Aided Design}}\/} ({San Diego California}, Oct. 2022), {ACM}, pp.~1--9.

\bibitem{Cavigelli2017}
{\sc Cavigelli, L., and Benini, L.}
\newblock {Origami: A 803-GOp/s/W Convolutional Network Accelerator}.
\newblock {\em IEEE Transactions on Circuits and Systems for Video Technology 27}, 11 (2017), 2461--2475.

\bibitem{Chang2020}
{\sc Chang, J.-W., Kang, K.-W., and Kang, S.-J.}
\newblock {An Energy-Efficient FPGA-Based Deconvolutional Neural Networks Accelerator for Single Image Super-Resolution}.
\newblock {\em IEEE Transactions on Circuits and Systems for Video Technology 30}, 1 (2020), 281--295.

\bibitem{karam2021hotchip}
{\sc Chatha, K.}
\newblock {Qualcomm® Cloud Al 100 : 12TOPS/W Scalable, High Performance and Low Latency Deep Learning Inference Accelerator}.
\newblock In {\em {2021 IEEE Hot Chips 33 Symposium (HCS)}\/} (2021), pp.~1--19.

\bibitem{chenXuantie910CommercialMultiCore2020}
{\sc Chen, C., Xiang, X., Liu, C., Shang, Y., Guo, R., Liu, D., Lu, Y., Hao, Z., Luo, J., Chen, Z., Li, C., Pu, Y., Meng, J., Yan, X., Xie, Y., and Qi, X.}
\newblock {Xuantie-910: A Commercial Multi-Core 12-Stage Pipeline Out-of-Order 64-Bit High Performance RISC-V Processor with Vector Extension : Industrial Product}.
\newblock In {\em 2020 {{ACM}}/{{IEEE}} 47th {{Annual International Symposium}} on {{Computer Architecture}} ({{ISCA}})\/} (May 2020), pp.~52--64.

\bibitem{chenEightCoreRISCVProcessor2022}
{\sc Chen, G.~K., Knag, P.~C., Tokunaga, C., and Krishnamurthy, R.~K.}
\newblock {An Eight-Core RISC-V Processor With Compute Near Last Level Cache in Intel 4 CMOS}.
\newblock {\em IEEE Journal of Solid-State Circuits\/} (2022), 1--12.

\bibitem{chen2016commacm}
{\sc Chen, Y., Chen, T., Xu, Z., Sun, N., and Temam, O.}
\newblock {DianNao Family: Energy-Efficient Hardware Accelerators for Machine Learning}.
\newblock {\em Commun. ACM 59}, 11 (Oct 2016), 105–112.

\bibitem{chen2020engineering}
{\sc Chen, Y., Xie, Y., Song, L., Chen, F., and Tang, T.}
\newblock {A Survey of Accelerator Architectures for Deep Neural Networks}.
\newblock {\em Engineering 6}, 3 (2020), 264--274.

\bibitem{Chen_2017}
{\sc Chen, Y.-H., Krishna, T., Emer, J.~S., and Sze, V.}
\newblock {Eyeriss: An Energy-Efficient Reconfigurable Accelerator for Deep Convolutional Neural Networks}.
\newblock {\em IEEE Journal of Solid-State Circuits 52}, 1 (Jan. 2017), 127--138.

\bibitem{eyerissv2:2019}
{\sc Chen, Y.-H., Yang, T.-J., Emer, J., and Sze, V.}
\newblock {Eyeriss v2: A Flexible Accelerator for Emerging Deep Neural Networks on Mobile Devices}.
\newblock {\em IEEE Journal on Emerging and Selected Topics in Circuits and Systems 9}, 2 (2019), 292--308.

\bibitem{Cheng_2004}
{\sc Cheng, C., and Parhi, K.}
\newblock {Hardware efficient fast parallel FIR filter structures based on iterated short convolution}.
\newblock {\em IEEE Transactions on Circuits and Systems I: Regular Papers 51}, 8 (2004), 1492--1500.

\bibitem{Cheng_2020}
{\sc Cheng, C., and Parhi, K.~K.}
\newblock {Fast 2D Convolution Algorithms for Convolutional Neural Networks}.
\newblock {\em IEEE Transactions on Circuits and Systems I: Regular Papers 67}, 5 (2020), 1678--1691.

\bibitem{chi2016isca}
{\sc Chi, P., Li, S., Xu, C., Zhang, T., Zhao, J., Liu, Y., Wang, Y., and Xie, Y.}
\newblock {PRIME: A Novel Processing-in-Memory Architecture for Neural Network Computation in ReRAM-Based Main Memory}.
\newblock In {\em {2016 ACM/IEEE 43rd Annual International Symposium on Computer Architecture (ISCA)}\/} (2016), pp.~27--39.

\bibitem{Chitty2022ACMSUR}
{\sc Chitty-Venkata, K.~T., and Somani, A.~K.}
\newblock {Neural Architecture Search Survey: A Hardware Perspective}.
\newblock {\em ACM Comput. Surv. 55}, 4 (nov 2022).

\bibitem{choquetteNVIDIAHopperH1002023}
{\sc Choquette, J.}
\newblock {NVIDIA Hopper H100 GPU: Scaling Performance}.
\newblock {\em IEEE Micro\/} (2023), 1--13.

\bibitem{choquetteVoltaPerformanceProgrammability2018}
{\sc Choquette, J., Giroux, O., and Foley, D.}
\newblock {Volta: Performance and Programmability}.
\newblock {\em IEEE Micro 38}, 2 (Mar. 2018), 42--52.

\bibitem{choquetteA100DatacenterGPU2021}
{\sc Choquette, J., Lee, E., Krashinsky, R., Balan, V., and Khailany, B.}
\newblock {3.2 The A100 Datacenter GPU and Ampere Architecture}.
\newblock In {\em 2021 {{IEEE International Solid- State Circuits Conference}} ({{ISSCC}})\/} (Feb. 2021), vol.~64, pp.~48--50.

\bibitem{Clarke2015}
{\sc Clarke, P.}
\newblock {Intel, Micron Launch “Bulk-Switching” ReRAM}.
\newblock \url{https://www.eetimes.com/intel-micron-launch-bulk-switching-reram/}, Jul. 2015.
\newblock Accessed: 18/04/2023.

\bibitem{cococcioniLightweightPositPorcessing2021}
{\sc Cococcioni, M., Rossi, F., Ruffaldi, E., and Saponara, S.}
\newblock {A Lightweight Posit Processing Unit for RISC-V Processors in Deep Neural Network Applications}.
\newblock {\em IEEE Transactions on Emerging Topics in Computing 10}, 4 (2022), 1898--1908.

\bibitem{Lau2011}
{\sc Conference, A. . P. R.~T., on~Packaging, E., of~Electronic, I., Photonic~Systems, M., and 1, N.~V.}, Eds.
\newblock {\em {The Most Cost-Effective Integrator (TSV Interposer) for 3D IC Integration System-in-Package (SiP)}\/} (07 2011), International Electronic Packaging Technical Conference and Exhibition.

\bibitem{conti22124TOPS2023}
{\sc Conti, F., Rossi, D., Paulin, G., Garofalo, A., Di~Mauro, A., Rutishauer, G., marco Ottavi, G., Eggimann, M., Okuhara, H., Huard, V., Montfort, O., Jure, L., Exibard, N., Gouedo, P., Louvat, M., Botte, E., and Benini, L.}
\newblock {22.1 A 12.4TOPS/W @ 136GOPS AI-IoT System-on-Chip with 16 RISC-V, 2-to-8b Precision-Scalable DNN Acceleration and 30\%-{{Boost Adaptive Body Biasing}}}.
\newblock In {\em 2023 {{IEEE International Solid- State Circuits Conference}} ({{ISSCC}})\/} (Feb. 2023), pp.~21--23.

\bibitem{Cordeiro2021}
{\sc Cordeiro, A.~S., Santos, S. R.~d., Moreira, F.~B., Santos, P.~C., Carro, L., and Alves, M. A.~Z.}
\newblock {Machine Learning Migration for Efficient Near-Data Processing}.
\newblock In {\em 2021 29th Euromicro International Conference on Parallel, Distributed and Network-Based Processing (PDP)\/} (2021), pp.~212--219.

\bibitem{D-WAVE2023}
{D-Wave Systems - The Practical Quantum Computing Company}, 2023.

\bibitem{davidsonCelerityOpenSource511Core2018}
{\sc Davidson, S., Xie, S., Torng, C., {Al-Hawai}, K., Rovinski, A., Ajayi, T., Vega, L., Zhao, C., Zhao, R., Dai, S., Amarnath, A., Veluri, B., Gao, P., Rao, A., Liu, G., Gupta, R.~K., Zhang, Z., Dreslinski, R., Batten, C., and Taylor, M.~B.}
\newblock {The Celerity Open-Source 511-Core RISC-V Tiered Accelerator Fabric: Fast Architectures and Design Methodologies for Fast Chips}.
\newblock {\em IEEE Micro 38}, 2 (Mar. 2018), 30--41.

\bibitem{davies2018micro}
{\sc Davies, M., Srinivasa, N., Lin, T.-H., Chinya, G., Cao, Y., Choday, S.~H., Dimou, G., Joshi, P., Imam, N., Jain, S., Liao, Y., Lin, C.-K., Lines, A., Liu, R., Mathaikutty, D., McCoy, S., Paul, A., Tse, J., Venkataramanan, G., Weng, Y.-H., Wild, A., Yang, Y., and Wang, H.}
\newblock {Loihi: A Neuromorphic Manycore Processor with On-Chip Learning}.
\newblock {\em IEEE Micro 38}, 1 (2018), 82--99.

\bibitem{davis2007}
{\sc Davis, T.}
\newblock {Wilkinson's Sparse Matrix Definition}.
\newblock {\em NA Digest 7}, 12 (2007), 379--401.

\bibitem{bittactical:2019}
{\sc Delmas~Lascorz, A., Judd, P., Stuart, D.~M., Poulos, Z., Mahmoud, M., Sharify, S., Nikolic, M., Siu, K., and Moshovos, A.}
\newblock {Bit-Tactical: A Software/Hardware Approach to Exploiting Value and Bit Sparsity in Neural Networks}.
\newblock In {\em Twenty-Fourth International Conference on Architectural Support for Programming Languages and Operating Systems\/} (2019), ASPLOS '19, pp.~749--763.

\bibitem{perm:micro2018}
{\sc Deng, C., Liao, S., Xie, Y., Parhi, K.~K., Qian, X., and Yuan, B.}
\newblock {PermDNN: Efficient Compressed DNN Architecture with Permuted Diagonal Matrices}.
\newblock In {\em 51st Annual IEEE/ACM International Symposium on Microarchitecture\/} (2018), MICRO-51, pp.~189--202.

\bibitem{Deng_2018}
{\sc Deng, Q., Jiang, L., Zhang, Y., Zhang, M., and Yang, J.}
\newblock {DrAcc: a DRAM based accelerator for accurate CNN inference}.
\newblock In {\em Proceedings of the 55th Annual Design Automation Conference\/} (New York, NY, USA, 2018), DAC '18, Association for Computing Machinery.

\bibitem{desoli2023isscc}
{\sc Desoli, G., Chawla, N., Boesch, T., Avodhyawasi, M., Rawat, H., Chawla, H., Abhijith, V., Zambotti, P., Sharma, A., Cappetta, C., Rossi, M., De~Vita, A., and Girardi, F.}
\newblock {A 40-310TOPS/W SRAM-Based All-Digital Up to 4b In-Memory Computing Multi-Tiled NN Accelerator in FD-SOI 18nm for Deep-Learning Edge Applications}.
\newblock In {\em 2023 IEEE International Solid- State Circuits Conference (ISSCC)\/} (2023), pp.~260--262.

\bibitem{desoli17isscc}
{\sc Desoli, G., Chawla, N., Boesch, T., Singh, S.-p., Guidetti, E., De~Ambroggi, F., Majo, T., Zambotti, P., Ayodhyawasi, M., Singh, H., and Aggarwal, N.}
\newblock {A 2.9TOPS/W deep convolutional neural network SoC in FD-SOI 28nm for intelligent embedded systems}.
\newblock In {\em {2017 IEEE International Solid-State Circuits Conference (ISSCC)}\/} (2017), pp.~238--239.

\bibitem{DHILLESWARARAO_2022}
{\sc Dhilleswararao, P., Boppu, S., Manikandan, M.~S., and Cenkeramaddi, L.~R.}
\newblock {Efficient Hardware Architectures for Accelerating Deep Neural Networks: Survey}.
\newblock {\em IEEE Access 10\/} (2022), 131788--131828.

\bibitem{dimauroKrakenDirectEvent2022}
{\sc Di~Mauro, A., Scherer, M., Rossi, D., and Benini, L.}
\newblock {Kraken: A Direct Event/Frame-Based Multi-sensor Fusion SoC for Ultra-Efficient Visual Processing in Nano-UAVs}, Aug. 2022.

\bibitem{ditzelAcceleratingMLRecommendation2022}
{\sc Ditzel, D.~R., and {team}, t.~E.}
\newblock {Accelerating ML Recommendation With Over 1,000 RISC-V/Tensor Processors on Esperanto's ET-SoC-1 Chip}.
\newblock {\em IEEE Micro 42}, 3 (May 2022), 31--38.

\bibitem{dossantos1412nmLinuxSMPCapable2024}
{\sc Dos~Santos, M.~C., Jia, T., Zuckerman, J., Cochet, M., Giri, D., Loscalzo, E.~J., Swaminathan, K., Tambe, T., Zhang, J.~J., Buyuktosunoglu, A., Chiu, K.-L., Guglielmo, G.~D., Mantovani, P., Piccolboni, L., Tombesi, G., Trilla, D., Wellman, J.-D., Yang, E.-Y., Amarnath, A., Jing, Y., Mishra, B., Park, J., Suresh, V., Adve, S., Bose, P., Brooks, D., Carloni, L.~P., Shepard, K.~L., and Wei, G.-Y.}
\newblock 14.5 {{A}} 12nm {{Linux-SMP-Capable RISC-V SoC}} with 14 {{Accelerator Types}}, {{Distributed Hardware Power Management}} and {{Flexible NoC-Based Data Orchestration}}.
\newblock In {\em 2024 {{IEEE International Solid-State Circuits Conference}} ({{ISSCC}})\/} (Feb. 2024), vol.~67, pp.~262--264.

\bibitem{dosovitskiyImageWorth16x162021}
{\sc Dosovitskiy, A., Beyer, L., Kolesnikov, A., Weissenborn, D., Zhai, X., Unterthiner, T., Dehghani, M., Minderer, M., Heigold, G., Gelly, S., Uszkoreit, J., and Houlsby, N.}
\newblock An {{Image}} is {{Worth}} 16x16 {{Words}}: {{Transformers}} for {{Image Recognition}} at {{Scale}}, June 2021.

\bibitem{Du_2018}
{\sc Du, L., Du, Y., Li, Y., and Chang, M.-C.~F.}
\newblock {A Reconfigurable Streaming Deep Convolutional Neural Network Accelerator for Internet of Things}.
\newblock {\em IEEE Transactions on Circuits and Systems I: Regular Papers PP\/} (07 2017).

\bibitem{hls4ml}
{\sc Duarte, J., et~al.}
\newblock {Fast inference of deep neural networks in FPGAs for particle physics}.
\newblock {\em JINST 13}, 07 (2018), P07027.

\bibitem{volta}
{\sc Durant, L., Giroux, O., Harris, M., and Stam, N.}
\newblock {Inside Volta: The World’s Most Advanced Data Center GPU}, May 2017.

\bibitem{elsterNvidiaHopperGPU2022}
{\sc Elster, A.~C., and Haugdahl, T.~A.}
\newblock {Nvidia Hopper GPU and Grace CPU Highlights}.
\newblock {\em Computing in Science \& Engineering 24}, 2 (Mar. 2022), 95--100.

\bibitem{esmaeilzadeh_isca11}
{\sc Esmaeilzadeh, H., Blem, E., Amant, R.~S., Sankaralingam, K., and Burger, D.}
\newblock {Dark silicon and the end of multicore scaling}.
\newblock In {\em 2011 38th Annual International Symposium on Computer Architecture (ISCA)\/} (2011), pp.~365--376.

\bibitem{Esposito_2017}
{\sc Esposito, D., Strollo, A. G.~M., and Alioto, M.}
\newblock {Low-power approximate MAC unit}.
\newblock In {\em 2017 13th Conference on Ph.D. Research in Microelectronics and Electronics (PRIME)\/} (2017), pp.~81--84.

\bibitem{fedus:jmlr2022}
{\sc Fedus, W., Zoph, B., and Shazeer, N.}
\newblock {Switch Transformers: Scaling to Trillion Parameter Models with Simple and Efficient Sparsity}.
\newblock {\em Journal of Machine Learning Research 23}, 120 (2022), 1--39.

\bibitem{Bambu}
{\sc Ferrandi, F., Castellana, V.~G., Curzel, S., Fezzardi, P., Fiorito, M., Lattuada, M., Minutoli, M., Pilato, C., and Tumeo, A.}
\newblock {Invited: Bambu: an Open-Source Research Framework for the High-Level Synthesis of Complex Applications}.
\newblock In {\em 2021 58th ACM/IEEE Design Automation Conference (DAC)\/} (2021), pp.~1327--1330.

\bibitem{filippone2017sparse}
{\sc Filippone, S., Cardellini, V., Barbieri, D., and Fanfarillo, A.}
\newblock {Sparse Matrix-Vector Multiplication on GPGPUs}.
\newblock {\em ACM Trans. Math. Softw. 43}, 4 (2017).

\bibitem{Finkbeiner_2017}
{\sc Finkbeiner, T., Hush, G., Larsen, T., Lea, P., Leidel, J., and Manning, T.}
\newblock {In-Memory Intelligence}.
\newblock {\em IEEE Micro 37}, 4 (2017), 30--38.

\bibitem{Brainwave}
{\sc Fowers, J., Ovtcharov, K., Papamichael, M., Massengill, T., Liu, M., Lo, D., Alkalay, S., Haselman, M., Adams, L., Ghandi, M., Heil, S., Patel, P., Sapek, A., Weisz, G., Woods, L., Lanka, S., Reinhardt, S.~K., Caulfield, A.~M., Chung, E.~S., and Burger, D.}
\newblock {A Configurable Cloud-Scale DNN Processor for Real-Time AI}.
\newblock In {\em ACM/IEEE 45th Annual International Symposium on Computer Architecture (ISCA)\/} (2018), pp.~1--14.

\bibitem{frenkel2021arxiv}
{\sc Frenkel, C., Bol, D., and Indiveri, G.}
\newblock {Bottom-Up and Top-Down Neural Processing Systems Design: Neuromorphic Intelligence as the Convergence of Natural and Artificial Intelligence}.
\newblock {\em CoRR abs/2106.01288\/} (2021).

\bibitem{frenkel2018tbiocas}
{\sc Frenkel, C., Lefebvre, M., Legat, J.-D., and Bol, D.}
\newblock {A 0.086-mm$^2$ 12.7-pJ/SOP 64k-Synapse 256-Neuron Online-Learning Digital Spiking Neuromorphic Processor in 28-nm CMOS}.
\newblock {\em IEEE Transactions on Biomedical Circuits and Systems 13}, 1 (2019), 145--158.

\bibitem{frenkel2019tbiocas}
{\sc Frenkel, C., Legat, J.-D., and Bol, D.}
\newblock {MorphIC: A 65-nm 738k-Synapse/mm$^2$ Quad-Core Binary-Weight Digital Neuromorphic Processor With Stochastic Spike-Driven Online Learning}.
\newblock {\em IEEE Transactions on Biomedical Circuits and Systems 13}, 5 (2019), 999--1010.

\bibitem{Frustaci2020}
{\sc Frustaci, F., Perri, S., Corsonello, P., and Alioto, M.}
\newblock Approximate multipliers with dynamic truncation for energy reduction via graceful quality degradation.
\newblock {\em IEEE Transactions on Circuits and Systems II: Express Briefs 67}, 12 (2020), 3427--3431.

\bibitem{gallo202264}
{\sc Gallo, M.~L., Khaddam-Aljameh, R., Stanisavljevic, M., Vasilopoulos, A., Kersting, B., Dazzi, M., Karunaratne, G., Braendli, M., Singh, A., Mueller, S.~M., et~al.}
\newblock {A 64-core mixed-signal in-memory compute chip based on phase-change memory for deep neural network inference}.
\newblock {\em arXiv preprint arXiv:2212.02872\/} (2022).

\bibitem{gaoDECADES67mm246TOPS2023a}
{\sc Gao, F., Chang, T.-J., Li, A., {Orenes-Vera}, M., Giri, D., Jackson, P.~J., Ning, A., Tziantzioulis, G., Zuckerman, J., Tu, J., Xu, K., Chirkov, G., Tombesi, G., Balkind, J., Martonosi, M., Carloni, L., and Wentzlaff, D.}
\newblock {{DECADES}}: {{A}} 67mm2, 1.{{46TOPS}}, 55 {{Giga Cache-Coherent}} 64-bit {{RISC-V Instructions}} per second, {{Heterogeneous Manycore SoC}} with 109 {{Tiles}} including {{Accelerators}}, {{Intelligent Storage}}, and {{eFPGA}} in 12nm {{FinFET}}.
\newblock In {\em 2023 {{IEEE Custom Integrated Circuits Conference}} ({{CICC}})\/} (Apr. 2023), pp.~1--2.

\bibitem{gao2023acm}
{\sc Gao, J., Ji, W., Chang, F., Han, S., Wei, B., Liu, Z., and Wang, Y.}
\newblock {A Systematic Survey of General Sparse Matrix-Matrix Multiplication}.
\newblock {\em ACM Comput. Surv. 55}, 12 (2023).

\bibitem{Gao2017}
{\sc Gao, M., Pu, J., Yang, X., Horowitz, M., and Kozyrakis, C.}
\newblock {TETRIS: Scalable and Efficient Neural Network Acceleration with 3D Memory}.
\newblock {\em SIGARCH Comput. Archit. News 45}, 1 (2017), 751–764.

\bibitem{tangram}
{\sc Gao, M., Yang, X., Pu, J., Horowitz, M., and Kozyrakis, C.}
\newblock {TANGRAM: Optimized Coarse-Grained Dataflow for Scalable NN Accelerators}.
\newblock In {\em Twenty-Fourth International Conference on Architectural Support for Programming Languages and Operating Systems\/} (New York, NY, USA, 2019), ASPLOS '19, Association for Computing Machinery, p.~807–820.

\bibitem{garofaloHeterogeneousInMemoryComputing2022}
{\sc Garofalo, A., Ottavi, G., Conti, F., Karunaratne, G., Boybat, I., Benini, L., and Rossi, D.}
\newblock {A Heterogeneous In-Memory Computing Cluster for Flexible End-to-End Inference of Real-World Deep Neural Networks}.
\newblock {\em IEEE Journal on Emerging and Selected Topics in Circuits and Systems 12}, 2 (June 2022), 422--435.

\bibitem{garofaloDARKSIDEHeterogeneousRISCV2022}
{\sc Garofalo, A., Tortorella, Y., Perotti, M., Valente, L., Nadalini, A., Benini, L., Rossi, D., and Conti, F.}
\newblock {DARKSIDE: A Heterogeneous RISC-V Compute Cluster for Extreme-Edge On-Chip DNN Inference and Training}.
\newblock {\em IEEE Open Journal of the Solid-State Circuits Society 2\/} (2022), 231--243.

\bibitem{gencGemminiEnablingSystematic2021}
{\sc Genc, H., Kim, S., Amid, A., {Haj-Ali}, A., Iyer, V., Prakash, P., Zhao, J., Grubb, D., Liew, H., Mao, H., Ou, A., Schmidt, C., Steffl, S., Wright, J., Stoica, I., {Ragan-Kelley}, J., Asanovic, K., Nikolic, B., and Shao, Y.~S.}
\newblock {Gemmini: Enabling Systematic Deep-Learning Architecture Evaluation via Full-Stack Integration}.
\newblock In {\em 2021 58th {{ACM}}/{{IEEE Design Automation Conference}} ({{DAC}})\/} (Dec. 2021), pp.~769--774.

\bibitem{quant2021}
{\sc Gholami, A., Kim, S., Dong, Z., Yao, Z., Mahoney, M.~W., and Keutzer, K.}
\newblock {A Survey of Quantization Methods for Efficient Neural Network Inference}, 2021.

\bibitem{giriESP4MLPlatformBasedDesign2020}
{\sc Giri, D., Chiu, K.-L., Di~Guglielmo, G., Mantovani, P., and Carloni, L.~P.}
\newblock {ESP4ML: Platform-Based Design of Systems-on-Chip for Embedded Machine Learning}.
\newblock In {\em 2020 {{Design}}, {{Automation}} \& {{Test}} in {{Europe Conference}} \& {{Exhibition}} ({{DATE}})\/} (Mar. 2020), pp.~1049--1054.

\bibitem{giriAcceleratorIntegrationOpenSource2021}
{\sc Giri, D., Chiu, K.-L., Eichler, G., Mantovani, P., and Carloni, L.~P.}
\newblock {Accelerator Integration for Open-Source SoC Design}.
\newblock {\em IEEE Micro 41}, 4 (July 2021), 8--14.

\bibitem{gobieskiMANIC19Mu2023}
{\sc Gobieski, G., Atli, O., Erbagci, C., Mai, K., Beckmann, N., and Lucia, B.}
\newblock {{MANIC}}: {{A}} 19{\textbackslash}mu{\textbackslash}{{mathrmW}} @ {{4MHz}}, 256 {{MOPS}}/{{mW}}, {{RISC-V}} microcontroller with embedded {{MRAM}} main memory and vector-dataflow co-processor in 22nm bulk {{finFET CMOS}}.
\newblock In {\em 2023 {{IEEE International Symposium}} on {{Circuits}} and {{Systems}} ({{ISCAS}})\/} (May 2023), pp.~1--4.

\bibitem{sparten:micro2019}
{\sc Gondimalla, A., Chesnut, N., Thottethodi, M., and Vijaykumar, T.~N.}
\newblock {SparTen: A Sparse Tensor Accelerator for Convolutional Neural Networks}.
\newblock In {\em 52nd Annual IEEE/ACM International Symposium on Microarchitecture\/} (2019), MICRO'52, pp.~151--165.

\bibitem{gonzalez16mm106GOPS2021}
{\sc Gonzalez, A., Zhao, J., Korpan, B., Genc, H., Schmidt, C., Wright, J., Biswas, A., Amid, A., Sheikh, F., Sorokin, A., Kale, S., Yalamanchi, M., Yarlagadda, R., Flannigan, M., Abramowitz, L., Alon, E., Shao, Y.~S., Asanovic, K., and Nikolic, B.}
\newblock {A 16mm \textsuperscript{2} 106.1 GOPS/W Heterogeneous RISC-V Multi-Core Multi-Accelerator SoC in Low-Power 22nm FinFET}.
\newblock In {\em {{ESSCIRC}} 2021 - {{IEEE}} 47th {{European Solid State Circuits Conference}} ({{ESSCIRC}})\/} ({Grenoble, France}, Sept. 2021), {IEEE}, pp.~259--262.

\bibitem{Goodfellow_2016}
{\sc Goodfellow, I., Bengio, Y., and Courville, A.}
\newblock {\em {Deep Learning}}.
\newblock MIT Press, 2016.
\newblock \url{http://www.deeplearningbook.org}.

\bibitem{gap9Product}
{\sc {GreenWaves Technologies GAP9 Processor}}.
\newblock \url{https://greenwaves-technologies.com/gap9_processor/}, 2023.
\newblock Accessed: 2023-04-18.

\bibitem{Gu2020}
{\sc Gu, P., Xie, X., Ding, Y., Chen, G., Zhang, W., Niu, D., and Xie, Y.}
\newblock {iPIM: Programmable In-Memory Image Processing Accelerator Using Near-Bank Architecture}.
\newblock In {\em 2020 ACM/IEEE 47th Annual International Symposium on Computer Architecture (ISCA)\/} (2020), pp.~804--817.

\bibitem{Guan2017}
{\sc Guan, Y., Liang, H., Xu, N., Wang, W., Shi, S., Chen, X., Sun, G., Zhang, W., and Cong, J.}
\newblock {FP-DNN: An Automated Framework for Mapping Deep Neural Networks onto FPGAs with RTL-HLS Hybrid Templates}.
\newblock In {\em 2017 IEEE 25th Annual International Symposium on Field-Programmable Custom Computing Machines (FCCM)\/} (2017), pp.~152--159.

\bibitem{guo:sc2020}
{\sc Guo, C., Hsueh, B.~Y., Leng, J., Qiu, Y., Guan, Y., Wang, Z., Jia, X., Li, X., Guo, M., and Zhu, Y.}
\newblock {Accelerating Sparse DNN Models without Hardware-Support via Tile-Wise Sparsity}.
\newblock In {\em International Conference for High Performance Computing, Networking, Storage and Analysis\/} (2020), SC'20.

\bibitem{Guo_Online}
{\sc Guo, K., Li, W., Zhong, K., Zhu, Z., Zeng, S., Han, S., Xie, Y., Debacker, P., Verhelst, M., and Wang, Y.}
\newblock Neural network accelerator comparison, 2023.

\bibitem{guo2019dl}
{\sc Guo, K., Zeng, S., Yu, J., Wang, Y., and Yang, H.}
\newblock {[DL] A survey of FPGA-based neural network inference accelerators}.
\newblock {\em ACM Transactions on Reconfigurable Technology and Systems (TRETS) 12}, 1 (2019), 1--26.

\bibitem{Gupta2015}
{\sc Gupta, S., Agrawal, A., Gopalakrishnan, K., and Narayanan, P.}
\newblock {Deep Learning with Limited Numerical Precision}.
\newblock In {\em 32nd International Conference on Machine Learning\/} (07-09 Jul 2015), vol.~37 of {\em Proceedings of Machine Learning Research}, pp.~1737--1746.

\bibitem{guthmullerXvpfloatRISCVISA2024}
{\sc Guthmuller, E., Fuguet, C., Bocco, A., Fereyre, J., Alidori, R., Tahir, I., and Durand, Y.}
\newblock Xvpfloat: {{RISC-V ISA Extension}} for {{Variable Extended Precision Floating Point Computation}}.
\newblock {\em IEEE Transactions on Computers 73}, 7 (July 2024), 1683--1697.

\bibitem{GYONGYOSI201951}
{\sc Gyongyosi, L., and Imre, S.}
\newblock {A Survey on quantum computing technology}.
\newblock {\em Computer Science Review 31\/} (2019), 51--71.

\bibitem{Ha_2018}
{\sc Ha, M., and Lee, S.}
\newblock {Multipliers With Approximate 4–2 Compressors and Error Recovery Modules}.
\newblock {\em IEEE Embedded Systems Letters 10}, 1 (2018), 6--9.

\bibitem{hager11MetisAIPU2024}
{\sc Hager, P.~A., Moons, B., Cosemans, S., Papistas, I.~A., Rooseleer, B., Loon, J.~V., Uytterhoeven, R., Zaruba, F., Koumousi, S., Stanisavljevic, M., Mach, S., Mutsaards, S., Aljameh, R.~K., Khov, G.~H., Machiels, B., Olar, C., Psarras, A., Geursen, S., Vermeeren, J., Lu, Y., Maringanti, A., Ameta, D., Katselas, L., H{\"u}tter, N., Schmuck, M., Sivadas, S., Sharma, K., Oliveira, M., Aerne, R., Sharma, N., Soni, T., Bussolino, B., Pesut, D., Pallaro, M., Podlesnii, A., Lyrakis, A., Ruffiner, Y., Dazzi, M., Thiele, J., Goetschalckx, K., Bruschi, N., Doevenspeck, J., Verhoef, B., Linz, S., Garcea, G., Ferguson, J., Koltsidas, I., and Eleftheriou, E.}
\newblock 11.3 {{Metis AIPU}}: {{A}} 12nm {{15TOPS}}/{{W}} 209.{{6TOPS SoC}} for {{Cost-}} and {{Energy-Efficient Inference}} at the {{Edge}}.
\newblock In {\em 2024 {{IEEE International Solid-State Circuits Conference}} ({{ISSCC}})\/} (Feb. 2024), vol.~67, pp.~212--214.

\bibitem{Hajinazar_2021}
{\sc Hajinazar, N., Oliveira, G.~F., Gregorio, S., Ferreira, J.~a.~D., Ghiasi, N.~M., Patel, M., Alser, M., Ghose, S., G\'{o}mez-Luna, J., and Mutlu, O.}
\newblock {SIMDRAM: a framework for bit-serial SIMD processing using DRAM}.
\newblock In {\em Proceedings of the 26th ACM International Conference on Architectural Support for Programming Languages and Operating Systems\/} (New York, NY, USA, 2021), ASPLOS '21, Association for Computing Machinery, p.~329–345.

\bibitem{Hamanaka2023}
{\sc Hamanaka, F., Odan, T., Kise, K., and Chu, T.~V.}
\newblock {An Exploration of State-of-the-Art Automation Frameworks for FPGA-Based DNN Acceleration}.
\newblock {\em IEEE Access 11\/} (2023), 5701--5713.

\bibitem{han2016isca}
{\sc Han, S., Liu, X., Mao, H., Pu, J., Pedram, A., Horowitz, M.~A., and Dally, W.~J.}
\newblock {EIE: Efficient Inference Engine on Compressed Deep Neural Network}.
\newblock In {\em {43rd International Symposium on Computer Architecture}\/} (2016), ISCA'16, p.~243–254.

\bibitem{hao:arxiv2015}
{\sc Han, S., Mao, H., and Dally, W.~J.}
\newblock {Deep Compression: Compressing Deep Neural Networks with Pruning, Trained Quantization and Huffman Coding}.
\newblock {\em CoRR abs/1510.00149\/} (2015).

\bibitem{han:nips2015}
{\sc Han, S., Pool, J., Tran, J., and Dally, W.}
\newblock {Learning both Weights and Connections for Efficient Neural Network}.
\newblock In {\em Advances in Neural Information Processing Systems\/} (2015), vol.~28, Curran Associates, Inc.

\bibitem{pascal}
{\sc Harris, M.}
\newblock {Inside Pascal: NVIDIA’s Newest Computing Platform}, Apr. 2016.

\bibitem{Hashemi_2015}
{\sc Hashemi, S., Bahar, R.~I., and Reda, S.}
\newblock {DRUM: A Dynamic Range Unbiased Multiplier for approximate applications}.
\newblock In {\em 2015 IEEE/ACM International Conference on Computer-Aided Design (ICCAD)\/} (2015), pp.~418--425.

\bibitem{Hassanpour2022}
{\sc Hassanpour, M., Riera, M., and González, A.}
\newblock {A Survey of Near-Data Processing Architectures for Neural Networks}.
\newblock {\em Machine Learning and Knowledge Extraction 4}, 1 (2022), 66--102.

\bibitem{He2020}
{\sc He, M., Song, C., Kim, I., Jeong, C., Kim, S., Park, I., Thottethodi, M., and Vijaykumar, T.~N.}
\newblock {Newton: A DRAM-maker’s Accelerator-in-Memory (AiM) Architecture for Machine Learning}.
\newblock In {\em 2020 53rd Annual IEEE/ACM International Symposium on Microarchitecture (MICRO)\/} (2020), pp.~372--385.

\bibitem{ucnn:isca2018}
{\sc Hegde, K., Yu, J., Agrawal, R., Yan, M., Pellauer, M., and Fletcher, C.~W.}
\newblock {UCNN: Exploiting Computational Reuse in Deep Neural Networks via Weight Repetition}.
\newblock In {\em 45th Annual International Symposium on Computer Architecture\/} (2018), ISCA'18, p.~674–687.

\bibitem{Heyman2022}
{\sc Heyman, K.}
\newblock {DRAM Thermal Issues Reach Crisis Point}.
\newblock \url{https://semiengineering.com/dram-thermal-issues-reach-crisis-point/}, Jun. 2022.
\newblock Accessed: 18/04/2023.

\bibitem{houshmandDIANAEndtoEndHybrid2023}
{\sc Houshmand, P., Sarda, G.~M., Jain, V., Ueyoshi, K., Papistas, I.~A., Shi, M., Zheng, Q., Bhattacharjee, D., Mallik, A., Debacker, P., Verkest, D., and Verhelst, M.}
\newblock {DIANA: An End-to-End Hybrid DIgital and ANAlog Neural Network SoC for the Edge}.
\newblock {\em IEEE Journal of Solid-State Circuits 58}, 1 (Jan. 2023), 203--215.

\bibitem{HPCWIRE2022}
{Quantum computers emerging as accelerators in HPC}, 2022.

\bibitem{hu_micro18}
{\sc Hu, X., Stow, D., and Xie, Y.}
\newblock {Die Stacking Is Happening}.
\newblock {\em IEEE Micro 38}, 1 (2018), 22--28.

\bibitem{HMC2014}
{\sc {Hybrid Memory Cube Consortium}}.
\newblock {Hybrid Memory Cube specification 2.1}, Nov. 2014.

\bibitem{ielmini2018nature}
{\sc Ielmini, D., and Wong, H.-S.~P.}
\newblock {In-memory computing with resistive switching devices}.
\newblock {\em Nature Electronics 1}, 6 (Jun 2018), 333--343.

\bibitem{lunarlake}
{\sc Intel}.
\newblock {Lunar Lake processor specifications}.

\bibitem{arc770}
{\sc Intel}.
\newblock {Intel Arc A770 Graphics 16GB}, Jul 2022.

\bibitem{IntelHLS2022}
{\sc {Intel}}.
\newblock {Intel High Level Synthesis Compiler Reference Manual}, 2022.

\bibitem{jainTinyVersTinyVersatile2023}
{\sc Jain, V., Giraldo, S., Roose, J.~D., Mei, L., Boons, B., and Verhelst, M.}
\newblock {TinyVers: A Tiny Versatile System-on-Chip With State-Retentive eMRAM for ML Inference at the Extreme Edge}.
\newblock {\em IEEE Journal of Solid-State Circuits\/} (2023), 1--12.

\bibitem{Jeddeloh2012}
{\sc Jeddeloh, J., and Keeth, B.}
\newblock {Hybrid memory cube new DRAM architecture increases density and performance}.
\newblock In {\em 2012 Symposium on VLSI Technology (VLSIT)\/} (2012), pp.~87--88.

\bibitem{JEDEC2023}
{\sc {JEDEC Solid State Technology Association}}.
\newblock {High Bandwidth Memory DRAM (HBM3), JESD238A}, Jan. 2023.

\bibitem{jia12nmAgileDesignedSoC2022}
{\sc Jia, T., Mantovani, P., Dos~Santos, M.~C., Giri, D., Zuckerman, J., Loscalzo, E.~J., Cochet, M., Swaminathan, K., Tombesi, G., Zhang, J.~J., Chandramoorthy, N., Wellman, J.-D., Tien, K., Carloni, L., Shepard, K., Brooks, D., Wei, G.-Y., and Bose, P.}
\newblock {A 12nm Agile-Designed SoC for Swarm-Based Perception with Heterogeneous IP Blocks, a Reconfigurable Memory Hierarchy, and an 800MHz Multi-Plane NoC}.
\newblock In {\em {{ESSCIRC}} 2022- {{IEEE}} 48th {{European Solid State Circuits Conference}} ({{ESSCIRC}})\/} (Sept. 2022), pp.~269--272.

\bibitem{Caffe}
{\sc Jia, Y., Shelhamer, E., Donahue, J., Karayev, S., Long, J., Girshick, R., Guadarrama, S., and Darrell, T.}
\newblock {Caffe: Convolutional Architecture for Fast Feature Embedding}.
\newblock {\em arXiv preprint arXiv:1408.5093\/} (2014).

\bibitem{jiaDissectingGraphcoreIPU2019}
{\sc Jia, Z., Tillman, B., Maggioni, M., and Scarpazza, D.~P.}
\newblock {Dissecting the Graphcore IPU Architecture via Microbenchmarking}, Dec. 2019.

\bibitem{JiaoRISCVExtensionTransformer2021}
{\sc Jiao, Q., Hu, W., Liu, F., and Dong, Y.}
\newblock {RISC-VTF: RISC-V Based Extended Instruction Set for Transformer}.
\newblock In {\em 2021 IEEE International Conference on Systems, Man, and Cybernetics (SMC)\/} (2021), pp.~1565--1570.

\bibitem{jiao2020isscc}
{\sc Jiao, Y., Han, L., Jin, R., Su, Y.-J., Ho, C., Yin, L., Li, Y., Chen, L., Chen, Z., Liu, L., He, Z., Yan, Y., He, J., Mao, J., Zai, X., Wu, X., Zhou, Y., Gu, M., Zhu, G., Zhong, R., Lee, W., Chen, P., Chen, Y., Li, W., Xiao, D., Yan, Q., Zhuang, M., Chen, J., Tian, Y., Lin, Y., Wu, W., Li, H., and Dou, Z.}
\newblock {A 12nm Programmable Convolution-Efficient Neural-Processing-Unit Chip Achieving 825TOPS}.
\newblock In {\em {2020 IEEE International Solid- State Circuits Conference - (ISSCC)}\/} (2020), pp.~136--140.

\bibitem{jin2022fnet}
{\sc Jin, Q., Ren, J., Zhuang, R., Hanumante, S., Li, Z., Chen, Z., Wang, Y., Yang, K., and Tulyakov, S.}
\newblock {F8Net: Fixed-Point 8-bit Only Multiplication for Network Quantization}.
\newblock In {\em International Conference on Learning Representations\/} (2022).

\bibitem{shalf_ptrs19}
{\sc John, S.}
\newblock {The future of computing beyond Moore's Law}.
\newblock {\em Phil. Trans. R. Soc.\/} (2020).

\bibitem{MinJou_1999}
{\sc Jou, J.~M., Kuang, S.~R., and Chen, R.~D.}
\newblock {Design of low-error fixed-width multipliers for DSP applications}.
\newblock {\em IEEE Transactions on Circuits and Systems II: Analog and Digital Signal Processing 46}, 6 (1999), 836--842.

\bibitem{Jouppi_2017}
{\sc Jouppi, N., Borchers, A., Boyle, R., Cantin, P.-l., Chao, C., Clark, C., Coriell, J., Daley, M., Dau, M., Dean, J., Gelb, B., Young, C., Ghaemmaghami, T., Gottipati, R., Gulland, W., Hagmann, R., Ho, C., Hogberg, D., Hu, J., and Boden, N.}
\newblock {In-Datacenter Performance Analysis of a Tensor Processing Unit}.
\newblock In {\em 44th {{Annual International Symposium}} on {{Computer Architecture}}\/} (06 2017), {Association for Computing Machinery}, pp.~1--12.

\bibitem{jouppi2023isca}
{\sc Jouppi, N., Kurian, G., Li, S., Ma, P., Nagarajan, R., Nai, L., Patil, N., Subramanian, S., Swing, A., Towles, B., Young, C., Zhou, X., Zhou, Z., and Patterson, D.~A.}
\newblock {TPU v4: An Optically Reconfigurable Supercomputer for Machine Learning with Hardware Support for Embeddings}.
\newblock In {\em {50th Annual International Symposium on Computer Architecture}\/} (2023).

\bibitem{jouppiMotivationEvaluationFirst2018}
{\sc Jouppi, N., Young, C., Patil, N., and Patterson, D.}
\newblock {Motivation for and Evaluation of the First Tensor Processing Unit}.
\newblock {\em IEEE Micro 38}, 3 (May 2018), 10--19.

\bibitem{jouppiDomainspecificSupercomputerTraining2020}
{\sc Jouppi, N.~P., Yoon, D.~H., Kurian, G., Li, S., Patil, N., Laudon, J., Young, C., and Patterson, D.}
\newblock {A Domain-Specific Supercomputer for Training Deep Neural Networks}.
\newblock {\em Communications of the ACM 63}, 7 (June 2020), 67--78.

\bibitem{juSystolicNeuralCPU2023}
{\sc Ju, Y., and Gu, J.}
\newblock {A Systolic Neural CPU Processor Combining Deep Learning and General-Purpose Computing With Enhanced Data Locality and End-to-End Performance}.
\newblock {\em IEEE Journal of Solid-State Circuits 58}, 1 (Jan. 2023), 216--226.

\bibitem{Kada2015}
{\sc Kada, M.}
\newblock {\em {Research and Development History of Three-Dimensional Integration Technology}}.
\newblock Springer International Publishing, Cham, 2015, pp.~1--23.

\bibitem{ajaykumar_micro15}
{\sc Kannan, A., Jerger, N.~E., and Loh, G.~H.}
\newblock {Enabling interposer-based disintegration of multi-core processors}.
\newblock In {\em 2015 48th Annual IEEE/ACM International Symposium on Microarchitecture (MICRO)\/} (2015), pp.~546--558.

\bibitem{kaplanScalingLawsNeural2020}
{\sc Kaplan, J., McCandlish, S., Henighan, T., Brown, T.~B., Chess, B., Child, R., Gray, S., Radford, A., Wu, J., and Amodei, D.}
\newblock Scaling {{Laws}} for {{Neural Language Models}}, Jan. 2020.

\bibitem{khaddam2022hermes}
{\sc Khaddam-Aljameh, R., Stanisavljevic, M., Mas, J.~F., Karunaratne, G., Br{\"a}ndli, M., Liu, F., Singh, A., M{\"u}ller, S.~M., Egger, U., Petropoulos, A., et~al.}
\newblock {HERMES-core—A 1.59-TOPS/mm 2 PCM on 14-nm CMOS in-memory compute core using 300-ps/LSB linearized CCO-based ADCs}.
\newblock {\em IEEE Journal of Solid-State Circuits 57}, 4 (2022), 1027--1038.

\bibitem{khairy2019}
{\sc Khairy, M., Wassal, A.~G., and Zahran, M.}
\newblock {A survey of architectural approaches for improving GPGPU performance, programmability and heterogeneity}.
\newblock {\em Journal of Parallel and Distributed Computing 127\/} (2019), 65--88.

\bibitem{Kim2016}
{\sc Kim, D., Kung, J., Chai, S., Yalamanchili, S., and Mukhopadhyay, S.}
\newblock {Neurocube: A Programmable Digital Neuromorphic Architecture with High-Density 3D Memory}.
\newblock In {\em 2016 ACM/IEEE 43rd Annual International Symposium on Computer Architecture (ISCA)\/} (2016), pp.~380--392.

\bibitem{Kim2022}
{\sc Kim, D., Yu, C., Xie, S., Chen, Y., Kim, J.-Y., Kim, B., Kulkarni, J.~P., and Kim, T. T.-H.}
\newblock {An Overview of Processing-in-Memory Circuits for Artificial Intelligence and Machine Learning}.
\newblock {\em IEEE Journal on Emerging and Selected Topics in Circuits and Systems 12}, 2 (2022), 338--353.

\bibitem{Joonyoung2014}
{\sc Kim, J., and Kim, Y.}
\newblock {HBM: Memory solution for bandwidth-hungry processors}.
\newblock In {\em 2014 IEEE Hot Chips 26 Symposium (HCS)\/} (2014), pp.~1--24.

\bibitem{kim2023stackoptimizationtransformerinference}
{\sc Kim, S., Hooper, C., Wattanawong, T., Kang, M., Yan, R., Genc, H., Dinh, G., Huang, Q., Keutzer, K., Mahoney, M.~W., Shao, Y.~S., and Gholami, A.}
\newblock {Full Stack Optimization of Transformer Inference: a Survey}, 2023.

\bibitem{kim2022vlsi}
{\sc Kim, S., Kim, S., Um, S., Kim, S., Kim, K., and Yoo, H.-J.}
\newblock {Neuro-CIM: A 310.4 TOPS/W Neuromorphic Computing-in-Memory Processor with Low WL/BL activity and Digital-Analog Mixed-mode Neuron Firing}.
\newblock In {\em 2022 IEEE Symposium on VLSI Technology and Circuits (VLSI Technology and Circuits)\/} (2022), pp.~38--39.

\bibitem{Knechtel2017}
{\sc Knechtel, J., Sinanoglu, O., Elfadel, I. A.~M., Lienig, J., and Sze, C. C.~N.}
\newblock {Large-Scale 3D Chips: Challenges and Solutions for Design Automation, Testing, and Trustworthy Integration}.
\newblock {\em IPSJ Transactions on System and LSI Design Methodology 10\/} (2017), 45--62.

\bibitem{knight2018frontiers}
{\sc Knight, J.~C., and Nowotny, T.}
\newblock {GPUs Outperform Current HPC and Neuromorphic Solutions in Terms of Speed and Energy When Simulating a Highly-Connected Cortical Model}.
\newblock {\em Frontiers in Neuroscience 12\/} (2018).

\bibitem{knowlesGraphcore2021}
{\sc Knowles, S.}
\newblock {Graphcore}.
\newblock In {\em 2021 {{IEEE Hot Chips}} 33 {{Symposium}} ({{HCS}})\/} (Aug. 2021), pp.~1--25.

\bibitem{Kong_2021}
{\sc Kong, T., and Li, S.}
\newblock {Design and Analysis of Approximate 4–2 Compressors for High-Accuracy Multipliers}.
\newblock {\em IEEE Transactions on Very Large Scale Integration (VLSI) Systems 29}, 10 (2021), 1771--1781.

\bibitem{Ampere}
{\sc Krashinsky, R., Giroux, O., Jones, S., Stam, N., and Ramaswamy, S.}
\newblock {NVIDIA Ampere Architecture In-Depth}, May 2020.

\bibitem{Krizhevsky2012}
{\sc Krizhevsky, A., Sutskever, I., and Hinton, G.~E.}
\newblock {ImageNet Classification with Deep Convolutional Neural Networks}.
\newblock In {\em Advances in Neural Information Processing Systems 25}, F.~Pereira, C.~J.~C. Burges, L.~Bottou, and K.~Q. Weinberger, Eds. Curran Associates, Inc., 2012, pp.~1097--1105.

\bibitem{Kulkarni_2011}
{\sc Kulkarni, P., Gupta, P., and Ercegovac, M.}
\newblock {Trading Accuracy for Power with an Underdesigned Multiplier Architecture}.
\newblock In {\em 2011 24th Internatioal Conference on VLSI Design\/} (2011), pp.~346--351.

\bibitem{kwon_iceic23}
{\sc Kwon, Y., Han, J., Cho, Y.~P., Kim, J., Chung, J., Choi, J., Park, S., Kim, I., Kwon, H., Kim, J., Kim, H., Jeon, W., Jeon, Y., Cho, M., and Choi, M.}
\newblock {Chiplet Heterogeneous-Integration AI Processor}.
\newblock In {\em 2023 International Conference on Electronics, Information, and Communication (ICEIC)\/} (2023).

\bibitem{Kwon2021}
{\sc Kwon, Y.-C., Lee, S.~H., Lee, J., Kwon, S.-H., Ryu, J.~M., Son, J.-P., Seongil, O., Yu, H.-S., Lee, H., Kim, S.~Y., Cho, Y., Kim, J.~G., Choi, J., Shin, H.-S., Kim, J., Phuah, B., Kim, H., Song, M.~J., Choi, A., Kim, D., Kim, S., Kim, E.-B., Wang, D., Kang, S., Ro, Y., Seo, S., Song, J., Youn, J., Sohn, K., and Kim, N.~S.}
\newblock {25.4 A 20nm 6GB Function-In-Memory DRAM, Based on HBM2 with a 1.2TFLOPS Programmable Computing Unit Using Bank-Level Parallelism, for Machine Learning Applications}.
\newblock In {\em 2021 IEEE International Solid- State Circuits Conference (ISSCC)\/} (2021), vol.~64, pp.~350--352.

\bibitem{lan_eptc21}
{\sc Lan, J., Nambiar, V.~P., Sabapathy, R., Rotaru, M.~D., and Do, A.~T.}
\newblock {Chiplet-based Architecture Design for Multi-Core Neuromorphic Processor}.
\newblock In {\em 2021 IEEE 23rd Electronics Packaging Technology Conference (EPTC)\/} (2021), pp.~410--412.

\bibitem{lattner2021mlir}
{\sc Lattner, C., Amini, M., Bondhugula, U., Cohen, A., Davis, A., Pienaar, J., Riddle, R., Shpeisman, T., Vasilache, N., and Zinenko, O.}
\newblock {MLIR: Scaling Compiler Infrastructure for Domain Specific Computation}.
\newblock In {\em IEEE/ACM International Symposium on Code Generation and Optimization (CGO)\/} (2021), pp.~2--14.

\bibitem{lazoAdaptableRegisterFile2022}
{\sc Lazo, C.~R., Reggiani, E., Morales, C.~R., Figueras~Bagu{\'e}, R., Villa~Vargas, L.~A., Ram{\'i}rez~Salinas, M.~A., Cort{\'e}s, M.~V., Sabri~{\"U}nsal, O., and Cristal, A.}
\newblock {Adaptable Register File Organization for Vector Processors}.
\newblock In {\em 2022 {{IEEE International Symposium}} on {{High-Performance Computer Architecture}} ({{HPCA}})\/} (Apr. 2022), pp.~786--799.

\bibitem{lecun2015}
{\sc LeCun, Y., Bengio, Y., and Hinton, G.}
\newblock {Deep learning}.
\newblock {\em Nature 521}, 7553 (2015), 436.

\bibitem{Lecun1998}
{\sc Lecun, Y., Bottou, L., Bengio, Y., and Haffner, P.}
\newblock {Gradient-based learning applied to document recognition}.
\newblock {\em Proceedings of the IEEE 86}, 11 (1998), 2278--2324.

\bibitem{lee2022jsscc}
{\sc Lee, S.~K., Agrawal, A., Silberman, J., Ziegler, M., Kang, M., Venkataramani, S., Cao, N., Fleischer, B., Guillorn, M., Cohen, M., Mueller, S.~M., Oh, J., Lutz, M., Jung, J., Koswatta, S., Zhou, C., Zalani, V., Kar, M., Bonanno, J., Casatuta, R., Chen, C.-Y., Choi, J., Haynie, H., Herbert, A., Jain, R., Kim, K.-H., Li, Y., Ren, Z., Rider, S., Schaal, M., Schelm, K., Scheuermann, M.~R., Sun, X., Tran, H., Wang, N., Wang, W., Zhang, X., Shah, V., Curran, B., Srinivasan, V., Lu, P.-F., Shukla, S., Gopalakrishnan, K., and Chang, L.}
\newblock {A 7-nm Four-Core Mixed-Precision AI Chip With 26.2-TFLOPS Hybrid-FP8 Training, 104.9-TOPS INT4 Inference, and Workload-Aware Throttling}.
\newblock {\em IEEE Journal of Solid-State Circuits 57}, 1 (2022), 182--197.

\bibitem{lee64TOPSEnergyEfficientTensor2022}
{\sc Lee, S.~M., Kim, H., Yeon, J., Lee, J., Choi, Y., Kim, M., Park, C., Jang, K., Kim, Y., Kim, Y., Lee, C., Han, H., Kim, W.~E., Tang, R., and Baek, J.~H.}
\newblock {A 64-TOPS Energy-Efficient Tensor Accelerator in 14nm With Reconfigurable Fetch Network and Processing Fusion for Maximal Data Reuse}.
\newblock {\em IEEE Open Journal of the Solid-State Circuits Society 2\/} (2022), 219--230.

\bibitem{lee201445nm}
{\sc Lee, Y., Waterman, A., Avizienis, R., Cook, H., Sun, C., Stojanovi{\'c}, V., and Asanovi{\'c}, K.}
\newblock {A 45nm 1.3 GHz 16.7 double-precision GFLOPS/W RISC-V processor with vector accelerators}.
\newblock In {\em ESSCIRC 2014-40th European Solid State Circuits Conference (ESSCIRC)\/} (2014), IEEE, pp.~199--202.

\bibitem{lepri2023jeds}
{\sc Lepri, N., Glukhov, A., Cattaneo, L., Farronato, M., Mannocci, P., and Ielmini, D.}
\newblock {In-memory computing for machine learning and deep learning}.
\newblock {\em {IEEE Journal of the Electron Devices Society}\/} (2023), 1--1.

\bibitem{liCIFERCacheCoherent12nm2023}
{\sc Li, A., Chang, T.-J., Gao, F., Ta, T., Tziantzioulis, G., Ou, Y., Wang, M., Tu, J., Xu, K., Jackson, P., Ning, A., Chirkov, G., {Orenes-Vera}, M., Agwa, S., Yan, X., Tang, E., Balkind, J., Batten, C., and Wentzlaff, D.}
\newblock {{CIFER}}: {{A Cache-Coherent}} 12-nm 16-mm2 {{SoC With Four}} 64-{{Bit RISC-V Application Cores}}, 18 32-{{Bit RISC-V Compute Cores}}, and a 1541 {{LUT6}}/mm2 {{Synthesizable eFPGA}}.
\newblock {\em IEEE Solid-State Circuits Letters 6\/} (2023), 229--232.

\bibitem{Li2021}
{\sc Li, G., Liu, Z., Li, F., and Cheng, J.}
\newblock {Block Convolution: Towards Memory-Efficient Inference of Large-Scale CNNs on FPGA}.
\newblock {\em CoRR abs/2105.08937\/} (2021).

\bibitem{squeezeflow:2019}
{\sc Li, J., Jiang, S., Gong, S., Wu, J., Yan, J., Yan, G., and Li, X.}
\newblock {SqueezeFlow: A Sparse CNN Accelerator Exploiting Concise Convolution Rules}.
\newblock {\em IEEE Transactions on Computers 68}, 11 (2019), 1663--1677.

\bibitem{Li_2021}
{\sc Li, L., Hammad, I., and El-Sankary, K.}
\newblock {Dual segmentation approximate multiplier}.
\newblock {\em Electronics Letters 57}, 19 (2021), 718--720.

\bibitem{Li_2017}
{\sc Li, S., Niu, D., Malladi, K.~T., Zheng, H., Brennan, B., and Xie, Y.}
\newblock {DRISA: a DRAM-based Reconfigurable In-Situ Accelerator}.
\newblock In {\em Proceedings of the 50th Annual IEEE/ACM International Symposium on Microarchitecture\/} (New York, NY, USA, 2017), MICRO-50 '17, Association for Computing Machinery, p.~288–301.

\bibitem{li2021sprint}
{\sc Li, Y., Louri, A., and Karanth, A.}
\newblock {SPRINT: A high-performance, energy-efficient, and scalable chiplet-based accelerator with photonic interconnects for CNN inference}.
\newblock {\em IEEE Transactions on Parallel and Distributed Systems 33}, 10 (2021), 2332--2345.

\bibitem{lin2020isscc}
{\sc Lin, C.-H., Cheng, C.-C., Tsai, Y.-M., Hung, S.-J., Kuo, Y.-T., Wang, P.~H., Tsung, P.-K., Hsu, J.-Y., Lai, W.-C., Liu, C.-H., Wang, S.-Y., Kuo, C.-H., Chang, C.-Y., Lee, M.-H., Lin, T.-Y., and Chen, C.-C.}
\newblock {A 3.4-to-13.3TOPS/W 3.6TOPS Dual-Core Deep-Learning Accelerator for Versatile AI Applications in 7nm 5G Smartphone SoC}.
\newblock In {\em {2020 IEEE International Solid- State Circuits Conference - (ISSCC)}\/} (2020), pp.~134--136.

\bibitem{lin_vlsi19}
{\sc Lin, M.-S., Huang, T.-C., Tsai, C.-C., Tam, K.-H., Hsieh, C.-H., Chen, T., Huang, W.-H., Hu, J., Chen, Y.-C., Goel, S.~K., Fu, C.-M., Rusu, S., Li, C.-C., Yang, S.-Y., Wong, M., Yang, S.-C., and Lee, F.}
\newblock {A 7nm 4GHz Arm-core-based CoWoS Chiplet Design for High Performance Computing}.
\newblock In {\em 2019 Symposium on VLSI Circuits\/} (2019).

\bibitem{liu2018frontiers}
{\sc Liu, C., Bellec, G., Vogginger, B., Kappel, D., Partzsch, J., Neumärker, F., Höppner, S., Maass, W., Furber, S.~B., Legenstein, R., and Mayr, C.~G.}
\newblock {Memory-Efficient Deep Learning on a SpiNNaker 2 Prototype}.
\newblock {\em Frontiers in Neuroscience 12\/} (2018).

\bibitem{Leibo2019ACMSUR}
{\sc Liu, L., Zhu, J., Li, Z., Lu, Y., Deng, Y., Han, J., Yin, S., and Wei, S.}
\newblock {A Survey of Coarse-Grained Reconfigurable Architecture and Design: Taxonomy, Challenges, and Applications}.
\newblock {\em ACM Comput. Surv. 52}, 6 (2019).

\bibitem{liu2015dac}
{\sc Liu, X., Mao, M., Liu, B., Li, H., Chen, Y., Li, B., Wang, Y., Jiang, H., Barnell, M., Wu, Q., and Yang, J.}
\newblock {RENO: A high-efficient reconfigurable neuromorphic computing accelerator design}.
\newblock In {\em {2015 52nd ACM/EDAC/IEEE Design Automation Conference (DAC)}\/} (2015), pp.~1--6.

\bibitem{Liu2022}
{\sc Liu, Z., Cheng, K.-T., Huang, D., Xing, E., and Shen, Z.}
\newblock {Nonuniform-to-Uniform Quantization: Towards Accurate Quantization via Generalized Straight-Through Estimation}.
\newblock In {\em 2022 IEEE/CVF Conference on Computer Vision and Pattern Recognition (CVPR)\/} (2022), pp.~4932--4942.

\bibitem{Ma2018}
{\sc Ma, Y., Cao, Y., Vrudhula, S., and Seo, J.-s.}
\newblock {Optimizing the Convolution Operation to Accelerate Deep Neural Networks on FPGA}.
\newblock {\em IEEE Transactions on Very Large Scale Integration (VLSI) Systems 26}, 7 (2018), 1354--1367.

\bibitem{Suda2016}
{\sc Ma, Y., Suda, N., Cao, Y., Seo, J.-s., and Vrudhula, S.}
\newblock {Scalable and modularized RTL compilation of Convolutional Neural Networks onto FPGA}.
\newblock In {\em 2016 26th International Conference on Field Programmable Logic and Applications (FPL)\/} (2016), pp.~1--8.

\bibitem{Machupalli_2022}
{\sc Machupalli, R., Hossain, M., and Mandal, M.}
\newblock {Review of ASIC accelerators for deep neural network}.
\newblock {\em Microprocessors and Microsystems 89\/} (2022), 104441.

\bibitem{Machura2022}
{\sc Machura, M., Danilowicz, M., and Kryjak, T.}
\newblock {Embedded Object Detection with Custom LittleNet, FINN and Vitis AI DCNN Accelerators}.
\newblock {\em Journal of Low Power Electronics and Applications 12}, 2 (2022).

\bibitem{mallasenPERCIVAL2022}
{\sc Mallasén, D., Murillo, R., Barrio, A. A.~D., Botella, G., Piñuel, L., and Prieto-Matias, M.}
\newblock {PERCIVAL: Open-Source Posit RISC-V Core With Quire Capability}.
\newblock {\em IEEE Transactions on Emerging Topics in Computing 10}, 3 (2022), 1241--1252.

\bibitem{martinez_vlsi20}
{\sc Martinez, P.~Y., Beilliard, Y., Godard, M., Danovitch, D., Drouin, D., Charbonnier, J., Coudrain, P., Garnier, A., Lattard, D., Vivet, P., Cheramy, S., Guthmuller, E., Tortolero, C.~F., Mengue, V., Durupt, J., Philippe, A., and Dutoit, D.}
\newblock {ExaNoDe: Combined Integration of Chiplets on Active Interposer with Bare Dice in a Multi-Chip-Module for Heterogeneous and Scalable High Performance Compute Nodes}.
\newblock In {\em 2020 IEEE Symposium on VLSI Technology\/} (2020).

\bibitem{Mathur2021}
{\sc Mathur, R., Kumar, A. K.~A., John, L., and Kulkarni, J.~P.}
\newblock {Thermal-Aware Design Space Exploration of 3-D Systolic ML Accelerators}.
\newblock {\em IEEE Journal on Exploratory Solid-State Computational Devices and Circuits 7}, 1 (2021), 70--78.

\bibitem{mlperf-training}
{\sc Mattson, P., Cheng, C., Diamos, G., Coleman, C., Micikevicius, P., Patterson, D., Tang, H., Wei, G.-Y., Bailis, P., Bittorf, V., et~al.}
\newblock {Mlperf training benchmark}.
\newblock {\em Proceedings of Machine Learning and Systems 2\/} (2020), 336--349.

\bibitem{mlperf}
{\sc Mattson, P., Reddi, V.~J., Cheng, C., Coleman, C., Diamos, G., Kanter, D., Micikevicius, P., Patterson, D., Schmuelling, G., Tang, H., Wei, G.-Y., and Wu, C.-J.}
\newblock {MLPerf: An Industry Standard Benchmark Suite for Machine Learning Performance}.
\newblock {\em IEEE Micro 40}, 2 (2020), 8--16.

\bibitem{medina2019hotchips}
{\sc Medina, E.}
\newblock {[Habana Labs presentation]}.
\newblock In {\em 2019 IEEE Hot Chips 31 Symposium (HCS)\/} (2019), pp.~1--29.

\bibitem{Micron2018}
{\sc {Micron}}.
\newblock {Hybrid Memory Cube – HMC Gen2 HMC Memory Features}, 2018.

\bibitem{Min2019}
{\sc Min, C., Mao, J., Li, H., and Chen, Y.}
\newblock {NeuralHMC: An Efficient HMC-Based Accelerator for Deep Neural Networks}.
\newblock In {\em 24th Asia and South Pacific Design Automation Conference\/} (2019), ACM, p.~394–399.

\bibitem{minerviniVitruviusAreaEfficientRISCV2023}
{\sc Minervini, F., Palomar, O., Unsal, O., Reggiani, E., Quiroga, J., Marimon, J., Rojas, C., Figueras, R., Ruiz, A., Gonzalez, A., Mendoza, J., Vargas, I., Hernandez, C., Cabre, J., Khoirunisya, L., Bouhali, M., Pavon, J., Moll, F., Olivieri, M., Kovac, M., Kovac, M., Dragic, L., Valero, M., and Cristal, A.}
\newblock {Vitruvius+: An Area-Efficient RISC-V Decoupled Vector Coprocessor for High Performance Computing Applications}.
\newblock {\em ACM Transactions on Architecture and Code Optimization 20}, 2 (Mar. 2023), 28:1--28:25.

\bibitem{miro-panadesSamurAIVersatileIoT2022}
{\sc {Miro-Panades}, I., Tain, B., Christmann, J.-F., Coriat, D., Lemaire, R., Jany, C., Martineau, B., Chaix, F., Waltener, G., Pluchart, E., Noel, J.-P., Makosiej, A., Montoya, M., {Bacles-Min}, S., Briand, D., Philippe, J.-M., Thonnart, Y., Valentian, A., Heitzmann, F., and Clermidy, F.}
\newblock {SamurAI: A Versatile IoT Node With Event-Driven Wake-Up and Embedded ML Acceleration}.
\newblock {\em IEEE Journal of Solid-State Circuits\/} (2022), 1--0.

\bibitem{sparsetensorcore:2021}
{\sc Mishra, A.~K., Latorre, J.~A., Pool, J., Stosic, D., Stosic, D., Venkatesh, G., Yu, C., and Micikevicius, P.}
\newblock {Accelerating Sparse Deep Neural Networks}.
\newblock {\em CoRR abs/2104.08378\/} (2021).

\bibitem{mounce_ac16}
{\sc Mounce, G., Lyke, J., Horan, S., Powell, W., Doyle, R., and Some, R.}
\newblock {Chiplet based approach for heterogeneous processing and packaging architectures}.
\newblock In {\em 2016 IEEE Aerospace Conference\/} (2016), pp.~1--12.

\bibitem{flexagon:aplos2023}
{\sc Mu\~{n}oz Mart\'{\i}nez, F., Garg, R., Pellauer, M., Abell\'{a}n, J.~L., Acacio, M.~E., and Krishna, T.}
\newblock {Flexagon: A Multi-Dataflow Sparse-Sparse Matrix Multiplication Accelerator for Efficient DNN Processing}.
\newblock In {\em 28th ACM International Conference on Architectural Support for Programming Languages and Operating Systems, Volume 3\/} (2023), ASPLOS 2023, pp.~252--265.

\bibitem{nadaliniTOPSRISCVParallel2023a}
{\sc Nadalini, A., Rutishauser, G., Burrello, A., Bruschi, N., Garofalo, A., Benini, L., Conti, F., and Rossi, D.}
\newblock A 3 {{TOPS}}/{{W RISC-V Parallel Cluster}} for {{Inference}} of {{Fine-Grain Mixed-Precision Quantized Neural Networks}}.
\newblock In {\em 2023 {{IEEE Computer Society Annual Symposium}} on {{VLSI}} ({{ISVLSI}})\/} (June 2023), pp.~1--6.

\bibitem{nahmias_2020}
{\sc Nahmias, M.~A., de~Lima, T.~F., Tait, A.~N., Peng, H.-T., Shastri, B.~J., and Prucnal, P.~R.}
\newblock {Photonic Multiply-Accumulate Operations for Neural Networks}.
\newblock {\em IEEE Journal of Selected Topics in Quantum Electronics 26}, 1 (2020), 1--18.

\bibitem{Narayanamoorthy_2015}
{\sc Narayanamoorthy, S., Moghaddam, H.~A., Liu, Z., Park, T., and Kim, N.~S.}
\newblock {Energy-Efficient Approximate Multiplication for Digital Signal Processing and Classification Applications}.
\newblock {\em IEEE Transactions on Very Large Scale Integration (VLSI) Systems 23}, 6 (2015), 1180--1184.

\bibitem{narayanan2021ted}
{\sc Narayanan, P., Ambrogio, S., Okazaki, A., Hosokawa, K., Tsai, H., Nomura, A., Yasuda, T., Mackin, C., Lewis, S.~C., Friz, A., Ishii, M., Kohda, Y., Mori, H., Spoon, K., Khaddam-Aljameh, R., Saulnier, N., Bergendahl, M., Demarest, J., Brew, K.~W., Chan, V., Choi, S., Ok, I., Ahsan, I., Lie, F.~L., Haensch, W., Narayanan, V., and Burr, G.~W.}
\newblock {Fully On-Chip MAC at 14 nm Enabled by Accurate Row-Wise Programming of PCM-Based Weights and Parallel Vector-Transport in Duration-Format}.
\newblock {\em IEEE Transactions on Electron Devices 68}, 12 (2021), 6629--6636.

\bibitem{Nitin2018}
{\sc Nitin, ., Thottethodi, M., and Vijaykumar, T.~N.}
\newblock {Millipede: Die-Stacked Memory Optimizations for Big Data Machine Learning Analytics}.
\newblock In {\em 2018 IEEE International Parallel and Distributed Processing Symposium (IPDPS)\/} (2018), pp.~160--171.

\bibitem{nurvitadhi_fccm19}
{\sc Nurvitadhi, E., Kwon, D., Jafari, A., Boutros, A., Sim, J., Tomson, P., Sumbul, H., Chen, G., Knag, P., Kumar, R., Krishnamurthy, R., Gribok, S., Pasca, B., Langhammer, M., Marr, D., and Dasu, A.}
\newblock {Why Compete When You Can Work Together: FPGA-ASIC Integration for Persistent RNNs}.
\newblock In {\em 2019 IEEE 27th Annual International Symposium on Field-Programmable Custom Computing Machines (FCCM)\/} (2019), pp.~199--207.

\bibitem{fermi}
{\sc {NVIDIA}}.
\newblock {Fermi}, 2009.

\bibitem{kepler}
{\sc {NVIDIA}}.
\newblock {Kepler GK110}, 2012.

\bibitem{eos}
{\sc NVidia}.
\newblock {NVIDIA Announces DGX H100 Systems – World’s Most Advanced Enterprise AI Infrastructure}, Mar. 2022.

\bibitem{dgx}
{\sc NVidia}.
\newblock {NVIDIA DGX Platform The best of NVIDIA AI—all in one place}, May 2023.

\bibitem{nvlink}
{\sc NVidia}.
\newblock {NVLink and NVSwitch}, May 2023.

\bibitem{oh2020vlsic}
{\sc Oh, J., Lee, S.~K., Kang, M., Ziegler, M., Silberman, J., Agrawal, A., Venkataramani, S., Fleischer, B., Guillorn, M., Choi, J., Wang, W., Mueller, S., Ben-Yehuda, S., Bonanno, J., Cao, N., Casatuta, R., Chen, C.-Y., Cohen, M., Erez, O., Fox, T., Gristede, G., Haynie, H., Ivanov, V., Koswatta, S., Lo, S.-H., Lutz, M., Maier, G., Mesh, A., Nustov, Y., Rider, S., Schaal, M., Scheuermann, M., Sun, X., Wang, N., Yee, F., Zhou, C., Shah, V., Curran, B., Srinivasan, V., Lu, P.-F., Shukla, S., Gopalakrishnan, K., and Chang, L.}
\newblock {A 3.0 TFLOPS 0.62V Scalable Processor Core for High Compute Utilization AI Training and Inference}.
\newblock In {\em 2020 IEEE Symposium on VLSI Circuits\/} (2020), pp.~1--2.

\bibitem{Oliveira2017}
{\sc Oliveira, G.~F., Santos, P.~C., Alves, M. A.~Z., and Carro, L.}
\newblock {NIM: An HMC-Based Machine for Neuron Computation}.
\newblock In {\em Applied Reconfigurable Computing\/} (Cham, 2017), S.~Wong, A.~C. Beck, K.~Bertels, and L.~Carro, Eds., Springer International Publishing, pp.~28--35.

\bibitem{openaiGPT4TechnicalReport2023}
{\sc OpenAI}.
\newblock {{GPT-4 Technical Report}}, Mar. 2023.

\bibitem{ottaviDustin16CoresParallel2023}
{\sc Ottavi, G., Garofalo, A., Tagliavini, G., Conti, F., Mauro, A.~D., Benini, L., and Rossi, D.}
\newblock {Dustin: A 16-Cores Parallel Ultra-Low-Power Cluster With 2b-to-32b Fully Flexible Bit-Precision and Vector Lockstep Execution Mode}.
\newblock {\em IEEE Transactions on Circuits and Systems I: Regular Papers\/} (2023), 1--14.

\bibitem{painkras2013jsscc}
{\sc Painkras, E., Plana, L.~A., Garside, J., Temple, S., Galluppi, F., Patterson, C., Lester, D.~R., Brown, A.~D., and Furber, S.~B.}
\newblock {SpiNNaker: A 1-W 18-Core System-on-Chip for Massively-Parallel Neural Network Simulation}.
\newblock {\em IEEE Journal of Solid-State Circuits 48}, 8 (2013), 1943--1953.

\bibitem{SCNN:isca2017}
{\sc Parashar, A., Rhu, M., Mukkara, A., Puglielli, A., Venkatesan, R., Khailany, B., Emer, J., Keckler, S.~W., and Dally, W.~J.}
\newblock {SCNN: An Accelerator for Compressed-Sparse Convolutional Neural Networks}.
\newblock In {\em Proceedings of the 2017 ACM/IEEE 44th Annual International Symposium on Computer Architecture\/} (2017), ISCA'17, pp.~27--40.

\bibitem{Park_2021}
{\sc Park, G., Kung, J., and Lee, Y.}
\newblock {Design and Analysis of Approximate Compressors for Balanced Error Accumulation in MAC Operator}.
\newblock {\em IEEE Transactions on Circuits and Systems I: Regular Papers 68}, 7 (2021), 2950--2961.

\bibitem{park:arxiv2016}
{\sc Park, J., Li, S.~R., Wen, W., Li, H., Chen, Y., and Dubey, P.}
\newblock {Holistic SparseCNN: Forging the Trident of Accuracy, Speed, and Size}.
\newblock {\em CoRR abs/1608.01409\/} (2016).

\bibitem{park-facebook:arxiv2018}
{\sc Park, J., Naumov, M., Basu, P., Deng, S., Kalaiah, A., Khudia, D.~S., Law, J., Malani, P., Malevich, A., Satish, N., Pino, J.~M., Schatz, M., Sidorov, A., Sivakumar, V., Tulloch, A., Wang, X., Wu, Y., Yuen, H., Diril, U., Dzhulgakov, D., Hazelwood, K.~M., Jia, B., Jia, Y., Qiao, L., Rao, V., Rotem, N., Yoo, S., and Smelyanskiy, M.}
\newblock {Deep Learning Inference in Facebook Data Centers: Characterization, Performance Optimizations and Hardware Implications}.
\newblock {\em CoRR abs/1811.09886\/} (2018).

\bibitem{park2021isscc}
{\sc Park, J.-S., Jang, J.-W., Lee, H., Lee, D., Lee, S., Jung, H., Lee, S., Kwon, S., Jeong, K., Song, J.-H., Lim, S., and Kang, I.}
\newblock {A 6K-MAC Feature-Map-Sparsity-Aware Neural Processing Unit in 5nm Flagship Mobile SoC}.
\newblock In {\em {2021 IEEE International Solid- State Circuits Conference (ISSCC)}\/} (2021), vol.~64, pp.~152--154.

\bibitem{park2022isscc}
{\sc Park, J.-S., Park, C., Kwon, S., Kim, H.-S., Jeon, T., Kang, Y., Lee, H., Lee, D., Kim, J., Lee, Y., Park, S., Jang, J.-W., Ha, S., Kim, M., Bang, J., Lim, S.~H., and Kang, I.}
\newblock {A Multi-Mode 8K-MAC HW-Utilization-Aware Neural Processing Unit with a Unified Multi-Precision Datapath in 4nm Flagship Mobile SoC}.
\newblock In {\em {2022 IEEE International Solid- State Circuits Conference (ISSCC)}\/} (2022), vol.~65, pp.~246--248.

\bibitem{Park2015}
{\sc Park, S.-W., Park, J., Bong, K., Shin, D., Lee, J., Choi, S., and Yoo, H.-J.}
\newblock {An Energy-Efficient and Scalable Deep Learning/Inference Processor With Tetra-Parallel MIMD Architecture for Big Data Applications}.
\newblock {\em IEEE Transactions on Biomedical Circuits and Systems 9}, 6 (2015), 838--848.

\bibitem{torch}
{\sc Paszke, A., Gross, S., Massa, F., Lerer, A., Bradbury, J., Chanan, G., Killeen, T., Lin, Z., Gimelshein, N., Antiga, L., Desmaison, A., Kopf, A., Yang, E., DeVito, Z., Raison, M., Tejani, A., Chilamkurthy, S., Steiner, B., Fang, L., Bai, J., and Chintala, S.}
\newblock {PyTorch: An Imperative Style, High-Performance Deep Learning Library}.
\newblock In {\em Advances in Neural Information Processing Systems 32}. Curran Associates, Inc., 2019, pp.~8024--8035.

\bibitem{paulinRNNBasedRadioResource2021}
{\sc Paulin, G., Andri, R., Conti, F., and Benini, L.}
\newblock {RNN-Based Radio Resource Management on Multicore RISC-V Accelerator Architectures}.
\newblock {\em IEEE Transactions on Very Large Scale Integration (VLSI) Systems 29}, 9 (Sept. 2021), 1624--1637.

\bibitem{paulinOccamy432Core282024}
{\sc Paulin, G., Scheffler, P., Benz, T., Cavalcante, M., Fischer, T., Eggimann, M., Zhang, Y., Wistoff, N., Bertaccini, L., Colagrande, L., Ottavi, G., G{\"u}rkaynak, F.~K., Rossi, D., and Benini, L.}
\newblock Occamy: {{A}} 432-{{Core}} 28.1 {{DP-GFLOP}}/s/{{W}} 83\% {{FPU Utilization Dual-Chiplet}}, {{Dual-HBM2E RISC-V-based Accelerator}} for {{Stencil}} and {{Sparse Linear Algebra Computations}} with 8-to-64-bit {{Floating-Point Support}} in 12nm {{FinFET}}, June 2024.

\bibitem{Pawlowski2011}
{\sc Pawlowski, J.~T.}
\newblock {Hybrid memory cube (HMC)}.
\newblock In {\em 2011 IEEE Hot Chips 23 Symposium (HCS)\/} (2011), pp.~1--24.

\bibitem{Peemen2013}
{\sc Peemen, M., Setio, A. A.~A., Mesman, B., and Corporaal, H.}
\newblock {Memory-centric accelerator design for Convolutional Neural Networks}.
\newblock In {\em 2013 IEEE 31st International Conference on Computer Design (ICCD)\/} (2013), pp.~13--19.

\bibitem{perottiYunOpenSource64Bit2023}
{\sc Perotti, M., Cavalcante, M., Ottaviano, A., Liu, J., and Benini, L.}
\newblock Yun: {{An Open-Source}}, 64-{{Bit RISC-V-Based Vector Processor With Multi-Precision Integer}} and {{Floating-Point Support}} in 65-nm {{CMOS}}.
\newblock {\em IEEE Transactions on Circuits and Systems II: Express Briefs 70}, 10 (Oct. 2023), 3732--3736.

\bibitem{perottiNewAraVector2022}
{\sc Perotti, M., Cavalcante, M., Wistoff, N., Andri, R., Cavigelli, L., and Benini, L.}
\newblock {A ``New Ara'' for Vector Computing: An Open Source Highly Efficient RISC-V V 1.0 Vector Processor Design}.
\newblock In {\em 2022 {{IEEE}} 33rd {{International Conference}} on {{Application-specific Systems}}, {{Architectures}} and {{Processors}} ({{ASAP}})\/} (July 2022), pp.~43--51.

\bibitem{Perri2020}
{\sc Perri, S., Sestito, C., Spagnolo, F., and Corsonello, P.}
\newblock {Efficient Deconvolution Architecture for Heterogeneous Systems-on-Chip}.
\newblock {\em Journal of Imaging 6}, 9 (2020).

\bibitem{Petra_2011}
{\sc Petra, N., De~Caro, D., Garofalo, V., Napoli, E., and Strollo, A. G.~M.}
\newblock {Design of Fixed-Width Multipliers With Linear Compensation Function}.
\newblock {\em IEEE Transactions on Circuits and Systems I: Regular Papers 58}, 5 (2011), 947--960.

\bibitem{prabhu2022chimera}
{\sc Prabhu, K., Gural, A., Khan, Z.~F., Radway, R.~M., Giordano, M., Koul, K., Doshi, R., Kustin, J.~W., Liu, T., Lopes, G.~B., et~al.}
\newblock {CHIMERA: A 0.92-TOPS, 2.2-TOPS/W edge AI accelerator with 2-MByte on-chip foundry resistive RAM for efficient training and inference}.
\newblock {\em IEEE Journal of Solid-State Circuits 57}, 4 (2022), 1013--1026.

\bibitem{prasadSpecializationMeetsFlexibility}
{\sc Prasad, A., Benini, L., and Conti, F.}
\newblock {Specialization Meets Flexibility: A Heterogeneous Architecture for High-Efficiency, High-flexibility AR/VR Processing}.
\newblock In {\em Proceedings of the 2023 {{Design Automation Conference}} ({{DAC}} 2023), to Appear\/} (2023).

\bibitem{prasadSiracusa16Nm2024}
{\sc Prasad, A.~S., Scherer, M., Conti, F., Rossi, D., Mauro, A.~D., Eggimann, M., G{\'o}mez, J.~T., Li, Z., Sarwar, S.~S., Wang, Z., Salvo, B.~D., and Benini, L.}
\newblock Siracusa: {{A}} 16 nm {{Heterogenous RISC-V SoC}} for {{Extended Reality With At-MRAM Neural Engine}}.
\newblock {\em IEEE Journal of Solid-State Circuits\/} (2024), 1--15.

\bibitem{Prickett2022}
{\sc Prickett, M.}
\newblock {The HBM3 roadmap is just getting started}.
\newblock \url{https://www.nextplatform.com/2022/04/06/the-hbm3-roadmap-is-just-getting-started/}, Apr. 2022.
\newblock Accessed: 18/04/2023.

\bibitem{smx}
{\sc Proud, M.}
\newblock {ACHIEVING MAXIMUM COMPUTE THROUGHPUT: PCIE VS. SXM2}, May 2018.

\bibitem{sigma:2020}
{\sc Qin, E., Samajdar, A., Kwon, H., Nadella, V., Srinivasan, S., Das, D., Kaul, B., and Krishna, T.}
\newblock {SIGMA: A Sparse and Irregular GEMM Accelerator with Flexible Interconnects for DNN Training}.
\newblock In {\em 2020 IEEE International Symposium on High Performance Computer Architecture\/} (2020), HPCA'20, pp.~58--70.

\bibitem{BNN2020}
{\sc Qin, H., Gong, R., Liu, X., Bai, X., Song, J., and Sebe, N.}
\newblock {Binary neural networks: A survey}.
\newblock {\em Pattern Recognition 105\/} (2020), 107281.

\bibitem{QISKIT2023}
{IBM Qiskit Simulator}, 2023.

\bibitem{Qiu2016}
{\sc Qiu, J., Wang, J., Yao, S., Guo, K., Li, B., Zhou, E., Yu, J., Tang, T., Xu, N., Song, S., Wang, Y., and Yang, H.}
\newblock {Going Deeper with Embedded FPGA Platform for Convolutional Neural Network}.
\newblock {\em 2016 ACM/SIGDA International Symposium on Field-Programmable Gate Arrays\/} (2016).

\bibitem{Rahman2017}
{\sc Rahman, A., Oh, S., Lee, J., and Choi, K.}
\newblock {Design space exploration of FPGA accelerators for convolutional neural networks}.
\newblock In {\em Design, Automation \& Test in Europe Conference \& Exhibition (DATE)\/} (2017), pp.~1147--1152.

\bibitem{rathi2023acm}
{\sc Rathi, N., Chakraborty, I., Kosta, A., Sengupta, A., Ankit, A., Panda, P., and Roy, K.}
\newblock {Exploring Neuromorphic Computing Based on Spiking Neural Networks: Algorithms to Hardware}.
\newblock {\em ACM Comput. Surv. 55}, 12 (2023).

\bibitem{northpole}
{\sc Research, I.}
\newblock {A new chip architecture points to faster, more energy-efficient AI}.

\bibitem{reuther_hpec22}
{\sc Reuther, A., Michaleas, P., Jones, M., Gadepally, V., Samsi, S., and Kepner, J.}
\newblock {AI and ML Accelerator Survey and Trends}.
\newblock In {\em 2022 IEEE High Performance Extreme Computing Conference (HPEC)\/} (2022), pp.~1--10.

\bibitem{Robinson2022}
{\sc Robinson, C.}
\newblock {NVIDIA H100 Hopper Details at HC34 as it Waits for Next-Gen CPUs}.
\newblock \url{https://www.servethehome.com/nvidia-h100-hopper-details-at-hc34-as-it-waits-for-next-gen-cpus/}, Aug. 2022.
\newblock Accessed: 18/04/2023.

\bibitem{Rosenblatt1957}
{\sc Rosenblatt, F.}
\newblock {The perceptron - A perceiving and recognizing automaton}.
\newblock Tech. Rep. 85-460-1, Cornell Aeronautical Laboratory, Ithaca, New York, January 1957.

\bibitem{rossiVegaTenCoreSoC2022}
{\sc Rossi, D., Conti, F., Eggiman, M., Mauro, A.~D., Tagliavini, G., Mach, S., Guermandi, M., Pullini, A., Loi, I., Chen, J., Flamand, E., and Benini, L.}
\newblock {Vega: A Ten-Core SoC for IoT Endnodes With DNN Acceleration and Cognitive Wake-Up From MRAM-Based State-Retentive Sleep Mode}.
\newblock {\em IEEE Journal of Solid-State Circuits 57}, 1 (Jan. 2022), 127--139.

\bibitem{Roy_2021}
{\sc Roy, S., Ali, M., and Raghunathan, A.}
\newblock {PIM-DRAM: Accelerating Machine Learning Workloads Using Processing in Commodity DRAM}.
\newblock {\em IEEE Journal on Emerging and Selected Topics in Circuits and Systems 11}, 4 (2021), 701--710.

\bibitem{Sankaradas2009}
{\sc Sankaradas, M., Jakkula, V., Cadambi, S., Chakradhar, S., Durdanovic, I., Cosatto, E., and Graf, H.~P.}
\newblock {A Massively Parallel Coprocessor for Convolutional Neural Networks}.
\newblock In {\em 20th IEEE International Conference on Application-Specific Systems, Architectures and Processors\/} (USA, 2009), IEEE Computer Society, p.~53–60.

\bibitem{Schmidhuber14}
{\sc Schmidhuber, J.}
\newblock {Deep Learning in Neural Networks: An Overview}.
\newblock {\em CoRR abs/1404.7828\/} (2014).

\bibitem{Schmidhuber_2015}
{\sc Schmidhuber, J.}
\newblock {Deep learning in neural networks: An overview}.
\newblock {\em Neural Networks 61\/} (2015), 85--117.

\bibitem{seo2011cicc}
{\sc Seo, J.-s., Brezzo, B., Liu, Y., Parker, B.~D., Esser, S.~K., Montoye, R.~K., Rajendran, B., Tierno, J.~A., Chang, L., Modha, D.~S., and Friedman, D.~J.}
\newblock {A 45nm CMOS neuromorphic chip with a scalable architecture for learning in networks of spiking neurons}.
\newblock In {\em {2011 IEEE Custom Integrated Circuits Conference (CICC)}\/} (2011), pp.~1--4.

\bibitem{Seshadri_2013}
{\sc Seshadri, V., Kim, Y., Fallin, C., Lee, D., Ausavarungnirun, R., Pekhimenko, G., Luo, Y., Mutlu, O., Gibbons, P.~B., Kozuch, M.~A., and Mowry, T.~C.}
\newblock {RowClone: Fast and energy-efficient in-DRAM bulk data copy and initialization}.
\newblock In {\em 2013 46th Annual IEEE/ACM International Symposium on Microarchitecture (MICRO)\/} (2013), pp.~185--197.

\bibitem{Seshadri_2017}
{\sc Seshadri, V., Lee, D., Mullins, T., Hassan, H., Boroumand, A., Kim, J., Kozuch, M.~A., Mutlu, O., Gibbons, P.~B., and Mowry, T.~C.}
\newblock {Ambit: In-Memory Accelerator for Bulk Bitwise Operations Using Commodity DRAM Technology}.
\newblock In {\em 2017 50th Annual IEEE/ACM International Symposium on Microarchitecture (MICRO)\/} (2017), pp.~273--287.

\bibitem{Sestito2021}
{\sc Sestito, C., Spagnolo, F., and Perri, S.}
\newblock {Design of Flexible Hardware Accelerators for Image Convolutions and Transposed Convolutions}.
\newblock {\em Journal of Imaging 7}, 10 (2021).

\bibitem{shafiee2016isca}
{\sc Shafiee, A., Nag, A., Muralimanohar, N., Balasubramonian, R., Strachan, J.~P., Hu, M., Williams, R.~S., and Srikumar, V.}
\newblock {ISAAC: A Convolutional Neural Network Accelerator with in-Situ Analog Arithmetic in Crossbars}.
\newblock In {\em {43rd International Symposium on Computer Architecture}\/} (2016), p.~14–26.

\bibitem{Junnan2019}
{\sc Shan, J., Casu, M.~R., Cortadella, J., Lavagno, L., and Lazarescu, M.~T.}
\newblock {Exact and Heuristic Allocation of MuIti-kernel Applications to Multi-FPGA Platforms}.
\newblock In {\em 2019 56th ACM/IEEE Design Automation Conference (DAC)\/} (2019), pp.~1--6.

\bibitem{shao_micro19}
{\sc Shao, Y.~S., Clemons, J., Venkatesan, R., Zimmer, B., Fojtik, M., Jiang, N., Keller, B., Klinefelter, A., Pinckney, N., Raina, P., Tell, S.~G., Zhang, Y., Dally, W.~J., Emer, J., Gray, C.~T., Khailany, B., and Keckler, S.~W.}
\newblock {Simba: Scaling Deep-Learning Inference with Multi-Chip-Module-Based Architecture}.
\newblock In {\em 52nd Annual IEEE/ACM International Symposium on Microarchitecture\/} (Oct. 2019), Association for Computing Machinery, pp.~14--27.

\bibitem{sharma2022swap}
{\sc Sharma, H., Mandal, S.~K., Doppa, J.~R., Ogras, U.~Y., and Pande, P.~P.}
\newblock {SWAP: A Server-Scale Communication-Aware Chiplet-Based Manycore PIM Accelerator}.
\newblock {\em IEEE Transactions on Computer-Aided Design of Integrated Circuits and Systems 41}, 11 (2022), 4145--4156.

\bibitem{CatapultHLS2022}
{\sc {Siemens}}.
\newblock {\em {Catapult C++/Systemc Synthesis}}, 2022.

\bibitem{smagulova2023pieee}
{\sc Smagulova, K., Fouda, M.~E., Kurdahi, F., Salama, K.~N., and Eltawil, A.}
\newblock {Resistive Neural Hardware Accelerators}.
\newblock {\em Proceedings of the IEEE 111}, 5 (2023), 500--527.

\bibitem{smith_isscc24}
{\sc Smith, A., Chapman, E., Patel, C., Swaminathan, R., Wuu, J., Huang, T., Jung, W., Kaganov, A., McIntyre, H., and Mangaser, R.}
\newblock 11.1 amd instincttm mi300 series modular chiplet package – hpc and ai accelerator for exa-class systems.
\newblock In {\em 2024 IEEE International Solid-State Circuits Conference (ISSCC)\/} (2024), vol.~67, pp.~490--492.

\bibitem{snelgroveSpeedAI2402Petaflop30Teraflops2023}
{\sc Snelgrove, M., and Beachler, R.}
\newblock {{speedAI240}}: {{A}} 2-{{Petaflop}}, 30-{{Teraflops}}/{{W At-Memory Inference Acceleration Device With}} 1456 {{RISC-V Cores}}.
\newblock {\em IEEE Micro 43}, 3 (May 2023), 58--63.

\bibitem{Sohn2017}
{\sc Sohn, K., Yun, W.-J., Oh, R., Oh, C.-S., Seo, S.-Y., Park, M.-S., Shin, D.-H., Jung, W.-C., Shin, S.-H., Ryu, J.-M., Yu, H.-S., Jung, J.-H., Lee, H., Kang, S.-Y., Sohn, Y.-S., Choi, J.-H., Bae, Y.-C., Jang, S.-J., and Jin, G.}
\newblock {A 1.2 V 20 nm 307 GB/s HBM DRAM With At-Speed Wafer-Level IO Test Scheme and Adaptive Refresh Considering Temperature Distribution}.
\newblock {\em IEEE Journal of Solid-State Circuits 52}, 1 (2017), 250--260.

\bibitem{song2019isscc}
{\sc Song, J., Cho, Y., Park, J.-S., Jang, J.-W., Lee, S., Song, J.-H., Lee, J.-G., and Kang, I.}
\newblock {An 11.5TOPS/W 1024-MAC Butterfly Structure Dual-Core Sparsity-Aware Neural Processing Unit in 8nm Flagship Mobile SoC}.
\newblock In {\em {2019 IEEE International Solid- State Circuits Conference - (ISSCC)}\/} (2019), pp.~130--132.

\bibitem{song2017hpca}
{\sc Song, L., Qian, X., Li, H., and Chen, Y.}
\newblock {PipeLayer: A Pipelined ReRAM-Based Accelerator for Deep Learning}.
\newblock In {\em {2017 IEEE International Symposium on High Performance Computer Architecture (HPCA)}\/} (2017), pp.~541--552.

\bibitem{Soussan2008}
{\sc Soussan, P., Sabuncuoglu~Tezcan, D., Iker, F., Ruythooren, W., Swinnen, B., Majeed, B., and Beyne, E.}
\newblock {3D Wafer Level Packaging: Processes and Materials for Trough Silicon Vias \& Thin Die Embedding}.
\newblock In {\em MRS Online Proceedings Library\/} (01 2008), vol.~1112.

\bibitem{Spagnolo2022_3}
{\sc Spagnolo, F., Corsonello, P., Frustaci, F., and Perri, S.}
\newblock {Design of a Low-Power Super-Resolution Architecture for Virtual Reality Wearable Devices}.
\newblock {\em IEEE Sensors Journal 23}, 8 (2023), 9009--9016.

\bibitem{SPAGNOLO20201}
{\sc Spagnolo, F., Perri, S., and Corsonello, P.}
\newblock {Design of a real-time face detection architecture for heterogeneous systems-on-chips}.
\newblock {\em Integration 74\/} (2020), 1--10.

\bibitem{Spagnolo2022_1}
{\sc Spagnolo, F., Perri, S., and Corsonello, P.}
\newblock {Aggressive Approximation of the SoftMax Function for Power-Efficient Hardware Implementations}.
\newblock {\em IEEE Transactions on Circuits and Systems II: Express Briefs 69}, 3 (2022), 1652--1656.

\bibitem{Spagnolo2022_2}
{\sc Spagnolo, F., Perri, S., and Corsonello, P.}
\newblock {Approximate Down-Sampling Strategy for Power-Constrained Intelligent Systems}.
\newblock {\em IEEE Access 10\/} (2022), 7073--7081.

\bibitem{srinivas:2015}
{\sc Srinivas, S., and Babu, R.~V.}
\newblock {Data-free Parameter Pruning for Deep Neural Networks}.
\newblock {\em CoRR abs/1507.06149\/} (2015).

\bibitem{Sriam2011}
{\sc Sriram, V., Cox, D., Tsoi, K., and Luk, W.}
\newblock {Towards an embedded biologically-inspired machine vision processor}.
\newblock In {\em 2010 International Conference on Field-Programmable Technology\/} (01 2011), pp.~273--278.

\bibitem{stow_iccad16}
{\sc Stow, D., Akgun, I., Barnes, R., Gu, P., and Xie, Y.}
\newblock {Cost analysis and cost-driven IP reuse methodology for SoC design based on 2.5D/3D integration}.
\newblock In {\em 2016 IEEE/ACM International Conference on Computer-Aided Design (ICCAD)\/} (2016).

\bibitem{stow_iccad17}
{\sc Stow, D., Xie, Y., Siddiqua, T., and Loh, G.~H.}
\newblock {Cost-effective design of scalable high-performance systems using active and passive interposers}.
\newblock In {\em 2017 IEEE/ACM International Conference on Computer-Aided Design (ICCAD)\/} (2017), pp.~728--735.

\bibitem{Strollo_2020}
{\sc Strollo, A. G.~M., Napoli, E., De~Caro, D., Petra, N., and Di~Meo, G.}
\newblock {Comparison and Extension of Approximate 4-2 Compressors for Low-Power Approximate Multipliers}.
\newblock {\em IEEE Transactions on Circuits and Systems I: Regular Papers 67}, 9 (2020), 3021--3034.

\bibitem{Strollo_2022}
{\sc Strollo, A. G.~M., Napoli, E., De~Caro, D., Petra, N., Saggese, G., and Di~Meo, G.}
\newblock {Approximate Multipliers Using Static Segmentation: Error Analysis and Improvements}.
\newblock {\em IEEE Transactions on Circuits and Systems I: Regular Papers 69}, 6 (2022), 2449--2462.

\bibitem{Sze_2017}
{\sc Sze, V., Chen, Y.~H., Yang, T.~J., and Emer, J.~S.}
\newblock {Efficient Processing of Deep Neural Networks: A Tutorial and Survey}.
\newblock {\em Proceedings of the IEEE 105}, 12 (2017), 2295--2329.

\bibitem{Szegedy2015}
{\sc Szegedy, C., Liu, W., Jia, Y., Sermanet, P., Reed, S., Anguelov, D., Erhan, D., Vanhoucke, V., and Rabinovich, A.}
\newblock {Going deeper with convolutions}.
\newblock In {\em 2015 IEEE Conference on Computer Vision and Pattern Recognition (CVPR)\/} (2015), pp.~1--9.

\bibitem{talpes_micro23}
{\sc Talpes, E., Sarma, D.~D., Williams, D., Arora, S., Kunjan, T., Floering, B., Jalote, A., Hsiong, C., Poorna, C., Samant, V., Sicilia, J., Nivarti, A.~K., Ramachandran, R., Fischer, T., Herzberg, B., McGee, B., Venkataramanan, G., and Banon, P.}
\newblock The microarchitecture of dojo, tesla’s exa-scale computer.
\newblock {\em IEEE Micro 43}, 3 (2023), 31--39.

\bibitem{tambe2212nm182023}
{\sc Tambe, T., Zhang, J., Hooper, C., Jia, T., Whatmough, P.~N., Zuckerman, J., Santos, M. C.~D., Loscalzo, E.~J., Giri, D., Shepard, K., Carloni, L., Rush, A., Brooks, D., and Wei, G.-Y.}
\newblock {22.9 A 12nm 18.1TFLOPs/W Sparse Transformer Processor with Entropy-Based Early Exit, Mixed-Precision Predication and Fine-Grained Power Management}.
\newblock In {\em 2023 {{IEEE International Solid- State Circuits Conference}} ({{ISSCC}})\/} ({San Francisco, CA, USA}, Feb. 2023), {IEEE}, pp.~342--344.

\bibitem{tang_jssc24}
{\sc Tang, W., Cho, S.-G., Hoang, T.~T., Botimer, J., Zhu, W.~Q., Chang, C.-C., Lu, C.-H., Zhu, J., Tao, Y., Wei, T., Motwani, N.~K., Yalamanchi, M., Yarlagadda, R., Kale, S.~R., Flanigan, M., Chan, A., Tran, T., Shumarayev, S., and Zhang, Z.}
\newblock Arvon: A heterogeneous system-in-package integrating fpga and dsp chiplets for versatile workload acceleration.
\newblock {\em IEEE Journal of Solid-State Circuits 59}, 4 (2024), 1235--1245.

\bibitem{tangGeneralPurposeGraphConvolutionNeuralAccelerator2022}
{\sc Tang, W., and Zhang, P.}
\newblock {GPGCN: A General-Purpose Graph Convolution Neural Network Accelerator Based on RISC-V ISA Extension}.
\newblock {\em Electronics 11}, 22 (2022).

\bibitem{tortorellaRedMuleMixedPrecisionMatrixMatrix2023}
{\sc Tortorella, Y., Bertaccini, L., Benini, L., Rossi, D., and Conti, F.}
\newblock {RedMule: A Mixed-Precision Matrix-Matrix Operation Engine for Flexible and Energy-Efficient On-Chip Linear Algebra and TinyML Training Acceleration}.
\newblock {\em CoRR abs/2301.03904}, arXiv:2301.03904 (Jan. 2023).

\bibitem{tortorellaRedMulECompactFP162022a}
{\sc Tortorella, Y., Bertaccini, L., Rossi, D., Benini, L., and Conti, F.}
\newblock {RedMulE: A Compact FP16 Matrix-Multiplication Accelerator for Adaptive Deep Learning on RISC-V-based Ultra-Low-Power SoCs}.
\newblock In {\em 2022 {{Conference}} \& {{Exhibition}} on {{Design}}, {{Automation}} \& {{Test}} in {{Europe}}\/} (May 2022), {European Design and Automation Association}, pp.~1099--1102.

\bibitem{touvronLLaMAOpenEfficient2023}
{\sc Touvron, H., Lavril, T., Izacard, G., Martinet, X., Lachaux, M.-A., Lacroix, T., Rozi{\`e}re, B., Goyal, N., Hambro, E., Azhar, F., Rodriguez, A., Joulin, A., Grave, E., and Lample, G.}
\newblock {{LLaMA}}: {{Open}} and {{Efficient Foundation Language Models}}, Feb. 2023.

\bibitem{Ajili2022}
{\sc Trabelsi~Ajili, M., and Hara-Azumi, Y.}
\newblock {Multimodal Neural Network Acceleration on a Hybrid CPU-FPGA Architecture: A Case Study}.
\newblock {\em IEEE Access 10\/} (2022), 9603--9617.

\bibitem{Tsao_2012}
{\sc Tsao, Y.-C., and Choi, K.}
\newblock {Area-Efficient VLSI Implementation for Parallel Linear-Phase FIR Digital Filters of Odd Length Based on Fast FIR Algorithm}.
\newblock {\em IEEE Transactions on Circuits and Systems II: Express Briefs 59}, 6 (2012), 371--375.

\bibitem{Umuroglu2017}
{\sc Umuroglu, Y., Fraser, N.~J., Gambardella, G., Blott, M., Leong, P., Jahre, M., and Vissers, K.}
\newblock {FINN: A Framework for Fast, Scalable Binarized Neural Network Inference}.
\newblock In {\em 2017 ACM/SIGDA International Symposium on Field-Programmable Gate Arrays\/} (2017), ACM, p.~65–74.

\bibitem{upama2022}
{\sc Upama, P.~B., Faruk, M. J.~H., Nazim, M., Masum, M., Shahriar, H., Uddin, G., Barzanjeh, S., Ahamed, S.~I., and Rahman, A.}
\newblock {Evolution of Quantum Computing: A Systematic Survey on the Use of Quantum Computing Tools}, 2022.

\bibitem{Ushiroyama2022}
{\sc Ushiroyama, A., Watanabe, M., Watanabe, N., and Nagoya, A.}
\newblock {Convolutional neural network implementations using Vitis AI}.
\newblock In {\em 2022 IEEE 12th Annual Computing and Communication Workshop and Conference (CCWC)\/} (2022), pp.~0365--0371.

\bibitem{Vahdat_2019}
{\sc Vahdat, S., Kamal, M., Afzali-Kusha, A., and Pedram, M.}
\newblock {TOSAM: An Energy-Efficient Truncation- and Rounding-Based Scalable Approximate Multiplier}.
\newblock {\em IEEE Transactions on Very Large Scale Integration (VLSI) Systems 27}, 5 (2019), 1161--1173.

\bibitem{valenteHeterogeneousRISCVBased2024}
{\sc Valente, L., Nadalini, A., Veeran, A. H.~C., Sinigaglia, M., S{\'a}, B., Wistoff, N., Tortorella, Y., Benatti, S., Psiakis, R., Kulmala, A., Mohammad, B., Pinto, S., Palossi, D., Benini, L., and Rossi, D.}
\newblock A {{Heterogeneous RISC-V Based SoC}} for {{Secure Nano-UAV Navigation}}.
\newblock {\em IEEE Transactions on Circuits and Systems I: Regular Papers 71}, 5 (May 2024), 2266--2279.

\bibitem{Vandendriessche2022}
{\sc Vandendriessche, J., Da~Silva, B., and Touhafi, A.}
\newblock {Frequency Evaluation of the Xilinx DPU Towards Energy Efficiency}.
\newblock In {\em IECON 2022 – 48th Annual Conference of the IEEE Industrial Electronics Society\/} (2022), pp.~1--6.

\bibitem{vasiljevicComputeSubstrateSoftware2021}
{\sc Vasiljevic, J., Bajic, L., Capalija, D., Sokorac, S., Ignjatovic, D., Bajic, L., Trajkovic, M., Hamer, I., Matosevic, I., Cejkov, A., Aydonat, U., Zhou, T., Gilani, S.~Z., Paiva, A., Chu, J., Maksimovic, D., Chin, S.~A., Moudallal, Z., Rakhmati, A., Nijjar, S., Bhullar, A., Drazic, B., Lee, C., Sun, J., Kwong, K.-M., Connolly, J., Dooley, M., Farooq, H., Chen, J. Y.~T., Walker, M., Dabiri, K., Mabee, K., Lal, R.~S., Rajatheva, N., Retnamma, R., Karodi, S., Rosen, D., Munoz, E., Lewycky, A., Knezevic, A., Kim, R., Rui, A., Drouillard, A., and Thompson, D.}
\newblock {Compute Substrate for Software 2.0}.
\newblock {\em IEEE Micro 41}, 2 (Mar. 2021), 50--55.

\bibitem{vaswaniAttentionAllYou2017}
{\sc Vaswani, A., Shazeer, N., Parmar, N., Uszkoreit, J., Jones, L., Gomez, A.~N., Kaiser, {\L}., and Polosukhin, I.}
\newblock Attention is {{All}} you {{Need}}.
\newblock In {\em Advances in {{Neural Information Processing Systems}} 30}, I.~Guyon, U.~V. Luxburg, S.~Bengio, H.~Wallach, R.~Fergus, S.~Vishwanathan, and R.~Garnett, Eds. {Curran Associates, Inc.}, 2017, pp.~5998--6008.

\bibitem{Venieris2019}
{\sc Venieris, S.~I., and Bouganis, C.-S.}
\newblock {fpgaConvNet: Mapping Regular and Irregular Convolutional Neural Networks on FPGAs}.
\newblock {\em IEEE Transactions on Neural Networks and Learning Systems 30}, 2 (2019), 326--342.

\bibitem{venieris2018toolflows}
{\sc Venieris, S.~I., Kouris, A., and Bouganis, C.-S.}
\newblock {Toolflows for mapping convolutional neural networks on FPGAs: A survey and future directions}.
\newblock {\em ACM Computing Surveys (CSUR) 51}, 3 (2018), 1--39.

\bibitem{venkataramaniRaPiDAIAccelerator2021}
{\sc Venkataramani, S., Srinivasan, V., Wang, W., Sen, S., Zhang, J., Agrawal, A., Kar, M., Jain, S., Mannari, A., Tran, H., Li, Y., Ogawa, E., Ishizaki, K., Inoue, H., Schaal, M., Serrano, M., Choi, J., Sun, X., Wang, N., Chen, C.-Y., Allain, A., Bonano, J., Cao, N., Casatuta, R., Cohen, M., Fleischer, B., Guillorn, M., Haynie, H., Jung, J., Kang, M., Kim, K.-h., Koswatta, S., Lee, S., Lutz, M., Mueller, S., Oh, J., Ranjan, A., Ren, Z., Rider, S., Schelm, K., Scheuermann, M., Silberman, J., Yang, J., Zalani, V., Zhang, X., Zhou, C., Ziegler, M., Shah, V., Ohara, M., Lu, P.-F., Curran, B., Shukla, S., Chang, L., and Gopalakrishnan, K.}
\newblock {RaPiD: AI Accelerator for Ultra-low Precision Training and Inference}.
\newblock In {\em 2021 {{ACM}}/{{IEEE}} 48th {{Annual International Symposium}} on {{Computer Architecture}} ({{ISCA}})\/} (June 2021), pp.~153--166.

\bibitem{ventanaProduct}
{\sc {Ventana Micro}}.
\newblock \url{https://www.ventanamicro.com/}, 2023.
\newblock Accessed: 2023-04-18.

\bibitem{verhelst2022ml}
{\sc Verhelst, M., Shi, M., and Mei, L.}
\newblock {ML Processors Are Going Multi-Core: A performance dream or a scheduling nightmare?}
\newblock {\em IEEE Solid-State Circuits Magazine 14}, 4 (2022), 18--27.

\bibitem{vijayaraghavan_hpca17}
{\sc Vijayaraghavan, T., Eckert, Y., Loh, G.~H., Schulte, M.~J., Ignatowski, M., Beckmann, B.~M., Brantley, W.~C., Greathouse, J.~L., Huang, W., Karunanithi, A., Kayiran, O., Meswani, M., Paul, I., Poremba, M., Raasch, S., Reinhardt, S.~K., Sadowski, G., and Sridharan, V.}
\newblock {Design and Analysis of an APU for Exascale Computing}.
\newblock In {\em 2017 IEEE International Symposium on High Performance Computer Architecture (HPCA)\/} (2017), pp.~85--96.

\bibitem{vivet_jssc21}
{\sc Vivet, P., Guthmuller, E., Thonnart, Y., Pillonnet, G., Fuguet, C., Miro-Panades, I., Moritz, G., Durupt, J., Bernard, C., Varreau, D., Pontes, J., Thuries, S., Coriat, D., Harrand, M., Dutoit, D., Lattard, D., Arnaud, L., Charbonnier, J., Coudrain, P., Garnier, A., Berger, F., Gueugnot, A., Greiner, A., Meunier, Q.~L., Farcy, A., Arriordaz, A., Chéramy, S., and Clermidy, F.}
\newblock {IntAct: A 96-Core Processor With Six Chiplets 3D-Stacked on an Active Interposer With Distributed Interconnects and Integrated Power Management}.
\newblock {\em IEEE Journal of Solid-State Circuits 56}, 1 (2021), 79--97.

\bibitem{wan2022nature}
{\sc Wan, W., Kubendran, R., Schaefer, C., Eryilmaz, S.~B., Zhang, W., Wu, D., Deiss, S., Raina, P., Qian, H., Gao, B., Joshi, S., Wu, H., Wong, H.-S.~P., and Cauwenberghs, G.}
\newblock {A compute-in-memory chip based on resistive random-access memory}.
\newblock {\em Nature 608}, 7923 (Aug 2022), 504--512.

\bibitem{Wang_2022}
{\sc Wang, H., Xu, W., Zhang, Z., You, X., and Zhang, C.}
\newblock {An Efficient Stochastic Convolution Architecture Based on Fast FIR Algorithm}.
\newblock {\em IEEE Transactions on Circuits and Systems II: Express Briefs 69}, 3 (2022), 984--988.

\bibitem{Wang2021}
{\sc Wang, J., and Gu, S.}
\newblock {FPGA Implementation of Object Detection Accelerator Based on Vitis-AI}.
\newblock In {\em 2021 11th International Conference on Information Science and Technology (ICIST)\/} (2021), pp.~571--577.

\bibitem{Wang_2018}
{\sc Wang, J., Lin, J., and Wang, Z.}
\newblock {Efficient Hardware Architectures for Deep Convolutional Neural Network}.
\newblock {\em IEEE Transactions on Circuits and Systems I: Regular Papers 65}, 6 (2018), 1941--1953.

\bibitem{wangWinogradBasedConvolution2021}
{\sc Wang, S., Zhu, J., Wang, Q., He, C., and Ye, T.~T.}
\newblock {Customized Instruction on RISC-V for Winograd-Based Convolution Acceleration}.
\newblock In {\em 2021 IEEE 32nd International Conference on Application-specific Systems, Architectures and Processors (ASAP)\/} (2021), pp.~65--68.

\bibitem{Wang_2019}
{\sc Wang, Y., Lin, J., and Wang, Z.}
\newblock {FPAP: A Folded Architecture for Energy-Quality Scalable Convolutional Neural Networks}.
\newblock {\em IEEE Transactions on Circuits and Systems I: Regular Papers 66\/} (2019), 288--301.

\bibitem{wang30Vecim2892024}
{\sc Wang, Y., Yang, M., Lo, C.-P., and Kulkarni, J.~P.}
\newblock 30.6 {{Vecim}}: {{A}} 289.{{13GOPS}}/{{W RISC-V Vector Co-Processor}} with {{Compute-in-Memory Vector Register File}} for {{Efficient High-Performance Computing}}.
\newblock In {\em 2024 {{IEEE International Solid-State Circuits Conference}} ({{ISSCC}})\/} (San Francisco, CA, USA, Feb. 2024), IEEE, pp.~492--494.

\bibitem{ward-foxtonAxeleraDemosAI}
{\sc {Ward-Foxton}, S.}
\newblock {Axelera Demos AI Test Chip After Taping Out in Four Months}, 2022.

\bibitem{Warden2019}
{\sc Warden, P., and Situnayake, D.}
\newblock {\em {TinyML}}.
\newblock O'Reilly Media, Inc., 2019.

\bibitem{Xuechao2017}
{\sc Wei, X., Yu, C.~H., Zhang, P., Chen, Y., Wang, Y., Hu, H., Liang, Y., and Cong, J.}
\newblock {Automated systolic array architecture synthesis for high throughput CNN inference on FPGAs}.
\newblock In {\em 2017 54th ACM/EDAC/IEEE Design Automation Conference (DAC)\/} (2017), pp.~1--6.

\bibitem{wen:nisp2016}
{\sc Wen, W., Wu, C., Wang, Y., Chen, Y., and Li, H.}
\newblock {Learning Structured Sparsity in Deep Neural Networks}.
\newblock In {\em Proceedings of the 30th International Conference on Neural Information Processing Systems\/} (Red Hook, NY, USA, 2016), NIPS'16, Curran Associates Inc., pp.~2082--2090.

\bibitem{vitishls}
{\sc {Xilinx Inc.}}
\newblock {\em {Vitis High-Level Synthesis User Guide}}, 2022.

\bibitem{xuan_isocc22}
{\sc Xuan, Z.~Y., Lee, C.-J., and Yeh, T.~T.}
\newblock {Lego: Dynamic Tensor-Splitting Multi-Tenant DNN Models on Multi-Chip-Module Architecture}.
\newblock In {\em 2022 19th International SoC Design Conference (ISOCC)\/} (2022), pp.~173--174.

\bibitem{xue2019isscc}
{\sc Xue, C.-X., Chen, W.-H., Liu, J.-S., Li, J.-F., Lin, W.-Y., Lin, W.-E., Wang, J.-H., Wei, W.-C., Chang, T.-W., Chang, T.-C., Huang, T.-Y., Kao, H.-Y., Wei, S.-Y., Chiu, Y.-C., Lee, C.-Y., Lo, C.-C., King, Y.-C., Lin, C.-J., Liu, R.-S., Hsieh, C.-C., Tang, K.-T., and Chang, M.-F.}
\newblock {A 1Mb Multibit ReRAM Computing-In-Memory Macro with 14.6ns Parallel MAC Computing Time for CNN Based AI Edge Processors}.
\newblock In {\em {2019 IEEE International Solid- State Circuits Conference - (ISSCC)}\/} (2019), pp.~388--390.

\bibitem{Yang_2015}
{\sc Yang, Z., Han, J., and Lombardi, F.}
\newblock {Approximate compressors for error-resilient multiplier design}.
\newblock In {\em 2015 IEEE International Symposium on Defect and Fault Tolerance in VLSI and Nanotechnology Systems (DFTS)\/} (2015), pp.~183--186.

\bibitem{Yazdanbakhsh2018}
{\sc Yazdanbakhsh, A., Samadi, K., Kim, N.~S., Esmaeilzadeh, H., Falahati, H., and Wolfe, P.~J.}
\newblock {GANAX: A Unified MIMD-SIMD Acceleration for Generative Adversarial Networks}.
\newblock In {\em 2018 ACM/IEEE 45th Annual International Symposium on Computer Architecture (ISCA)\/} (2018), pp.~650--661.

\bibitem{scalehls2022dac}
{\sc Ye, H., Jun, H., Jeong, H., Neuendorffer, S., and Chen, D.}
\newblock {ScaleHLS: A Scalable High-Level Synthesis Framework with Multi-Level Transformations and Optimizations}.
\newblock In {\em Proceedings of the 59th ACM/IEEE Design Automation Conference (DAC)\/} (2022), pp.~1355--1358.

\bibitem{yiRDCIMRISCVSupported2024d}
{\sc Yi, W., Mo, K., Wang, W., Zhou, Y., Zeng, Y., Yuan, Z., Cheng, B., and Pan, B.}
\newblock {{RDCIM}}: {{RISC-V Supported Full-Digital Computing-in-Memory Processor With High Energy Efficiency}} and {{Low Area Overhead}}.
\newblock {\em IEEE Transactions on Circuits and Systems I: Regular Papers 71}, 4 (Apr. 2024), 1719--1732.

\bibitem{Zacharelos_2022}
{\sc Zacharelos, E., Nunziata, I., Saggese, G., Strollo, A.~G., and Napoli, E.}
\newblock {Approximate Recursive Multipliers Using Low Power Building Blocks}.
\newblock {\em IEEE Transactions on Emerging Topics in Computing 10}, 3 (2022), 1315--1330.

\bibitem{zarubaManticore4096CoreRISCV2021}
{\sc Zaruba, F., Schuiki, F., and Benini, L.}
\newblock {Manticore: A 4096-Core RISC-V Chiplet Architecture for Ultraefficient Floating-Point Computing}.
\newblock {\em IEEE Micro 41}, 2 (Mar. 2021), 36--42.

\bibitem{Zervakis_2016}
{\sc Zervakis, G., Tsoumanis, K., Xydis, S., Soudris, D., and Pekmestzi, K.}
\newblock {Design-Efficient Approximate Multiplication Circuits Through Partial Product Perforation}.
\newblock {\em IEEE Transactions on Very Large Scale Integration (VLSI) Systems 24}, 10 (2016), 3105--3117.

\bibitem{Zhang2016}
{\sc Zhang, C., Wu, D., Sun, J., Sun, G., Luo, G., and Cong, J.}
\newblock {Energy-Efficient CNN Implementation on a Deeply Pipelined FPGA Cluster}.
\newblock In {\em 2016 International Symposium on Low Power Electronics and Design\/} (2016), ACM, p.~326–331.

\bibitem{zhang2019vlsi}
{\sc Zhang, J.-F., Lee, C.-E., Liu, C., Shao, Y.~S., Keckler, S.~W., and Zhang, Z.}
\newblock {SNAP: A 1.67 — 21.55TOPS/W Sparse Neural Acceleration Processor for Unstructured Sparse Deep Neural Network Inference in 16nm CMOS}.
\newblock In {\em {2019 Symposium on VLSI Circuits}\/} (2019), pp.~C306--C307.

\bibitem{zhang2016micro}
{\sc Zhang, S., Du, Z., Zhang, L., Lan, H., Liu, S., Li, L., Guo, Q., Chen, T., and Chen, Y.}
\newblock {Cambricon-X: An accelerator for sparse neural networks}.
\newblock In {\em 2016 49th Annual IEEE/ACM International Symposium on Microarchitecture (MICRO)\/} (2016), MICRO-49, pp.~1--12.

\bibitem{zhang_apr15}
{\sc Zhang, X., Lin, J.~K., Wickramanayaka, S., Zhang, S., Weerasekera, R., Dutta, R., Chang, K.~F., Chui, K.-J., Li, H.~Y., Wee~Ho, D.~S., Ding, L., Katti, G., Bhattacharya, S., and Kwong, D.-L.}
\newblock {Heterogeneous 2.5D integration on through silicon interposer}.
\newblock {\em Applied Physics Reviews 2}, 2 (2015).

\bibitem{zhaoCambriconQHybridArchitecture2021}
{\sc Zhao, Y., Liu, C., Du, Z., Guo, Q., Hu, X., Zhuang, Y., Zhang, Z., Song, X., Li, W., Zhang, X., Li, L., Xu, Z., and Chen, T.}
\newblock {Cambricon-Q: A Hybrid Architecture for Efficient Training}.
\newblock In {\em 2021 {{ACM}}/{{IEEE}} 48th {{Annual International Symposium}} on {{Computer Architecture}} ({{ISCA}})\/} (June 2021), pp.~706--719.

\bibitem{zhouRISCVBasedFullyParallel2023}
{\sc Zhou, H., Hong, H., Liu, D., Liu, H., Xia, Y., Li, K., Liu, J., Luo, S., Mao, W., and Yu, H.}
\newblock {{RISC-V}} based {{Fully-Parallel SRAM Computing-in-Memory Accelerator}} with {{High Hardware Utilization}} and {{Data Reuse Rate}}.
\newblock In {\em 2023 {{IEEE}} 5th {{International Conference}} on {{Artificial Intelligence Circuits}} and {{Systems}} ({{AICAS}})\/} (June 2023), pp.~1--5.

\bibitem{zimmer_jssc20}
{\sc Zimmer, B., Venkatesan, R., Shao, Y.~S., Clemons, J., Fojtik, M., Jiang, N., Keller, B., Klinefelter, A., Pinckney, N., Raina, P., Tell, S.~G., Zhang, Y., Dally, W.~J., Emer, J.~S., Gray, C.~T., Keckler, S.~W., and Khailany, B.}
\newblock {A 0.32–128 TOPS, Scalable Multi-Chip-Module-Based Deep Neural Network Inference Accelerator With Ground-Referenced Signaling in 16 nm}.
\newblock {\em IEEE Journal of Solid-State Circuits 55}, 4 (2020), 920--932.

\end{thebibliography}

\end{document}